\documentclass[twocolumn,english,amsmath,amssymb,floats,prl,longbibliography]{revtex4-2}

\usepackage[utf8]{inputenc} 
\usepackage[dvipsnames]{xcolor}
\usepackage{graphicx}
\usepackage{wasysym}
\usepackage{wrapfig}
\usepackage{multirow,booktabs}
\usepackage{braket}
\usepackage{comment}
\usepackage{array}
\usepackage{soul}
\usepackage{subfigure}
\usepackage{float}
\usepackage{bm}
\usepackage{dsfont}
\usepackage{cellspace}
\usepackage{mathtools}
\usepackage{colortbl}
\usepackage[normalem]{ulem}
\usepackage{lipsum}
\usepackage{tikz}
\usetikzlibrary{matrix, backgrounds}

\usepackage[colorlinks,
            citecolor=blue,
            linkcolor=black,
            urlcolor=blue]{hyperref}

\newcommand{\mbfk}{\mathbf{k}}
\newcommand{\mbfq}{\mathbf{q}}
\newcommand{\iwm}{i\omega_m}

\newcommand{\iOn}{i\Omega_{n}}

\allowdisplaybreaks

\begin{document}

\title{Kohn-Luttinger Superconductivity of Weyl Fermi Arcs in PtBi$_2$}

\author{Reuel Dsouza}
\thanks{These two authors contributed equally.}


\author{Nikolaos Parthenios}
\thanks{These two authors contributed equally.}

\author{Brian M. Andersen}

\author{Morten H. Christensen}
\email{mchriste@nbi.ku.dk}
\affiliation{Niels Bohr Institute, University of Copenhagen, 2100 Copenhagen, Denmark}

\begin{abstract}
Recent experimental observations in the noncentrosymmetric Weyl semimetal PtBi$_2$ indicate unconventional superconductivity hosted by topological surface states---Weyl Fermi arcs---with a node at the center of each arc. Focusing on these Fermi arcs, we calculate the electronically mediated pairing interaction using a Kohn-Luttinger approach and find that, in a large region of the phase diagram, the leading superconducting instability has an $i$-wave symmetry featuring precisely such an intra-arc node. We study the dependence of the leading superconducting instabilities on electronic interaction parameters and chemical potential and show that the $i$-wave state is robust to changes in the model parameters. Our results provide a possible mechanism for the observation of topological $i$-wave superconductivity on the surface of PtBi$_2$ and may have implications for the broader landscape of superconducting instabilities arising from repulsive interactions on the surfaces of Weyl semimetals. 
\end{abstract}

\maketitle

Condensed matter systems provide a versatile platform for realizing exotic particles~\cite{Klitizing1980New,Tsui1982Two,Novoselov2005Two,Haldane1988Model,Semenoff1984Condensed,Laughlin1983Anomalous,Arovas1984Fractional,Castelnovo2008Magnetic,Fu2008Superconducting,Wan2011Topological,Lv2015Experimental,Xu2015Discovery}. The collective electronic behavior gives rise to emergent quasiparticles whose properties may differ substantially from the constituent electrons. One prominent example of this is the appearance of Weyl fermions---massless chiral particles---in three-dimensional solids lacking either inversion symmetry or time-reversal symmetry~\cite{Wan2011Topological,Morimoto2014Weyl,Lv2015Experimental,Xu2015Discovery,Burkov2011Weyl,Armitage2018Weyl}. Weyl fermions are the dominant low-energy excitations near Weyl points, band crossings in momentum space which act as a source or drain of Berry curvature. Consequently, Weyl points always appear in pairs of opposite chirality~\cite{Nielsen1981Absence}. The surface projections of these points are connected by Fermi arcs leading to exotic transport characteristics including unconventional quantum oscillations and the chiral anomaly-induced negative longitudinal magnetoresistance~\cite{Son2013Chiral,Zhang2016Signatures,Potter2014Quantum,Moll2016Transport}. If these states are rendered superconducting, either through the proximity effect or by an intrinsic instability of the system, the spectrum can host Majorana quasiparticles with possible relevance for topological quantum computation~\cite{Meng2012Weyl,Sato2017Topological,Teo2010Topological}. Intrinsic pairing of such chiral fermions constitutes an interesting fundamental question which is complicated by the fact that the density of states is suppressed near the Fermi level in Weyl semimetals. Hence, only a few electrons are able to form Cooper pairs. This is different on the surface, where the presence of Fermi arcs lead to a finite density of states. Surface superconductivity in these systems is therefore expected to be distinct from bulk superconductivity and possibly set in at a higher temperature~\cite{Nomani2023Intrinsic}. However, the nature of the surface superconducting state remains enigmatic.

Recent angle-resolved photoemission spectroscopy (ARPES) and scanning tunneling microscopy and spectroscopy (STM/STS) experiments on trigonal PtBi$_2$---a noncentrosymmetric type I Weyl semimetal---indicate a regime of surface superconductivity with $T_c\sim 5-45$~K, and a non-superconducting bulk~\cite{Schimmel2024Surface,Kuibarov2024Evidence,Changdar2025Topological,Besproswanny2025Temperature, Mathisen2026Fermiology}. Indeed, PtBi$_2$ appears to satisfy many of the requirements typically associated with high-temperature superconductivity~\cite{Kuibarov2025Three}. Curiously, STM and charge transport measurements indicate a much lower critical temperature for bulk superconductivity, $T_{\rm c}^{\rm bulk}\sim 0.5$~K~\cite{Shipunov2020Polymorphic, Zhang2025Atomic}. A Berezinskii-Kosterlitz-Thouless transition is observed in thin flakes~\cite{Veyrat2023Berezinskii} which persists upon increasing sample thickness, and STS measurements indicate that the surface properties of PtBi$_2$ are mainly governed by the Fermi arcs~\cite{Hoffmann2025Fermi}. The nature of the surface superconducting state remains controversial. One ARPES study finds no evidence for superconductivity in the Fermi arcs down to 3~K~\cite{Oleary2025Topography} while another study reports evidence for a highly nodal $i$-wave superconducting state at 10~K~\cite{Changdar2025Topological}. The role of the Fermi arc states in any purported superconducting pairing also remains an open question: it is not clear if the instability could be electronically mediated by the Weyl Fermi-arc states, phonon driven~\cite{Maeland2025Phonon}, or an intricate combination thereof~\cite{Maeland2025mechanism,Buccheri2026Phonon-driven}. The above dichotomy is not the only unresolved puzzle. The superconducting gap reported by surface-sensitive probes varies strongly in terms of reported homogeneity and amplitude~\cite{Schimmel2024Surface,Shipunov2020Polymorphic, Zhang2025Atomic,Jose2025Robust,Guo2025Topological,Huang2025Sizable}. In addition, the relationship between the surface superconducting state and the much lower-in-temperature bulk superconductivity remains unresolved and has only recently been theoretically explored~\cite{Nomani2023Intrinsic,Bai2025Superconductivity,Trama2025Self}. Beyond PtBi$_2$, surface superconductivity has also been reported in the Weyl semimetals MoTe$_2$~\cite{Naidyuk2018Surface} and TaIrTe$_4$~\cite{Xing2020Surface}.

In this Letter, we study superconducting pairing of Weyl Fermi arcs from the perspective of Kohn-Luttinger theory. Within this approach, superconductivity is driven purely by Coulomb repulsion, leading to momentum-dependent screened interactions that enable the formation of Cooper pairs in high angular-momentum channels~\cite{Kohn1965New,Maiti2013Superconductivity,Raghu2011, Tavakol2026Pairing}. Starting from a phenomenological patch description, we examine the landscape of potential pairing instabilities due to the different zero-momentum scattering processes. This approach---relevant for any noncentrosymmetric Weyl material where superconductivity emerges due to pairing in Weyl Fermi arcs---leads to a rich phase diagram exhibiting both nodal and nodeless states. We then perform a full Kohn-Luttinger calculation for the case of PtBi$_2$, based on a minimal model that captures the 12 bulk Weyl points and the Fermi arcs~\cite{Vocaturo2024Electronic}. We numerically solve the linearized gap equation and find a dominant nodal $i$-wave state for small values of the bare repulsive interaction. At larger values of the interaction, a fully gapped $s$-wave state emerges, and the two are separated by a narrow region featuring a nodal $s$-wave state. As the chemical potential is moved away from the Weyl point, the nodal $s$-wave becomes increasingly prominent.

\begin{figure}[t]
    \centering
    \includegraphics[width=\linewidth]{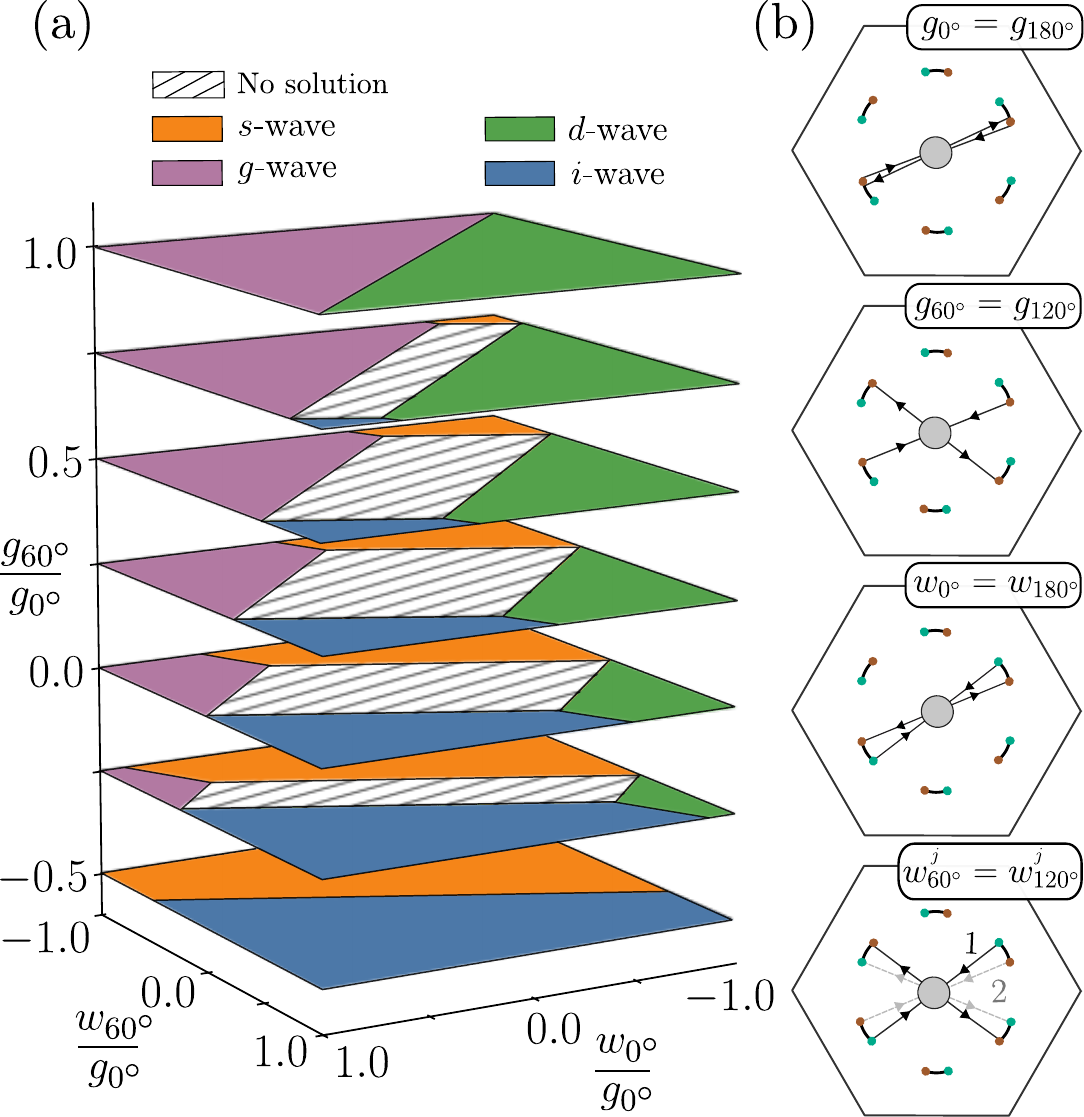}
    \caption{\textbf{Phenomenological model.} (a) Phase diagram of the 12-patch model assuming $g_{0^\circ}>0$. To visualize the parameter space we additionally set $w_{60^\circ}^1=w_{60^\circ}^2$. The region marked `No solution' is characterized by the absence of a positive eigenvalue in the 12-patch model. (b) Electronic interactions $g_\theta$ and $w_\theta$ of the 12-patch model. These correspond to intra- and inter-chirality scattering processes with fermionic indistinguishability taken into account. 
    }
    \label{fig:phase_space_of_solns}
\end{figure}

\textit{Phenomenological patch model.} Motivated by the experimental observation of superconducting Fermi arcs in PtBi$_2$, we consider a general phenomenological patch model as a starting point to establish the potential superconducting instabilities hosted by the surface states. Since each arc terminates at the surface projections of Weyl points with opposite chirality, we consider two patches per arc and label them by their corresponding Weyl point chirality in the bulk, leading to a total of twelve patches for the six Fermi arcs. As we are interested in studying instabilities in the superconducting channel, we consider scattering processes in the lab frame between an incoming pair of electrons with zero momentum since this can lead to a divergence in the Cooper channel~\cite{Cooper1956Bound}. This implies that the incoming pair constitutes an electronic state $|\mathbf{k}\rangle$ with momentum $\mathbf{k}$ and its time reversed partner $|-\mathbf{k}\rangle\equiv{\mathcal{T}}|\mathbf{k}\rangle$ with momentum $-\mathbf{k}$ and similarly for the outgoing pair with momenta $\mathbf{k}'$ and $-\mathbf{k}'$. 

For time-reversal symmetry preserving Weyl semimetals the chiralities satisfy $\nu(\mathbf{k}_0)=\nu(-\mathbf{k}_0)$~\cite{Vanderbilt2018Berry}, where $\mathbf{k}_0$ is the location of a Weyl point in the Brillouin zone (BZ). This leads to two classes of scattering processes. We label the first intra-chirality scattering and denote it $g_\theta$ corresponding to the case in which the outgoing pair has the same chirality as the incoming pair. The second is inter-chirality scattering, denoted by $w_\theta$, in which the outgoing pair has opposite chirality to the incoming pair. The subscript $\theta$ 
refers to the scattering angle between the incoming single electron state with momentum $\mathbf{k}$ and outgoing state with momentum $\mathbf{k'}$. For inter-chirality scattering $w_\theta$ the scattering angles are not exactly the subscript but in fact larger or smaller by an amount corresponding to the angle subtended by the arc to the $\Gamma$ point. This further splits $w_{60^\circ/120^\circ}$ into two cases labeled by the superscript $j$.

Having defined the relevant interactions $g_\theta$ and $w_\theta$ we construct a $12 \times 12$ pairing matrix that couples the pairing amplitudes on all patches via an eigenvalue equation
\begin{equation}
    -\begin{pmatrix}
        \mathbf{g} & \mathbf{w} \\
        \mathbf{w}^T & \mathbf{g}
    \end{pmatrix}\begin{pmatrix}
        \boldsymbol{\Delta}_+ \\
        \boldsymbol{\Delta}_-
    \end{pmatrix} = \lambda\begin{pmatrix}
        \boldsymbol{\Delta}_+ \\
        \boldsymbol{\Delta}_-
    \end{pmatrix}\,,\label{eq:patch_model_gap}
\end{equation}
where $\boldsymbol{\Delta}_{\pm}$ each has six components, the $\pm$ labeling the chirality. The above form is chosen to highlight the role of the intra- and inter-chirality interactions. The leading superconducting instability corresponds to the gap eigenvector $(\boldsymbol{\Delta}_+, \boldsymbol{\Delta}_-)^T$ with the largest positive eigenvalue $\lambda$. The matrix can be diagonalized exactly using properties of block circulant and symmetric matrices~\cite{SM}. The indistinguishability of fermions further requires that $(g,w)_{0^\circ}=(g,w)_{180^\circ}$ and $(g,w^{j})_{60^\circ}=(g,w^{j})_{120^\circ}$ which reduces the parameter space to a total of five parameters $g_{0^\circ},\,g_{60^\circ},\,w_{0^\circ}$ and $w^j_{60^\circ}$. The corresponding scattering processes are presented schematically in Fig.~\ref{fig:phase_space_of_solns}(b). Assuming $w^{1}_{60^{\circ}} = w^2_{60^{\circ}}$ and repulsive $g_{0^\circ}>0$, we obtain the three-dimensional phase diagram shown in Fig.~\ref{fig:phase_space_of_solns}(a), which shows all possible phases. If $g_{0^\circ}<0$ instead, the locations of the phases in Fig.~\ref{fig:phase_space_of_solns}(a) are inverted.

In the regime where the intra-chirality interactions are similar, $g_{60^\circ} / g_{0^\circ} \sim 1$, the phase diagram is dominated by the doubly degenerate $d$- and $g$-wave solutions. While Eq.~\eqref{eq:patch_model_gap} cannot pinpoint which linear combination condenses, it is likely that the chiral $d+id$ and $g+ig$ states are favored since these are the only combinations that gap out the Fermi surface~\cite{Sigrist1991Phenomenological,Nandkishore2012Chiral}. Reducing $g_{60^\circ}/g_{0^\circ}$ suppresses the $d$- and $g$-wave solutions and gives rise to both nodal $i$-wave and fully gapped $s$-wave pairing instabilities. This also creates a region where no solution exists, as the largest eigenvalue is not positive. The $i$-wave solution dominates the phase diagram when inter-chirality interactions $w_\theta$ are strongly repulsive. As we show below, we find this region to be the most relevant for the surface of PtBi$_2$. 

We can extend this procedure to three patches per Fermi arc by adding a central patch. This case is also amenable to an exact solution using the same methods as for the 12-patch model. Crucially, this finer resolution allows us to capture higher harmonics of the aforementioned states, such as a nodal $s$-wave solution that occupies the region labeled `No solution' in Fig.~\ref{fig:phase_space_of_solns}(a). However, such an extension does not lead to qualitative differences away from the phase boundaries. A detailed discussion is provided in the Supplementary Material (SM)~\cite{SM}.

As shown in Fig.~\ref{fig:phase_space_of_solns}(b), the odd-parity solutions are not favored in any region of parameter space. This is a direct consequence of fermionic indistinguishability in the time-reversed gauge which implies that the eigenvalues of the $p$-, $f$- and $h$-wave states in Eq.~\eqref{eq:patch_model_gap} all vanish~\cite{SM}. In fact, this conclusion is not specific to the patch model but holds for all noncentrosymmetric (intra-band) superconductors that preserve time-reversal symmetry. This is shown in Refs.~\cite{Samokhin2015Symmetry, Samokhin2017Noncentrosymmetric1D, Samokhin2008Gap3D} which demonstrate that, in noncentrosymmetric systems with time-reversal symmetry, the intra-band gap function is of even parity due to fermionic anti-commutativity on pairing time-reversed states. We further remark that in such systems the electronic bands are not spin degenerate, unlike the case of centrosymmetric superconductors with time-reversal symmetry. The existence of odd-parity gap solutions in centrosymmetric systems is facilitated by inter-spin-band pairing since the bands are Kramers degenerate~\cite{Anderson1984Structure}. 

Finally, we stress that the so-called singlet-triplet mixing frequently discussed in the noncentrosymmetric superconductivity literature does not imply that the intra-band gap function is a linear combination of even- and odd-parity components. Rather, the apparent mixing emerges entirely from the spin structure of the two-particle states during scattering. When an intra-band Cooper pair scatters from $\mathbf{k}$ to $\mathbf{k}'$, spin-orbit coupling in the inversion broken system enforces spin-momentum locking that necessitates a change in the spin amplitude of the pair. Therefore, singlet-triplet mixing in noncentrosymmetric systems is largely a kinematic consequence of projecting a helical pair onto a non-conserved spin basis, rather than a true mixing of parity-distinct gap functions~\cite{Samokhin2015Symmetry, Sigrist2009Introduction,Scheurer2016Mechanism}. 

The preceding discussion does not rely on any specific pairing mechanism and thus provides a general framework to model the intrinsic pairing on the Fermi arcs. In what follows, we will use the Kohn-Luttinger framework to compute the effective interactions in the Cooper channel from a microscopic model.

\textit{Electronically mediated pairing.}
\begin{figure}[t]
    \centering
    \includegraphics[width=\linewidth]{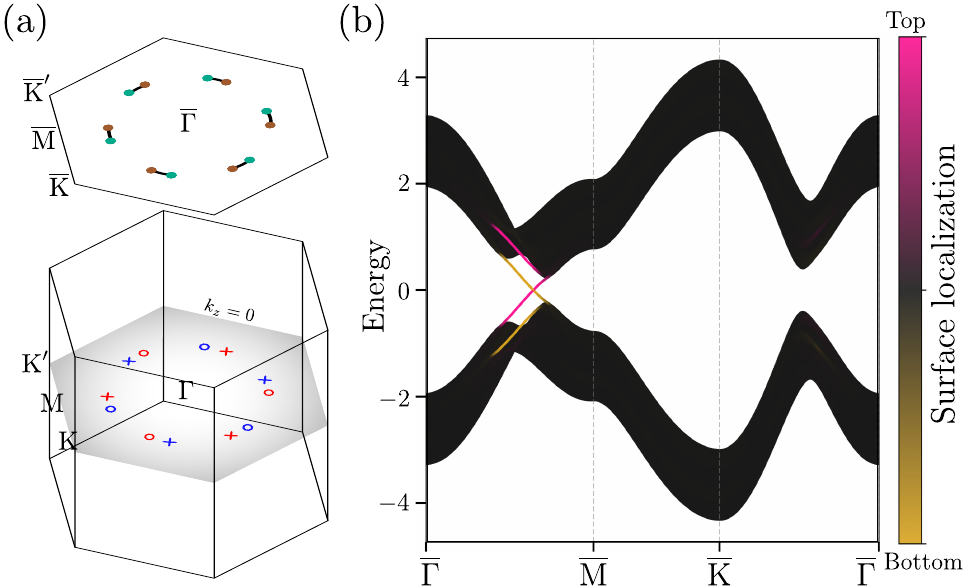}
    \caption{
    {\textbf{Electronic structure of minimal model.}
    (a) $E=0$ Fermi surfaces for the bulk/surface Brillouin zone (BZ). Bulk: circles/crosses denote Weyl points above/below the $k_z=0$ plane with blue being positive chirality and red negative chirality. Surface: Fermi arcs on the surface BZ with the projected Weyl points colored green and brown. (c) Band structure for the slab geometry along the $\overline{\Gamma}-\overline{\text{M}}-\overline{\text{K}}-\overline{\Gamma}$ path.}
    }
    \label{fig:Weyl_Fermi_arc_band_struct}
\end{figure}
For the microscopic calculation, we model the single particle dispersion by adopting an effective model for PtBi$_2$ with two spinful orbitals on the same site \cite{Vocaturo2024Electronic}. The Bloch Hamiltonian is invariant under the $C_{3v}$ point group, preserves time-reversal symmetry (TRS) and is given by
\begin{align}
     \widehat{H}_{\rm bulk}=\sum_{\substack{\mathbf{k}}}(h_0(\mathbf{k})+\alpha h_1(\mathbf{k}) + \gamma\tau_x \otimes \sigma_0)\mathbf{c}^\dagger_{\mathbf{k}}\mathbf{c}^{\phantom{\dagger}}_{\mathbf{k}}\,,
    \label{eq:WSM effective} 
\end{align}
where the spinor $\mathbf{c}_{\mathbf{k}}=(c_{\mathbf{k}A\uparrow},c_{\mathbf{k}A\downarrow},c_{\mathbf{k}B\uparrow},c_{\mathbf{k}B\downarrow})^T$ with $c_{\mathbf{k}\mu\sigma}$ ($c^{\dagger}_{\mathbf{k}\mu\sigma}$) annihilating (creating) an electron with momentum $\mathbf{k}$, orbital $\mu  \in\{A,B\}$ and spin $\sigma \in \{\uparrow,\downarrow\}$. The three terms $h_0(\mathbf{k})$, $h_1(\mathbf{k})$ and $\gamma\tau_x\otimes\sigma_0$, describe electron hopping, spin-orbit coupling (SOC) and inversion symmetry breaking respectively. 
For more details on the tight-binding model, see the End Matter. Henceforth, all energy parameters are measured in units of the in-plane, intra-orbital hopping which is set to unity.

To capture the Fermi arcs shown in Fig.~\ref{fig:Weyl_Fermi_arc_band_struct}(a), we diagonalize Eq.~\eqref{eq:WSM effective} for a finite number of layers $L$ along the $\hat{z}-$direction and focus on the electronic states of the topmost layer at the Fermi level. The electronic structure of this slab construction is shown in Fig.~\ref{fig:Weyl_Fermi_arc_band_struct}(b) with colors indicating the surface localization. We note that the Fermi arcs occur in $C_6$ symmetric locations as seen in Fig.~\ref{fig:Weyl_Fermi_arc_band_struct}(a). This is due to  an effective $C_6$ symmetry for scalar quantities on the surface BZ via the combination of $C_{3v}$ and TRS in 2D. We note that this is not true for scalars in the bulk BZ \cite{SM}. We model interactions on the bare level by considering the Hubbard-Hund-Kanamori Hamiltonian~\cite{Georges2013Strong} 
\begin{align}
    \widehat{H}_{\rm int} & = U \sum_{\mu}n_{\mu\uparrow}n_{\mu\downarrow} + {(U-2J)}\sum_{\mu\neq \nu}n_{\mu\uparrow}n_{\nu\downarrow}\nonumber\\ &+ {(U-3J)}\sum_{\mu<\nu,\sigma}n_{\mu\sigma}n_{\nu\sigma}  
      - {J}\sum_{\mu\neq \nu}c^\dagger_{\uparrow \mu}c_{\downarrow \mu}c^\dagger_{\downarrow \nu} c_{\uparrow \nu}\nonumber\\ & + {J}\sum_{\mu\neq \nu}c^\dagger_{\uparrow \mu}c^\dagger_{\downarrow \mu}c_{\downarrow \nu} c_{\uparrow \nu}\,,
     \label{eq:HK}
\end{align}
parametrized in terms of the on-site Hubbard repulsion $U$ and the exchange coupling $J$. We restrict their values such that $J/U<1/3$ in order for the interactions to be purely repulsive at the bare level. 

\begin{figure}[t]
    \centering
    \includegraphics[width=\linewidth]{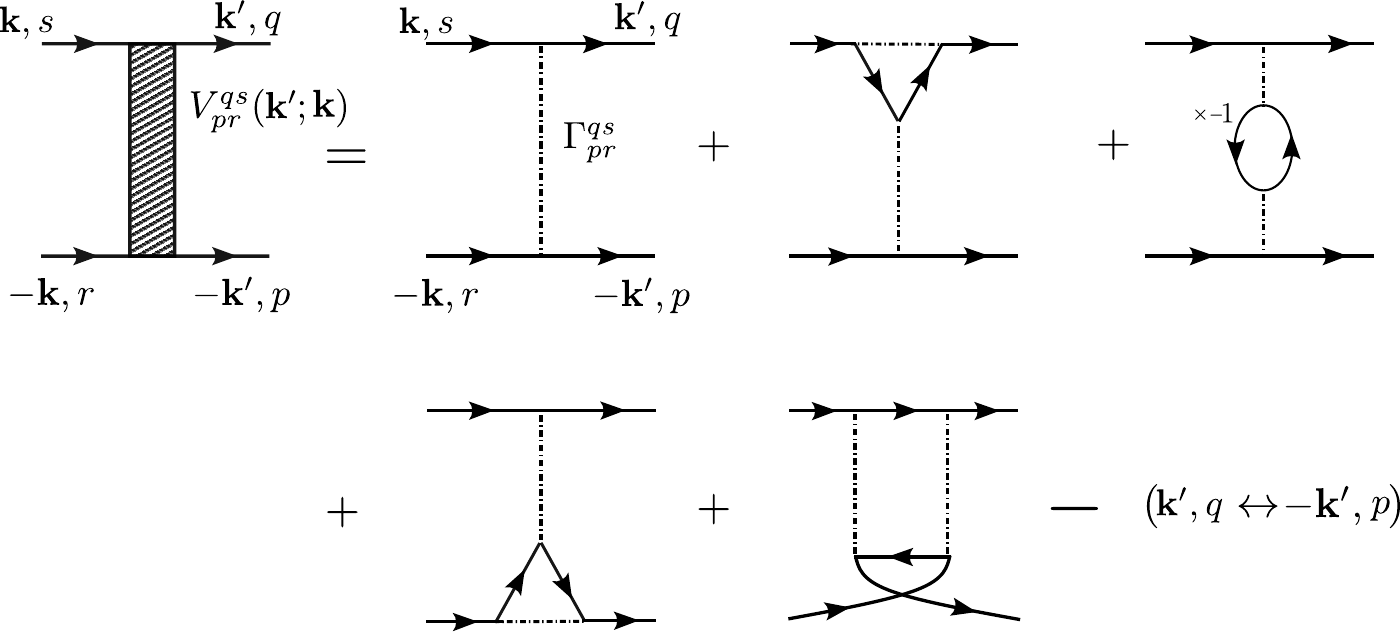}
    \caption{\textbf{Effective pairing interaction.} The Kohn-Luttinger diagrammatic contributions to the effective pairing vertex ${V}^{qs}_{pr}(\mathbf{k'};\mathbf{k})$. $\Gamma^{qs}_{pr}$ labels the bare, purely repulsive components of the Hubbard-Hund-Kanamori interaction, Eq~\eqref{eq:HK}. The indices $p,q,r,s$ are combined indices of both orbital and spin for electrons in the top layer.} 
    \label{fig:KL_diagrams}
\end{figure}

While the interaction is repulsive and momentum independent at the bare level, electronic fluctuations can induce non-trivial momentum dependence thereby potentially generating attractive components in higher angular-momentum pairing channels~\cite{Kohn1965New, Maiti2013Superconductivity}. We study the superconducting instabilities of the system arising from electronic fluctuations of the Fermi arc states.  Within a weak-coupling scenario, this can be captured by dressing the interaction in the particle-particle channel with particle-hole fluctuations \emph{via} the Kohn-Luttinger approach~\cite{Kohn1965New,Maiti2013Superconductivity}. To this end, we construct the effective pairing vertex ${V}^{qs}_{pr}(\mathbf{k}';\mathbf{k})$ by
\begin{equation}
    \widehat{H}_{\rm eff}
    = \sum_{\mathbf{k},\mathbf{k'}} \sum_{pqrs}
    {V}^{qs}_{pr}(\mathbf{k'}\,;\,\mathbf{k})\,
    c^\dagger_{-\mathbf{k'}p}
    c^\dagger_{\mathbf{k'}q}
    c_{\mathbf{k}s}
    c_{-\mathbf{k}r}\,,\label{eq:effective_interaction}
\end{equation}
using the diagrams in Fig.~\ref{fig:KL_diagrams}, consisting of four distinct second order diagrammatic contributions. The joint indices $p,q,r,s \in \{A,B\}\times\{\uparrow,\downarrow\}$ for electrons in the top layer. Further computational details are provided in the SM~\cite{SM}.

To study the momentum structure of the superconducting gap, we project the effective interaction vertex onto the states at the Fermi level, i.e., the Fermi arcs, and solve the linearized gap equation. This can be written as an eigenvalue problem when considering points along the Fermi surface,
\begin{align}
V_{\rm FS}(\mathbf{k}',\mathbf{k})
 &= \sum_{pqrs}
 \langle-\mathbf{k}'|_p
 \langle\mathbf{k}'|_q 
\,{V}^{qs}_{pr}(\mathbf{k}',\mathbf{k})\,
|-\mathbf{k}\rangle_r
 |\mathbf{k}\rangle_s,\\
\lambda\,\Delta(\mathbf{k}) &= -\frac{1}{{L}_{\rm FS}}\int_{\rm FS}\frac{{\rm d} {\mathbf{k}'}}{|v_{\rm F}(\mathbf{k'})|}\,{V}_{\rm FS}(\mathbf{k}',\mathbf{k})\,\Delta(\mathbf{k}')\,,
\end{align}
where $L_{\rm FS}$ is the length of a line-segment along the Fermi surface and $v_{\rm F}$ is the Fermi velocity. The symmetry of the superconducting state is obtained from the eigenvector associated to the leading eigenvalue.

\begin{figure}[t]
    \centering
        \includegraphics[width=\linewidth]{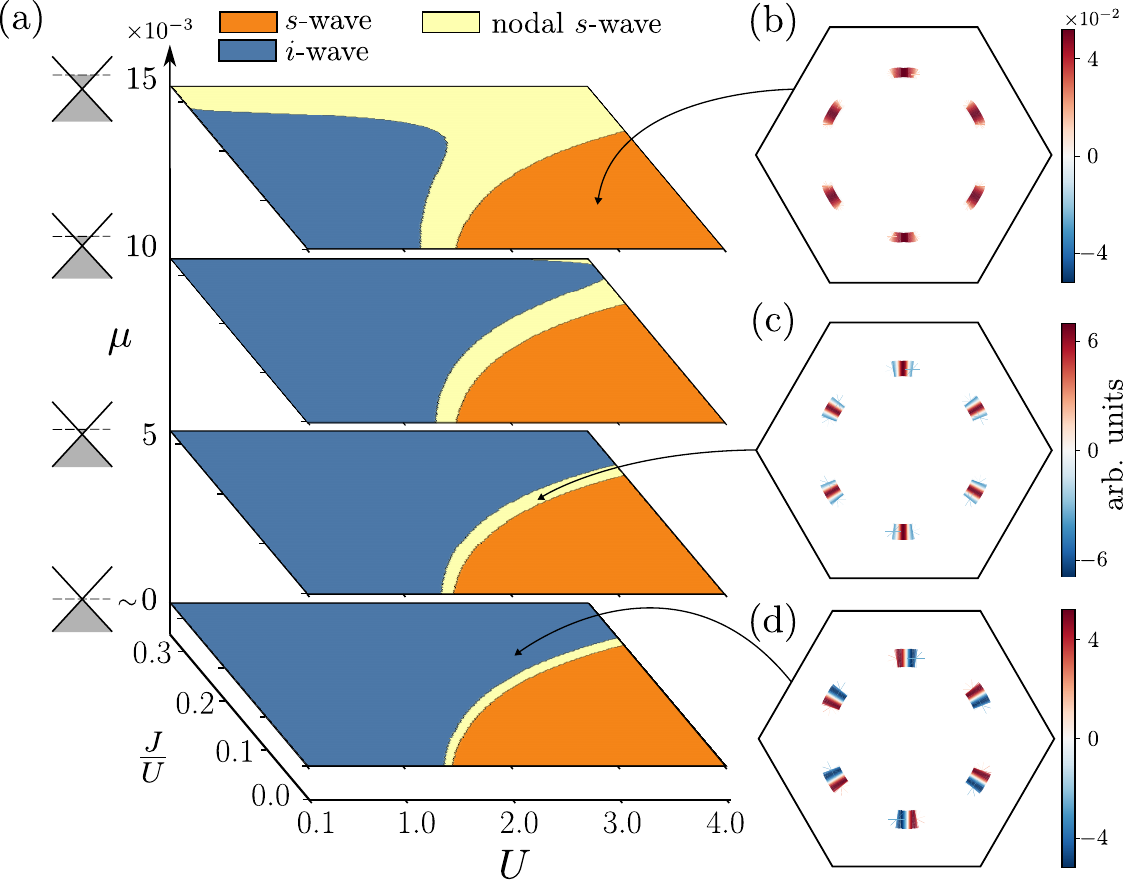}
    \caption{\label{fig:phase_diag}\textbf{Kohn-Luttinger phase diagram and superconducting order parameters.} (a) Symmetry of the superconducting order parameter as a function of $U$, $J/U$ and $\mu$, the chemical potential relative to the Weyl point. Near the Weyl point, the $i$-wave state is dominant, while the nodal $s$-wave state becomes increasingly favored as the Fermi level moves away from the Weyl point. Superconducting order parameters appearing in the phase diagram: (b) $s$-wave, (c) nodal $s$-wave, and (d) $i$-wave.
    }
\end{figure}

We evaluate $V_{\rm FS}(\mathbf{k},\mathbf{k}')$ for different values of the bare interactions $U$ and $J$, as well as different chemical potentials, $\mu$, and solve the corresponding linearized gap equation. This yields a phase diagram of the leading superconducting instability as a function of $U$, $J/U$, and $\mu$. Here, we focus on $\mu \geq 0$, as the negative case yields a nearly identical phase diagram~\cite{SM}. Within the Kohn-Luttinger scenario, the critical temperature, $T_c$, is typically quite small~\cite{Kohn1965New} and our method should not be understood as a quantitative prediction of $T_c$. At a chemical potential, $\mu=0$, corresponding to a Fermi level at the Weyl point, the leading superconducting instability is $i$-wave for $U<1.5$. This state exhibits a node at the midpoint of the Fermi arc, as shown in Fig.~\ref{fig:phase_diag}(d), consistent with recent ARPES observations~\cite{Changdar2025Topological}.

The $i$-wave state is accompanied by two distinct $s$-wave states. A fully gapped one and a nodal one. The fully gapped state appears at larger values of $U$ and is robust to changes in the chemical potential. In contrast, the nodal $s$-wave state, with two nodes per Fermi arc [see Fig.~\ref{fig:phase_diag}(c)], is very sensitive to the chemical potential. At $\mu=0$, it is represented by a small sliver but it grows rapidly as $\mu$ is increased. An example of the nodal $s$-wave state is shown in Fig.~\ref{fig:phase_diag}(c). The nodes appear in pairs symmetric about the midpoint of the arc, but we emphasize that their location is not constrained by symmetry and they can appear anywhere on the Fermi arc~\cite{SM}.

The nodal $s$-wave state is beyond the 12-patch model detailed above, but the extension to 18 patches captures the same phenomenology. As described above, within the 18-patch model, the gapped $s$-wave and the nodal $i$-wave state is separated by a region of nodal $s$-wave~\cite{SM}. Extracting the values for the phenomenological scattering amplitudes $g_{0^\circ},\,g_{60^\circ},\,w_{0^\circ}$ and $w^j_{60^\circ}$ of the patch model from the Kohn-Luttinger calculation, we find that the phase diagram from the microscopic calculation is quantitatively reproduced by the patch model calculation. This implies that the formation of the $i$-wave state is favored by repulsion between surface projections of Weyl points of opposite chirality. Finally, we checked how the phase diagram depends on SOC and the inversion symmetry-breaking parameter $\gamma$. For small variations, these simply lead to shifts of the phase boundaries but no qualitative changes of the phase diagram~\cite{SM}.

\textit{Discussion and conclusions.} We have demonstrated that the Kohn-Luttinger mechanism for electronically mediated superconductivity can give rise to $i$-wave superconductivity on the surface of a Weyl semimetal. Using a microscopic model which reproduces the six Fermi arcs on the surface of PtBi$_2$~\cite{Kuibarov2024Evidence,Oleary2025Topography} with prominent nesting at K as observed in quasiparticle interference experiments~\cite{Hoffmann2025Fermi}, we found this highly nodal state to be favored in a large region of the phase diagram, as shown in Fig.~\ref{fig:phase_diag}. A natural extension of our approach would be to adopt a more realistic tight-binding model obtained, e.g., from density-functional theory~\cite{Vocaturo2024Electronic}. This would address the interplay between surface and bulk states in the possible formation of a superconducting state, a question that is beyond the minimal model studied here. As the minimal model only includes a single site, it is also difficult to address how surface termination impacts superconductivity. However, as shown in Fig.~\ref{fig:phase_diag}, changes in the chemical potential can lead to changes in the superconducting ground state. Different surface terminations may lead to changes in the chemical potential, resulting in differences in both critical temperature and symmetry of the superconducting order parameter. Indeed, as demonstrated in Ref.~\cite{Vocaturo2026Engineering}, inclusion of a third nearest-neighbor hopping parameter can alter the shape of the Fermi arcs to mimic the one observed on the decorated honeycomb terminated surface. In this case, we find that a $g$-wave state becomes favored at high values of $J/U$~\cite{SM}. We note that differences in the symmetry of the order parameter can in principle be detected by application of a magnetic field~\cite{Waje2025Ginzburg-Landau}. In the case of $i$-wave superconductivity, the nodes are expected to gap out in an out-of-plane magnetic field while the nodes of a nodal $s$-wave state will persist~\cite{Waje2025Ginzburg-Landau}.

The microscopic calculations presented are further supported by a patch model approach, which shows that the $i$-wave state occupies a large region of the phase diagram, in particular when the effective interactions are repulsive, as shown in Fig.~\ref{fig:phase_space_of_solns}. Moreover, we find that projecting the microscopic dressed interactions onto the effective interactions of the patch model reproduces the phase diagram of Fig.~\ref{fig:phase_diag}. Our results have two important implications. First, the phase diagram of the patch model does not depend on any specific choice of pairing interaction and therefore evinces the variety of superconducting states that can appear on the surface of hexagonal noncentrosymmetric Weyl semimetals. These include the highly nodal topological $i$-wave and the chiral $g$- and $d$-wave states which could have implications for, e.g., the longitudinal magnetoresistance characteristic of these materials~\cite{Son2013Chiral}. Second, assuming an electronic pairing interaction, we find an $i$-wave superconducting state stabilized in a large region of the phase diagram. This is in agreement with recent ARPES observations on PtBi$_2$~\cite{Changdar2025Topological}. These results may have implications for the study and characterization of superconducting states on the surface of Weyl semimetals beyond PtBi$_2$.

\begin{acknowledgments}
We gratefully acknowledge D. F. Agterberg, J. Gondolf, P. J. Hirschfeld, A. Kreisel, and K. M{\ae}land for many helpful discussions.
R.D., N.P., and M.H.C. are supported by ERC grant project 101164202 -- SuperSOC. Funded by the European Union. Views and opinions expressed are however those of the authors only and do not necessarily reflect those of the European Union or the European Research Council Executive Agency. Neither the European Union nor the granting authority can be held responsible for them.
B.M.A. acknowledges support from the Independent Research Fund Denmark Grant No. 5241-00007B.
\end{acknowledgments}

\bibliographystyle{unsrtnat} 
\bibliography{weyl_sc}

@article{Anderson1984Structure,
  title = {Structure of ``triplet" superconducting energy gaps},
  author = {Anderson, P. W.},
  journal = {Phys. Rev. B},
  volume = {30},
  issue = {7},
  pages = {4000--4002},
  numpages = {0},
  year = {1984},
  month = {Oct},
  publisher = {American Physical Society},
  doi = {10.1103/PhysRevB.30.4000},
  url = {https://link.aps.org/doi/10.1103/PhysRevB.30.4000}
}

@article{Armitage2018Weyl,
  title = {{Weyl and Dirac semimetals in three-dimensional solids}},
  author = {Armitage, N. P. and Mele, E. J. and Vishwanath, Ashvin},
  journal = {Rev. Mod. Phys.},
  volume = {90},
  issue = {1},
  pages = {015001},
  numpages = {57},
  year = {2018},
  month = {Jan},
  publisher = {American Physical Society},
  doi = {10.1103/RevModPhys.90.015001},
  url = {https://link.aps.org/doi/10.1103/RevModPhys.90.015001}
}

@article{Arovas1984Fractional,
  title = {{Fractional Statistics and the Quantum Hall Effect}},
  author = {Arovas, Daniel and Schrieffer, J. R. and Wilczek, Frank},
  journal = {Phys. Rev. Lett.},
  volume = {53},
  issue = {7},
  pages = {722--723},
  numpages = {0},
  year = {1984},
  month = {Aug},
  publisher = {American Physical Society},
  doi = {10.1103/PhysRevLett.53.722},
  url = {https://link.aps.org/doi/10.1103/PhysRevLett.53.722}
}

@article{Bai2025Superconductivity,
	author = {Bai, Xuesong and LiMing, W. and Zhou, Tao},
	title = {{Superconductivity in Weyl semimetals with time reversal symmetry}},
	journal = {New J. Phys.},
	volume = {27},
	number = {1},
	pages = {013003},
	year = {2025},
	month = jan,
	issn = {1367-2630},
	publisher = {IOP Publishing},
	doi = {10.1088/1367-2630/ada574}
}

@ARTICLE{Besproswanny2025Temperature,
       author = {{Besproswanny}, Julia and {Schimmel}, Sebastian and {Fasano}, Yanina and {Shipunov}, Grigory and {Aswartham}, Saicharan and {Baumann}, Danny and {B{\"u}chner}, Bernd and {Hess}, Christian},
        title = {{Temperature dependence of surface superconductivity in t-PtBi$_2$}},
      journal = {arXiv:2507.10187},
         year = 2025,
        month = jul,
            url = {https://arxiv.org/abs/2507.10187},
}

@ARTICLE{Buccheri2026Phonon-driven,
       author = {{Buccheri}, Francesco and {de Martino}, Alessandro and {van den Brink}, Jeroen},
        title = {{Phonon-driven nodal surface superconductivity of Fermi arcs}},
      journal = {arXiv:2606.02371},
         year = 2026,
        month = jun,
        url = {https://arxiv.org/abs/2606.02371}
}

@article{Burkov2011Weyl,
  title = {{Weyl Semimetal in a Topological Insulator Multilayer}},
  author = {Burkov, A. A. and Balents, Leon},
  journal = {Phys. Rev. Lett.},
  volume = {107},
  issue = {12},
  pages = {127205},
  numpages = {4},
  year = {2011},
  month = {Sep},
  publisher = {American Physical Society},
  doi = {10.1103/PhysRevLett.107.127205},
  url = {https://link.aps.org/doi/10.1103/PhysRevLett.107.127205}
}

@article{Castelnovo2008Magnetic,
  title = {Magnetic monopoles in spin ice},
  volume = {451},
  ISSN = {1476-4687},
  url = {http://dx.doi.org/10.1038/nature06433},
  DOI = {10.1038/nature06433},
  number = {7174},
  journal = {Nature},
  publisher = {Springer Science and Business Media LLC},
  author = {Castelnovo,  C. and Moessner,  R. and Sondhi,  S. L.},
  year = {2008},
  month = Jan,
  pages = {42–45}
}

@article{Changdar2025Topological,
  title = {{Topological nodal i-wave superconductivity in PtBi$_2$}},
  volume = {647},
  ISSN = {1476-4687},
  url = {http://dx.doi.org/10.1038/s41586-025-09712-6},
  DOI = {10.1038/s41586-025-09712-6},
  number = {8090},
  journal = {Nature},
  publisher = {Springer Science and Business Media LLC},
  author = {Changdar,  Susmita and Suvorov,  Oleksandr and Kuibarov,  Andrii and Thirupathaiah,  Setti and Shipunov,  Grigory and Aswartham,  Saicharan and Wurmehl,  Sabine and Kovalchuk,  Iryna and Koepernik,  Klaus and Timm,  Carsten and B\"{u}chner,  Bernd and Fulga,  Ion Cosma and Borisenko,  Sergey and van den Brink,  Jeroen},
  year = {2025},
  month = nov,
  pages = {613--618}
}

@article{Cooper1956Bound,
  title = {{Bound Electron Pairs in a Degenerate Fermi Gas}},
  author = {Cooper, Leon N.},
  journal = {Phys. Rev.},
  volume = {104},
  issue = {4},
  pages = {1189--1190},
  numpages = {0},
  year = {1956},
  month = {Nov},
  publisher = {American Physical Society},
  doi = {10.1103/PhysRev.104.1189},
  url = {https://link.aps.org/doi/10.1103/PhysRev.104.1189}
}

@article{Fu2008Superconducting,
  title = {{Superconducting Proximity Effect and Majorana Fermions at the Surface of a Topological Insulator}},
  author = {Fu, Liang and Kane, C. L.},
  journal = {Phys. Rev. Lett.},
  volume = {100},
  issue = {9},
  pages = {096407},
  numpages = {4},
  year = {2008},
  month = {Mar},
  publisher = {American Physical Society},
  doi = {10.1103/PhysRevLett.100.096407},
  url = {https://link.aps.org/doi/10.1103/PhysRevLett.100.096407}
}

@article{Georges2013Strong,
author = {Georges, Antoine and de' Medici, Luca and Mravlje, Jernej},
title = {{Strong Correlations from Hund's Coupling}},
journal = {Annual Review of Condensed Matter Physics},
volume = {4},
number = {1},
pages = {137-178},
year = {2013},
doi = {10.1146/annurev-conmatphys-020911-125045},
URL = {https://doi.org/10.1146/annurev-conmatphys-020911-125045},
eprint = {https://doi.org/10.1146/annurev-conmatphys-020911-125045}
}

@ARTICLE{Guo2025Topological,
       author = {{Guo}, Yunkai and {Yan}, Jingming and {Dong}, Wen-Han and {Li}, Yongkai and {Peng}, Yucong and {Di}, Xuetao and {Li}, Caizhen and {Wang}, Zhiwei and {Xu}, Yong and {Tang}, Peizhe and {Yao}, Yugui and {Duan}, Wenhui and {Xue}, Qi-Kun and {Li}, Wei},
        title = {{Topological surface states in {\ensuremath{\gamma}}-PtBi$_2$ evidenced by scanning tunneling microscopy}},
      journal = {arXiv:2505.10808},
         year = 2025,
        month = may,
       url = {https://arxiv.org/abs/2505.10808},
}

@article{Haldane1988Model,
  title = {{Model for a Quantum Hall Effect without Landau Levels: Condensed-Matter Realization of the ``Parity Anomaly"}},
  author = {Haldane, F. D. M.},
  journal = {Phys. Rev. Lett.},
  volume = {61},
  issue = {18},
  pages = {2015--2018},
  numpages = {0},
  year = {1988},
  month = {Oct},
  publisher = {American Physical Society},
  doi = {10.1103/PhysRevLett.61.2015},
  url = {https://link.aps.org/doi/10.1103/PhysRevLett.61.2015}
}

@article{Hoffmann2025Fermi,
	author = {Hoffmann, Sven and Schimmel, Sebastian and Vocaturo, Riccardo and Puig, Joaquin and Shipunov, Grigory and Janson, Oleg and Aswartham, Saicharan and Baumann, Danny and B{\ifmmode\ddot{u}\else\"{u}\fi}chner, Bernd and van den Brink, Jeroen and Fasano, Yanina and Facio, Jorge I. and Hess, Christian},
	title = {{Fermi Arcs Dominating the Electronic Surface Properties of Trigonal PtBi2}},
	journal = {Adv. Phys. Res.},
	volume = {4},
	number = {5},
	pages = {2400150},
	year = {2025},
	month = may,
	issn = {2751-1200},
	publisher = {John Wiley {\&} Sons, Ltd},
	doi = {10.1002/apxr.202400150}
}

@article{Huang2025Sizable,
      title={{Sizable superconducting gap and anisotropic chiral topological superconductivity in the Weyl semimetal PtBi$_2$}}, 
      author={Xiaochun Huang and Lingxiao Zhao and Sebastian Schimmel and Julia Besproswanny and Patrick Härtl and Christian Hess and Bernd Büchner and Matthias Bode},
      journal = {arXiv:2507.13843},
      year={2025},
      url={https://arxiv.org/abs/2507.13843}, 
}

@ARTICLE{Jose2025Robust,
       author = {{Moreno}, Jose Antonio and {Garc{\'\i}a Talavera}, Pablo and {Herrera}, Edwin and {L{\'o}pez Valle}, Sara and {Li}, Zhuoqi and {Wang}, Lin-Lin and {Bud'ko}, Sergey and {Buzdin}, Alexander I. and {Guillam{\'o}n}, Isabel and {Canfield}, Paul C. and {Suderow}, Hermann},
        title = {{Robust two-dimensional surface superconductivity and vortex lattice in the Weyl semimetal $\gamma$-PtBi$_2$}},
      journal = {arXiv:2508.04867},
         year = 2025,
        month = aug,
       url = {https://arxiv.org/abs/2508.04867v2},
}

@article{Klitizing1980New,
  title = {{New Method for High-Accuracy Determination of the Fine-Structure Constant Based on Quantized Hall Resistance}},
  author = {Klitzing, K. v. and Dorda, G. and Pepper, M.},
  journal = {Phys. Rev. Lett.},
  volume = {45},
  issue = {6},
  pages = {494--497},
  numpages = {0},
  year = {1980},
  month = {Aug},
  publisher = {American Physical Society},
  doi = {10.1103/PhysRevLett.45.494},
  url = {https://link.aps.org/doi/10.1103/PhysRevLett.45.494}
}

@article{Kohn1965New,
  title = {{New Mechanism for Superconductivity}},
  author = {Kohn, W. and Luttinger, J. M.},
  journal = {Phys. Rev. Lett.},
  volume = {15},
  issue = {12},
  pages = {524--526},
  numpages = {0},
  year = {1965},
  month = {Sep},
  publisher = {American Physical Society},
  doi = {10.1103/PhysRevLett.15.524},
  url = {https://link.aps.org/doi/10.1103/PhysRevLett.15.524}
}

@article{Kuibarov2024Evidence,
	author = {Kuibarov, Andrii and Suvorov, Oleksandr and Vocaturo, Riccardo and Fedorov, Alexander and Lou, Rui and Merkwitz, Luise and Voroshnin, Vladimir and Facio, Jorge I. and Koepernik, Klaus and Yaresko, Alexander and Shipunov, Grigory and Aswartham, Saicharan and Brink, Jeroen van den and B{\ifmmode\ddot{u}\else\"{u}\fi}chner, Bernd and Borisenko, Sergey},
	title = {{Evidence of superconducting Fermi arcs}},
	journal = {Nature},
	volume = {626},
	pages = {294--299},
	year = {2024},
	month = feb,
	issn = {1476-4687},
	publisher = {Nature Publishing Group},
	doi = {10.1038/s41586-023-06977-7}
}

@ARTICLE{Kuibarov2025Three,
       author = {{Kuibarov}, Andrii and {Changdar}, Susmita and {Vocaturo}, Riccardo and {Suvorov}, Oleksandr and {Fedorov}, Alexander and {Lou}, Rui and {Krivenkov}, Maxim and {Harnagea}, Luminita and {Wurmehl}, Sabine and {van den Brink}, Jeroen and {B{\"u}chner}, Bernd and {Borisenko}, Sergey},
        title = {{Three prerequisites for high-temperature superconductivity in t-PtBi$_2$}},
      journal = {arXiv:2509.02178},
         year = 2025,
        month = sep,
            url = {https://arxiv.org/abs/2509.02178}
}

@article{Laughlin1983Anomalous,
  title = {{Anomalous Quantum Hall Effect: An Incompressible Quantum Fluid with Fractionally Charged Excitations}},
  author = {Laughlin, R. B.},
  journal = {Phys. Rev. Lett.},
  volume = {50},
  issue = {18},
  pages = {1395--1398},
  numpages = {0},
  year = {1983},
  month = {May},
  publisher = {American Physical Society},
  doi = {10.1103/PhysRevLett.50.1395},
  url = {https://link.aps.org/doi/10.1103/PhysRevLett.50.1395}
}

@article{Lv2015Experimental,
  title = {{Experimental Discovery of Weyl Semimetal TaAs}},
  author = {Lv, B. Q. and Weng, H. M. and Fu, B. B. and Wang, X. P. and Miao, H. and Ma, J. and Richard, P. and Huang, X. C. and Zhao, L. X. and Chen, G. F. and Fang, Z. and Dai, X. and Qian, T. and Ding, H.},
  journal = {Phys. Rev. X},
  volume = {5},
  issue = {3},
  pages = {031013},
  numpages = {8},
  year = {2015},
  month = {Jul},
  publisher = {American Physical Society},
  doi = {10.1103/PhysRevX.5.031013},
  url = {https://link.aps.org/doi/10.1103/PhysRevX.5.031013}
}

@article{Maeland2025Phonon,
	author = {M{\ae}land, Kristian and Bahari, Masoud and Trauzettel, Bj{\"o}rn},
	title = {{Phonon-mediated intrinsic topological superconductivity in Fermi arcs}},
	journal = {Phys. Rev. B},
	volume = {112},
	number = {10},
	pages = {104507},
	year = {2025},
	month = sep,
	publisher = {American Physical Society},
	doi = {10.1103/47vs-qgzk}
}

@ARTICLE{Maeland2025mechanism,
       author = {{M{\ae}land}, Kristian and {Sangiovanni}, Giorgio and {Trauzettel}, Bj{\"o}rn},
        title = {{Mechanism for Nodal Topological Superconductivity on PtBi$_2$ Surface}},
      journal = {arXiv:2512.09994},
         year = {2025},
        month = dec,
         url={https://arxiv.org/abs/2512.09994}
}

@inproceedings{Maiti2013Superconductivity,
   title={Superconductivity from repulsive interaction},
   ISSN={0094-243X},
   url={http://dx.doi.org/10.1063/1.4818400},
   DOI={10.1063/1.4818400},
   booktitle={AIP Conference Proceedings},
   publisher={AIP},
   author={Maiti, Saurabh and Chubukov, Andrey V.},
   year={2013} 
}

@misc{Mathisen2026Fermiology,
      title={{Fermiology and spin polarization of topological surface states in PtBi$_2$}}, 
      author={Anders Christian Mathisen and Xin Liang Tan and Stefanie Suzanne Brinkman and Kristian M{\ae}land and Fabian G{\"o}hler and {\O}yvind Finnseth and Grigory Shipunov and Falk Pabst and Manuel Alonso Lemos and Balasubramanian Thiagarajan and Craig Polley and Bj{\"o}rn Trauzettel and Anna Isaeva and Jorge I. Facio and Hendrik Bentmann},
      year={2026},
      eprint={2607.01947},
      archivePrefix={arXiv},
      primaryClass={cond-mat.supr-con},
      url={https://arxiv.org/abs/2607.01947}, 
}

@article{Meng2012Weyl,
	author = {Meng, Tobias and Balents, Leon},
	title = {{Weyl superconductors}},
	journal = {Phys. Rev. B},
	volume = {86},
	number = {5},
	pages = {054504},
	year = {2012},
	month = aug,
	publisher = {American Physical Society},
	doi = {10.1103/PhysRevB.86.054504}
}

@article{Moll2016Transport,
  title = {{Transport evidence for Fermi-arc-mediated chirality transfer in the Dirac semimetal Cd$_3$As$_2$}},
  volume = {535},
  ISSN = {1476-4687},
  url = {http://dx.doi.org/10.1038/nature18276},
  DOI = {10.1038/nature18276},
  number = {7611},
  journal = {Nature},
  publisher = {Springer Science and Business Media LLC},
  author = {Moll,  Philip J. W. and Nair,  Nityan L. and Helm,  Toni and Potter,  Andrew C. and Kimchi,  Itamar and Vishwanath,  Ashvin and Analytis,  James G.},
  year = {2016},
  month = July,
  pages = {266–270}
}

@article{Morimoto2014Weyl,
  title = {{Weyl and Dirac semimetals with ${\mathbb{Z}}_{2}$ topological charge}},
  author = {Morimoto, Takahiro and Furusaki, Akira},
  journal = {Phys. Rev. B},
  volume = {89},
  issue = {23},
  pages = {235127},
  numpages = {13},
  year = {2014},
  month = {Jun},
  publisher = {American Physical Society},
  doi = {10.1103/PhysRevB.89.235127},
  url = {https://link.aps.org/doi/10.1103/PhysRevB.89.235127}
}

@article{Naidyuk2018Surface,
	author = {Naidyuk, Yurii and Kvitnitskaya, Oksana and Bashlakov, Dmytro and Aswartham, Saicharan and Morozov, Igor and Chernyavskii, Ivan and Fuchs, G{\ifmmode\ddot{u}\else\"{u}\fi}nter and Drechsler, Stefan-L{\ifmmode\ddot{u}\else\"{u}\fi}dwig and H{\ifmmode\ddot{u}\else\"{u}\fi}hne, Ruben and Nielsch, Kornelius and B{\ifmmode\ddot{u}\else\"{u}\fi}chner, Bernd and Efremov, Dmitriy},
	title = {{Surface superconductivity in the Weyl semimetal MoTe$_2$ detected by point contact spectroscopy}},
	journal = {2D Mater.},
	volume = {5},
	number = {4},
	pages = {045014},
	year = {2018},
	month = aug,
	issn = {2053-1583},
	publisher = {IOP Publishing},
	doi = {10.1088/2053-1583/aad3e2}
}

@article{Nandkishore2012Chiral,
author={Nandkishore, Rahul
and Levitov, L. S.
and Chubukov, A. V.},
title={{Chiral superconductivity from repulsive interactions in doped graphene}},
journal={Nature Physics},
year={2012},
month={Feb},
day={01},
volume={8},
number={2},
pages={158-163},
issn={1745-2481},
doi={10.1038/nphys2208},
url={https://doi.org/10.1038/nphys2208}
}

@article{Nielsen1981Absence,
title = {{Absence of neutrinos on a lattice: (II). Intuitive topological proof}},
journal = {Nuclear Physics B},
volume = {193},
number = {1},
pages = {173-194},
year = {1981},
issn = {0550-3213},
doi = {https://doi.org/10.1016/0550-3213(81)90524-1},
url = {https://www.sciencedirect.com/science/article/pii/0550321381905241},
author = {H.B. Nielsen and M. Ninomiya}
}

@article{Nomani2023Intrinsic,
	author = {Nomani, Aymen and Hosur, Pavan},
	title = {{Intrinsic surface superconducting instability in type-I Weyl semimetals}},
	journal = {Phys. Rev. B},
	volume = {108},
	number = {16},
	pages = {165144},
	year = {2023},
	month = oct,
	publisher = {American Physical Society},
	doi = {10.1103/PhysRevB.108.165144}
}

@article{Novoselov2005Two,
  title = {{Two-dimensional gas of massless Dirac fermions in graphene}},
  volume = {438},
  ISSN = {1476-4687},
  url = {http://dx.doi.org/10.1038/nature04233},
  DOI = {10.1038/nature04233},
  number = {7065},
  journal = {Nature},
  publisher = {Springer Science and Business Media LLC},
  author = {Novoselov,  K. S. and Geim,  A. K. and Morozov,  S. V. and Jiang,  D. and Katsnelson,  M. I. and Grigorieva,  I. V. and Dubonos,  S. V. and Firsov,  A. A.},
  year = {2005},
  month = Nov,
  pages = {197–200}
}

@article{Oleary2025Topography,
  title = {{Topography of Fermi arcs in $t$-PtBi$_2$ using high-resolution angle-resolved photoemission spectroscopy}},
  author = {O'Leary, Evan and Li, Zhuoqi and Wang, Lin-Lin and Schrunk, Benjamin and Eaton, Andrew and Canfield, Paul C. and Kaminski, Adam},
  journal = {Phys. Rev. B},
  volume = {112},
  issue = {8},
  pages = {085154},
  numpages = {8},
  year = {2025},
  month = {Aug},
  publisher = {American Physical Society},
  doi = {10.1103/n5pz-j2sl},
  url = {https://link.aps.org/doi/10.1103/n5pz-j2sl}
}

@article{Potter2014Quantum,
  title = {{Quantum oscillations from surface Fermi arcs in Weyl and Dirac semimetals}},
  volume = {5},
  ISSN = {2041-1723},
  url = {http://dx.doi.org/10.1038/ncomms6161},
  DOI = {10.1038/ncomms6161},
  number = {1},
  journal = {Nature Communications},
  publisher = {Springer Science and Business Media LLC},
  author = {Potter,  Andrew C. and Kimchi,  Itamar and Vishwanath,  Ashvin},
  year = {2014},
  month = Oct 
}

@article{Raghu2011,
  title = {Superconductivity from repulsive interactions in the two-dimensional electron gas},
  author = {Raghu, S. and Kivelson, S. A.},
  journal = {Phys. Rev. B},
  volume = {83},
  issue = {9},
  pages = {094518},
  numpages = {8},
  year = {2011},
  month = {Mar},
  publisher = {American Physical Society},
  doi = {10.1103/PhysRevB.83.094518},
  url = {https://link.aps.org/doi/10.1103/PhysRevB.83.094518}
}

@article{Samokhin2017Noncentrosymmetric1D,
  title = {Noncentrosymmetric superconductors in one dimension},
  author = {Samokhin, K. V.},
  journal = {Phys. Rev. B},
  volume = {95},
  issue = {6},
  pages = {064504},
  numpages = {7},
  year = {2017},
  month = {Feb},
  publisher = {American Physical Society},
  doi = {10.1103/PhysRevB.95.064504},
  url = {https://link.aps.org/doi/10.1103/PhysRevB.95.064504}
}

@article{Samokhin2015Symmetry,
  title = {Symmetry and topology of two-dimensional noncentrosymmetric superconductors},
  author = {Samokhin, K. V.},
  journal = {Phys. Rev. B},
  volume = {92},
  issue = {17},
  pages = {174517},
  numpages = {15},
  year = {2015},
  month = {Nov},
  publisher = {American Physical Society},
  doi = {10.1103/PhysRevB.92.174517},
  url = {https://link.aps.org/doi/10.1103/PhysRevB.92.174517}
}

@article{Samokhin2008Gap3D,
  title = {Gap structure in noncentrosymmetric superconductors},
  author = {Samokhin, K. V. and Mineev, V. P.},
  journal = {Phys. Rev. B},
  volume = {77},
  issue = {10},
  pages = {104520},
  numpages = {9},
  year = {2008},
  month = {Mar},
  publisher = {American Physical Society},
  doi = {10.1103/PhysRevB.77.104520},
  url = {https://link.aps.org/doi/10.1103/PhysRevB.77.104520}
}

@article{Sato2017Topological,
doi = {10.1088/1361-6633/aa6ac7},
url = {https://doi.org/10.1088/1361-6633/aa6ac7},
year = {2017},
month = {may},
publisher = {IOP Publishing},
volume = {80},
number = {7},
pages = {076501},
author = {Sato, Masatoshi and Ando, Yoichi},
title = {Topological superconductors: a review},
journal = {Reports on Progress in Physics},
}

@article{Scheurer2016Mechanism,
  title = {{Mechanism, time-reversal symmetry, and topology of superconductivity in noncentrosymmetric systems}},
  author = {Scheurer, M. S.},
  journal = {Phys. Rev. B},
  volume = {93},
  issue = {17},
  pages = {174509},
  numpages = {25},
  year = {2016},
  month = {May},
  publisher = {American Physical Society},
  doi = {10.1103/PhysRevB.93.174509},
  url = {https://link.aps.org/doi/10.1103/PhysRevB.93.174509}
}

@article{Schimmel2024Surface,
  title = {{Surface superconductivity in the topological Weyl semimetal t-PtBi$_2$}},
  volume = {15},
  ISSN = {2041-1723},
  url = {http://dx.doi.org/10.1038/s41467-024-54389-6},
  DOI = {10.1038/s41467-024-54389-6},
  number = {1},
  journal = {Nature Communications},
  publisher = {Springer Science and Business Media LLC},
  author = {Schimmel,  Sebastian and Fasano,  Yanina and Hoffmann,  Sven and Besproswanny,  Julia and Corredor Bohorquez,  Laura Teresa and Puig,  Joaquín and Elshalem,  Bat-Chen and Kalisky,  Beena and Shipunov,  Grigory and Baumann,  Danny and Aswartham,  Saicharan and B\"{u}chner,  Bernd and Hess,  Christian},
  year = {2024},
  month = nov 
}

@article{Semenoff1984Condensed,
  title = {{Condensed-Matter Simulation of a Three-Dimensional Anomaly}},
  author = {Semenoff, Gordon W.},
  journal = {Phys. Rev. Lett.},
  volume = {53},
  issue = {26},
  pages = {2449--2452},
  numpages = {0},
  year = {1984},
  month = {Dec},
  publisher = {American Physical Society},
  doi = {10.1103/PhysRevLett.53.2449},
  url = {https://link.aps.org/doi/10.1103/PhysRevLett.53.2449}
}

@article{Shipunov2020Polymorphic,
	author = {Shipunov, G. and Kovalchuk, I. and Piening, B. R. and Labracherie, V. and Veyrat, A. and Wolf, D. and Lubk, A. and Subakti, S. and Giraud, R. and Dufouleur, J. and Shokri, S. and Caglieris, F. and Hess, C. and Efremov, D. V. and B{\ifmmode\ddot{u}\else\"{u}\fi}chner, B. and Aswartham, S.},
	title = {{Polymorphic ${\mathrm{PtBi}}_{2}$: Growth, structure, and superconducting properties}},
	journal = {Phys. Rev. Mater.},
	volume = {4},
	number = {12},
	pages = {124202},
	year = {2020},
	month = dec,
	publisher = {American Physical Society},
	doi = {10.1103/PhysRevMaterials.4.124202}
}

@article{Sigrist1991Phenomenological,
  title = {Phenomenological theory of unconventional superconductivity},
  author = {Sigrist, Manfred and Ueda, Kazuo},
  journal = {Rev. Mod. Phys.},
  volume = {63},
  issue = {2},
  pages = {239--311},
  numpages = {0},
  year = {1991},
  month = {Apr},
  publisher = {American Physical Society},
  doi = {10.1103/RevModPhys.63.239},
  url = {https://link.aps.org/doi/10.1103/RevModPhys.63.239}
}

@article{Sigrist2009Introduction,
	title = {Introduction to unconventional superconductivity in non‐centrosymmetric metals},
	volume = {1162},
	issn = {0094-243X},
	url = {https://doi.org/10.1063/1.3225489},
	doi = {10.1063/1.3225489},
	number = {1},
	journal = {AIP Conference Proceedings},
	author = {Sigrist, Manfred},
	month = aug,
	year = {2009},
}

@misc{SM,
      title="{See Supplementary Material.}",
      author={},
      year={},
      eprint={},
      archivePrefix={},
      primaryClass={}
}

@article{Son2013Chiral,
  title = {{Chiral anomaly and classical negative magnetoresistance of Weyl metals}},
  author = {Son, D. T. and Spivak, B. Z.},
  journal = {Phys. Rev. B},
  volume = {88},
  issue = {10},
  pages = {104412},
  numpages = {4},
  year = {2013},
  month = {Sep},
  publisher = {American Physical Society},
  doi = {10.1103/PhysRevB.88.104412},
  url = {https://link.aps.org/doi/10.1103/PhysRevB.88.104412}
}

@article{Tavakol2026Pairing,
  title = {{Pairing around a single Dirac point: A unifying view of Kohn-Luttinger superconductivity in Chern bands, quarter metals, and topological surface states}},
  author = {Tavakol, Omid and Scaffidi, Thomas},
  journal = {Phys. Rev. B},
  volume = {113},
  issue = {14},
  pages = {144502},
  numpages = {29},
  year = {2026},
  month = {Apr},
  publisher = {American Physical Society},
  doi = {10.1103/1tg5-qhtf},
  url = {https://link.aps.org/doi/10.1103/1tg5-qhtf}
}

@article{Teo2010Topological,
  title = {Topological defects and gapless modes in insulators and superconductors},
  author = {Teo, Jeffrey C. Y. and Kane, C. L.},
  journal = {Phys. Rev. B},
  volume = {82},
  issue = {11},
  pages = {115120},
  numpages = {26},
  year = {2010},
  month = {Sep},
  publisher = {American Physical Society},
  doi = {10.1103/PhysRevB.82.115120},
  url = {https://link.aps.org/doi/10.1103/PhysRevB.82.115120}
}

@article{Trama2025Self,
	author = {Trama, Mattia and K{\ifmmode\ddot{o}\else\"{o}\fi}nye, Viktor and Fulga, Ion Cosma and van den Brink, Jeroen},
	title = {{Self-consistent surface superconductivity in time-reversal symmetric Weyl semimetals}},
	journal = {Phys. Rev. B},
	volume = {112},
	number = {6},
	pages = {064514},
	year = {2025},
	month = aug,
	publisher = {American Physical Society},
	doi = {10.1103/bdtb-mb8c}
}

@article{Tsui1982Two,
  title = {{Two-Dimensional Magnetotransport in the Extreme Quantum Limit}},
  author = {Tsui, D. C. and Stormer, H. L. and Gossard, A. C.},
  journal = {Phys. Rev. Lett.},
  volume = {48},
  issue = {22},
  pages = {1559--1562},
  numpages = {0},
  year = {1982},
  month = {May},
  publisher = {American Physical Society},
  doi = {10.1103/PhysRevLett.48.1559},
  url = {https://link.aps.org/doi/10.1103/PhysRevLett.48.1559}
}

@article{Veyrat2023Berezinskii,
    author = {Veyrat, Arthur and Labracherie, Valentin and Bashlakov, Dima L. and Caglieris, Federico and Facio, Jorge I. and Shipunov, Grigory and Charvin, Titouan and Acharya, Rohith and Naidyuk, Yurii and Giraud, Romain and van den Brink, Jeroen and B{\"u}chner, Bernd and Hess, Christian and Aswartham, Saicharan and Dufouleur, Joseph},
    title = {{Berezinskii–Kosterlitz–Thouless Transition in the Type-I Weyl Semimetal PtBi$_2$}},
    journal = {Nano Letters},
    volume = {23},
    number = {4},
    pages = {1229-1235},
    year = {2023},
    doi = {10.1021/acs.nanolett.2c04297},
        note ={PMID: 36720048},
    URL = { https://doi.org/10.1021/acs.nanolett.2c04297 },
    eprint = {https://doi.org/10.1021/acs.nanolett.2c04297}
}

@book{Vanderbilt2018Berry, place={Cambridge}, title={{Berry Phases in Electronic Structure Theory: Electric Polarization, Orbital Magnetization and Topological Insulators}}, publisher={Cambridge University Press}, author={Vanderbilt, David}, year={2018}}

@article{Vocaturo2024Electronic,
  title = {{Electronic structure of the surface-superconducting Weyl semimetal ${\mathrm{PtBi}}_{2}$}},
  author = {Vocaturo, Riccardo and Koepernik, Klaus and Facio, Jorge I. and Timm, Carsten and Fulga, Ion Cosma and Janson, Oleg and van den Brink, Jeroen},
  journal = {Phys. Rev. B},
  volume = {110},
  issue = {5},
  pages = {054504},
  numpages = {12},
  year = {2024},
  month = {Aug},
  publisher = {American Physical Society},
  doi = {10.1103/PhysRevB.110.054504},
  url = {https://link.aps.org/doi/10.1103/PhysRevB.110.054504}
}

@misc{Vocaturo2026Engineering,
      title={{Engineering superconductivity on the surface of Weyl semimetals}}, 
      author={Riccardo Vocaturo and Mattia Trama},
      year={2026},
      eprint={2604.26859},
      archivePrefix={arXiv},
      primaryClass={cond-mat.supr-con},
      url={https://arxiv.org/abs/2604.26859}, 
}

@article{Waje2025Ginzburg-Landau,
  title = {{Ginzburg-Landau theory for unconventional surface superconductivity in ${\mathrm{PtBi}}_{2}$}},
  author = {Waje, Harald and Jakubczyk, Fabian and van den Brink, Jeroen and Timm, Carsten},
  journal = {Phys. Rev. B},
  volume = {112},
  issue = {14},
  pages = {144519},
  numpages = {12},
  year = {2025},
  month = {Oct},
  publisher = {American Physical Society},
  doi = {10.1103/kkqg-ntcz},
  url = {https://link.aps.org/doi/10.1103/kkqg-ntcz}
}

@article{Wan2011Topological,
  title = {{Topological semimetal and Fermi-arc surface states in the electronic structure of pyrochlore iridates}},
  author = {Wan, Xiangang and Turner, Ari M. and Vishwanath, Ashvin and Savrasov, Sergey Y.},
  journal = {Phys. Rev. B},
  volume = {83},
  issue = {20},
  pages = {205101},
  numpages = {9},
  year = {2011},
  month = {May},
  publisher = {American Physical Society},
  doi = {10.1103/PhysRevB.83.205101},
  url = {https://link.aps.org/doi/10.1103/PhysRevB.83.205101}
}

@article{Xing2020Surface,
	author = {Xing, Ying and Shao, Zhibin and Ge, Jun and Luo, Jiawei and Wang, Jinhua and Zhu, Zengwei and Liu, Jun and Wang, Yong and Zhao, Zhiying and Yan, Jiaqiang and Mandrus, David and Yan, Binghai and Liu, Xiong-Jun and Pan, Minghu and Wang, Jian},
	title = {{Surface superconductivity in the type II Weyl semimetal TaIrTe$_4$}},
	journal = {Natl. Sci. Rev.},
	volume = {7},
	number = {3},
	pages = {579--587},
	year = {2020},
	month = mar,
	issn = {2095-5138},
	publisher = {Oxford Academic},
	doi = {10.1093/nsr/nwz204}
}

@article{Xu2015Discovery,
author = {Su-Yang Xu  and Ilya Belopolski  and Nasser Alidoust  and Madhab Neupane  and Guang Bian  and Chenglong Zhang  and Raman Sankar  and Guoqing Chang  and Zhujun Yuan  and Chi-Cheng Lee  and Shin-Ming Huang  and Hao Zheng  and Jie Ma  and Daniel S. Sanchez  and BaoKai Wang  and Arun Bansil  and Fangcheng Chou  and Pavel P. Shibayev  and Hsin Lin  and Shuang Jia  and M. Zahid Hasan },
title = {{Discovery of a Weyl fermion semimetal and topological Fermi arcs}},
journal = {Science},
volume = {349},
number = {6248},
pages = {613-617},
year = {2015},
doi = {10.1126/science.aaa9297},
URL = {https://www.science.org/doi/abs/10.1126/science.aaa9297}
}

@article{Zhang2025Atomic,
	author = {Zhang, Hao and Chen, Hui and Huang, Zichen and Wang, Zi-Ang and Han, Guangyuan and Ma, Ruisong and Zhu, Xiangde and Ning, Wei and Shen, Chengmin and Huan, Qing and Gao, Hong-Jun},
	title = {{Atomic Visualization of Bulk and Surface Superconductivity in Weyl Semimetal {$\gamma$}-PtBi$_2$}},
	journal = {Chin. Phys. Lett.},
	volume = {42},
	number = {12},
	pages = {120708},
	year = {2025},
	month = dec,
	issn = {0256-307X},
	publisher = {Chinese Physical Society and IOP Publishing Ltd},
	doi = {10.1088/0256-307X/42/12/120708}
}

@article{Zhang2016Signatures,
	author = {Zhang, Cheng-Long and Xu, Su-Yang and Belopolski, Ilya and Yuan, Zhujun and Lin, Ziquan and Tong, Bingbing and Bian, Guang and Alidoust, Nasser and Lee, Chi-Cheng and Huang, Shin-Ming and Chang, Tay-Rong and Chang, Guoqing and Hsu, Chuang-Han and Jeng, Horng-Tay and Neupane, Madhab and Sanchez, Daniel S. and Zheng, Hao and Wang, Junfeng and Lin, Hsin and Zhang, Chi and Lu, Hai-Zhou and Shen, Shun-Qing and Neupert, Titus and Zahid Hasan, M. and Jia, Shuang},
	title = {{Signatures of the Adler{\textendash}Bell{\textendash}Jackiw chiral anomaly in a Weyl fermion semimetal}},
	journal = {Nat. Commun.},
	volume = {7},
	number = {10735},
	pages = {10735},
	year = {2016},
	month = feb,
	issn = {2041-1723},
	publisher = {Nature Publishing Group},
	doi = {10.1038/ncomms10735}
}

\setcounter{equation}{0}
\renewcommand{\theequation}{E\arabic{equation}}

\section*{End Matter}
\begin{table}
    \centering
    \renewcommand{\arraystretch}{1.2} 
    \setlength{\tabcolsep}{6pt}       
    \begin{tabular}{cccccccc}
    \toprule
    $m$ & $\alpha$ & $\beta$ &
    $\gamma$ & $\lambda$  \\
    0.4 & -0.25 & -0.65 & -0.25 & 1.2 \\
    \bottomrule
    \end{tabular}
    \caption{\label{tab:parameters} Parameters used in the tight binding model in units of the in-plane intra-orbital hopping. The slab model used to construct the effective interaction consists of $L=10$ layers. Additionally, the static particle-hole susceptibility used in the Kohn-Luttinger diagrams was evaluated at a temperature of $T=10^{-3}$.}
\end{table}

\textit{Tight binding model for PtBi$_2$.} The minimal tight-binding model describing the bulk electronic structure consisting of 12 Weyl nodes, as constructed in Ref.~\cite{Vocaturo2024Electronic} is written as:
\begin{equation}
H(\mathbf{k}) =h_0(\mathbf{k})+\alpha\,h_1(\mathbf{k}) + \gamma\,\tau_x\otimes\sigma_0.
    \label{Weyl point effective TB model}
\end{equation}
Here, $\gamma \otimes \tau_x\sigma_0$ is the inversion breaking term, $h_0(\mathbf{k})$ captures the dispersion and hopping between orbitals, and $h_1(\mathbf{k})$ introduces SOC and is explicitly given by
\begin{align}
        h_0(\mathbf{k}) &= [m-\cos k_1-\cos k_2 -\cos (k_1 + k_2) +\beta\cos k_3]\Gamma_1 \nonumber\\
        & +\{\beta\sin k_3  + \lambda[\sin k_1+\sin k_2 -\sin(k_1+k_2)]\}\Gamma_3 \\
     h_1(\mathbf{k}) &= (1-\cos k_3)[\sin k_1 \Gamma_2 + \sin k_2\Gamma_{2,1}\nonumber\\
     &-\sin(k_1+k_2)\Gamma_{2,2}].
\end{align}
The matrices $\Gamma$ represent orbital $\tau$ and spin $\sigma$ degrees of freedom, defined as
\begin{align}
    \Gamma_1 &= \tau_z\sigma_0 , \qquad
    \Gamma_2 = \tau_x\sigma_x,\qquad
    \Gamma_3 = \tau_y\sigma_0 \quad\nonumber\\
    \Gamma_{2,j} &=\mathcal{C}_3^j\Gamma_2\mathcal{C}_3^{-j}\;\;\left(\text{where  }\mathcal{C}_3\equiv\tau_0\exp\left\{-i\frac{\pi}{3}\sigma_z\right\} \right)
\end{align}
with $\Gamma_1$ corresponding to intra-orbital hoppings, $\Gamma_3$ to inter-orbital hoppings, and $\Gamma_2$ along with $\Gamma_{2,j}$ represent the spin-orbit coupling.

The basis vectors are parameterized as $k_i  \equiv \mathbf{k}\cdot \mathbf{a}_i$, with the lattice vectors $\mathbf{a}_1=(0,\,1,\,0)$, $\mathbf{a}_2=(\sqrt{3}/2,\,-1/2,\,0)$, and $\mathbf{a}_3=(0,\,0,\,1)$. As mentioned in the main text, we take the in-plane intra-orbital hopping as a our energy scale and set it to unity. The parameter $m$ acts as the mass term, while $\beta$ is the magnitude of the out-of-plane hopping for both inter and intra-orbital processes. The parameter $\lambda$ splits the nodal lines into Dirac crossings, yielding $12$ in total (six each for $k_z=0$ and $k_z=\pi$). Spin-orbit coupling is introduced via $\alpha$, which preserves inversion and time-reversal symmetries but splits the crossings at $k_z=\pi$, resulting in six Dirac crossings located at $k_z=0$. The inversion symmetry-breaking parameter denoted $\gamma$ hybridizes the orbitals (via the $\tau_x\sigma_0$ term) and splits the six Dirac crossings into $12$ Distinct Weyl points. The slab model is constructed by Fourier transforming along $k_3$, $H(\mathbf{k}) =  H_{\text{layer}}(k_1, k_2) \,+\, e^{ik_3}T(k_1, k_2) \, +\, e^{-ik_3}T^\dagger(k_1, k_2)$, for $L$ layers in the standard tridiagonal block form.

We note that the representation of inversion symmetry is given by $\mathcal{P}=\tau_z\sigma_0$, time-reversal symmetry by $\mathcal{T}=i\tau_0\sigma_y\mathcal{K}$ and one of the mirror symmetries is represented by $\mathcal{M}_1=i\tau_0\sigma_x$, satisfying $\mathcal{M}_1^{-1}H(k_1,k_2,k_3)\mathcal{M}_1=H(k_2,-k_1-k_2,k_3)$. The other two mirror planes can be generated by applying the threefold rotation operators $\mathcal{C}_3$ and $\mathcal{C}_3^2$.

\clearpage 
\onecolumngrid 

\setcounter{secnumdepth}{3}
\begin{center}
  \textbf{\large Supplementary Material:\\ Kohn-Luttinger Superconductivity of Weyl Fermi Arcs in PtBi$_2$}\\[.2cm]
\end{center}

\setcounter{equation}{0}
\setcounter{figure}{0}
\setcounter{table}{0}
\setcounter{page}{1}

\renewcommand{\thefigure}{S\arabic{figure}}
\renewcommand{\theequation}{S\arabic{equation}}
\renewcommand{\thetable}{S\arabic{table}}
\renewcommand{\thesection}{S.\Roman{section}}

\section{Surface Brillouin zone symmetry for scalar quantities}\label{sec:effective_c6v}

While the bulk Hamiltonian preserves all symmetries of the $C_{3v}$ point group, it manifestly breaks $C_{6v}$ due to spin--orbit--coupling terms, as reflected in the $k_z$ offset of the Weyl points [Fig.\ 2(a), main text]. In contrast, the calculated surface-projected Fermi arcs display an apparent $C_{6v}$ symmetry. This effective $C_{6v}$ symmetry in the surface Brillouin zone ($s$BZ) arises from the combination of $C_{3v}$ point-group operations and time-reversal symmetry (TRS), once these symmetries are restricted to a two-dimensional manifold.

Consider an arbitrary scalar quantity $\Phi(\mathbf{k}_s)$ defined on the surface-projected Brillouin zone, $\mathbf{k}_s \in s\mathrm{BZ}$, such as the surface energy dispersion $E(\mathbf{k}_s)$. If the cleave plane is invariant under $C_{3v}$, then the surface inherits this symmetry because the point-group operations do not involve the $z$-direction. The group elements are
$C_{3v} = \{E,\, 2C_3,\, 3\sigma_v\}$, where the three mirror lines $\sigma_v$ lie along the $\overline{\rm K}-\overline{\Gamma}-\overline{\rm K}'$ path and the rotations are given by $C_{\frac{2\pi}{3}}$ and $C_{\frac{4\pi}{3}}$ .\\

\paragraph{Rotations:}
From $C_3$ symmetry $\Phi(\mathbf{k}_s) = \Phi(C_3 \mathbf{k}_s) \equiv \Phi(C_{\frac{2\pi}{3}}\mathbf{k}_s)$. TRS further implies $
\Phi(\mathbf{k}_s) = \Phi(-\mathbf{k}_s)$. In two dimensions, $-\mathbf{k}_s$ is equivalent to a $C_{2z}$ rotation, i.e.\ $C_{\pm\pi}\mathbf{k}_s$. Thus,
$
\Phi(\mathbf{k}_s) = \Phi(C_{\pm\pi}\mathbf{k}_s)
$. On combining these rotations
\begin{align}
\Phi(\mathbf{k}_s)
= \Phi(C_{\frac{2\pi}3}\mathbf{k}_s)
&= \Phi(C_{\pi}C_{\frac{2\pi}3}\mathbf{k}_s)
= \Phi(C_{-\frac\pi3}\mathbf{k}_s)
= \Phi(C^{-1}_6\mathbf{k}_s)
\end{align}

\paragraph{Mirrors:}
The mirror sector follows analogously. Each of the three mirrors $\sigma_v\in C_{3v}$ combines with TRS in 2D to generate three additional mirror operations $\sigma_d$ along the  $\overline{\rm M}-\overline{\Gamma}-\overline{\rm M}$ path, completing the group structure of $C_{6v}$ for scalar quantities $\Phi$ defined over the $s$BZ.

\subsection{Comment on TRS-Breaking Weyl Semimetals}
One might attempt to generalize the discussion above to Weyl semimetals that preserve bulk inversion symmetry but break TRS by replacing TRS with inversion in the discussion above. However, the key distinction here is that TRS is a \emph{local} symmetry, whereas inversion is a \emph{spatial} symmetry. The action of the respective symmetries on momentum and coordinates are given by
\begin{align}
    \mathcal{T}(\mathbf{k}_s, z) = (-\mathbf{k}_s,\, z)\quad\text{and}\quad \mathcal{P}(\mathbf{k}_s, z) = (-\mathbf{k}_s,\,-z).
\end{align}
Thus, inversion necessarily exchanges the two surfaces of a slab model, while TRS does not. For a scalar defined on the surface BZ,
\begin{equation}
\Phi(\mathbf{k}_s, z) = \Phi(-\mathbf{k}_s, -z).
\label{Eq:supplementary_inversion_scalar}
\end{equation}
Following the previous construction would imply that the $C_6$ related state lies on the opposite surface. Additionally, Eq.~\eqref{Eq:supplementary_inversion_scalar} implies that zero momentum pairing is generically suppressed on the Fermi arcs of TRS breaking Weyl semimetals (WSMs)~\cite{Sato2017Topological}.

\section{Patch model for zero momentum pairing on the Fermi arcs}

Here, we illustrate the exact solutions to the patch models mentioned in the main text. We begin with the eigenvalues and eigenvectors of a $6\times 6$ circulant matrix in \ref{supp:circulant_primer}. Next, we use these eigenvectors and eigenvalues to solve the 12-patch model in \ref{supp:12-patch}. Initially, we assume $w^1_{60^\circ}=w^2_{60^\circ}$ which can be interpreted as the \emph{narrow arc} limit. We also examine the consequences of relaxing this assumption $w^1_{60^\circ}\neq w^2_{60^\circ}$. Finally, we study the 18-patch extension capturing higher harmonics in \ref{supp:18-patch}. Additionally, we explicitly demonstrate the vanishing eigenvalue of odd-parity solutions due to the indistinguishability of fermions on pairing time-reversed states \cite{Samokhin2015Symmetry,Samokhin2017Noncentrosymmetric1D,Samokhin2008Gap3D} in these models.

\subsection{Eigenvalues and eigenvectors of a circulant symmetric \texorpdfstring{$6\times6$}{} matrix}\label{supp:circulant_primer}

A general $n\times n$ circulant matrix takes the form
\begin{equation}
\mathbf{C} =
    \begin{pmatrix}
    c_{0}   & c_{1}   & \cdots & c_{n-2} & c_{n-1} \\
    c_{n-1} & c_{0}   & \cdots & c_{n-3} & c_{n-2} \\
    \vdots  & \vdots  & \ddots & \vdots  & \vdots  \\
    c_{2}   & c_{3}   & \cdots & c_{0}   & c_{1}   \\
    c_{1}   & c_{2}   & \cdots & c_{n-1} & c_{0}
    \end{pmatrix} \equiv \textbf{circ}\big(c_{0}, c_{1}, \dots,  c_{n-2}, c_{n-1} \big).
\end{equation}
To solve the eigenvalue problem $\mathbf{C}u^{(l)}=\lambda_l u^{(l)}$ the matrix is diagonalized using the discrete Fourier transform matrix and we obtain
\begin{align}
    u^{(l)}= \frac{1}{\sqrt{n}} \bigl(1,\; \omega^{l},\;\omega^{2l},\;\dots,\; \omega^{(n-1)l} \;\bigr)^T,\qquad
    \lambda_l = \sum_{k=0}^{n-1} c_k\,\omega^{lk}.
    \label{supp:eq:circulant_eigsystem}
\end{align}
Here, $l = 0,1,\dots,n-1$ and $\omega=\exp\left({\frac{2\pi i}{n}}\right)$ is the primitive $n^{\rm th}$ root of unity. 

\begin{figure}[t]
    \centering
    \includegraphics[width = \linewidth]{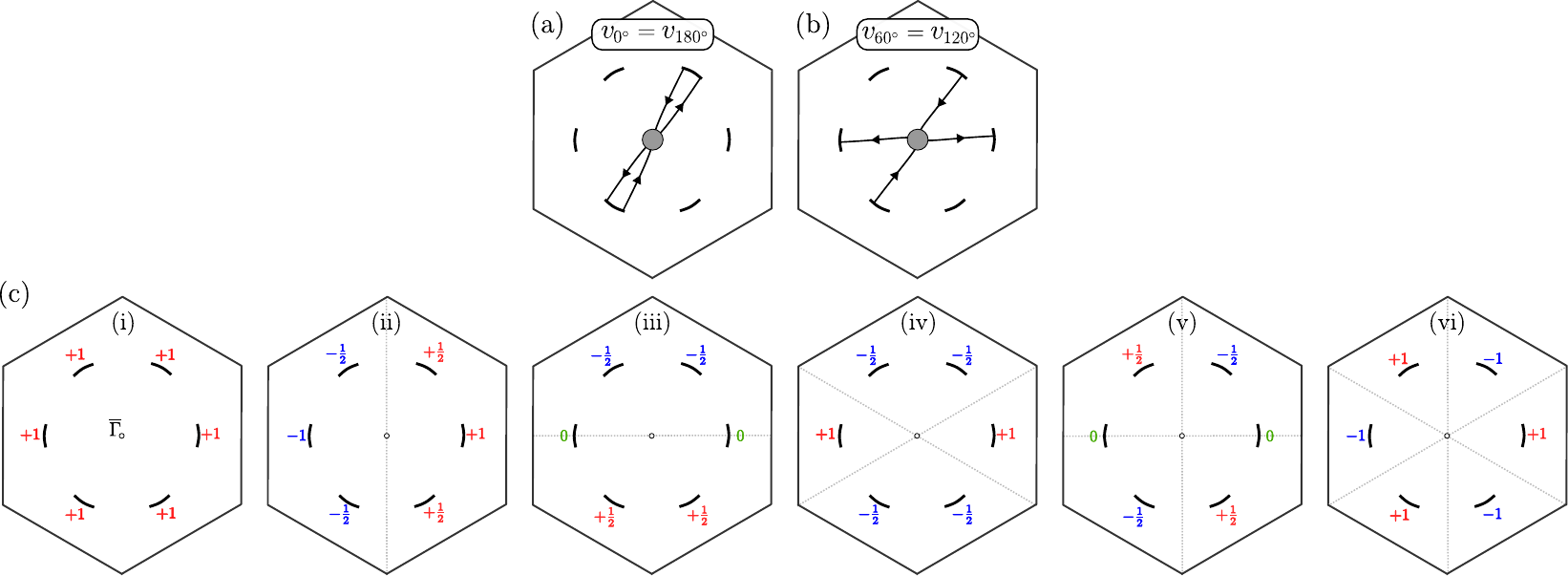}
    \caption{(a-b) Zero momentum scattering processes within the six-patch model. (c) Pictorial representation of the gap eigenvectors within the \textbf{6-patch model}: (i) $s$-wave; (ii)-(iii) $p$-wave; (iv)-(v) $d$-wave and (vi) $f$-wave solutions.}
    \label{supp:fig:6_patch_soln}
\end{figure}

We will employ this framework to construct $n$-patch models. Initially, we consider a 6-patch model which will then serve as a building block for all further generalizations. In this model, each Fermi arc is assigned a uniform gap value. Working in the time-reversed gauge implies that the Fermi surface-projected interaction is real and even in momenta (see Sec.~\ref{supp:TRS_gauge_section} for details) and so we have the 6-patch pairing matrix and gap equation given by
\begin{align}
    \mathbf{V}_{6\times6} = \textbf{circ}\big(v_{0^\circ},\,v_{60^\circ},\,v_{120^\circ},\,v_{180^\circ},\,v_{120^\circ},\,v_{60^\circ}\big),
    \qquad - \mathbf{V}_{6\times6}\mathbf{\Delta}^{(l)} = \lambda_l\,\mathbf{\Delta}^{(l)}.
    \label{supp:6_patch_eigsystem}
\end{align}
The interactions used in Eq.~\eqref{supp:6_patch_eigsystem} are depicted in Fig.~\ref{supp:fig:6_patch_soln}(a) and (b). The superconducting instability corresponds to the \emph{largest positive eigenvalue} of $-\mathbf{V}$.
The pairing matrix defined above is a circulant matrix that is real and symmetric since $v_{\theta}\in \mathbb{R}$ and $c_{i}=c_{n-i}$. Using Eq.~\eqref{supp:eq:circulant_eigsystem} the eigenvalues are real as well, i.e., $\lambda_l\in\mathbb{R}$. Note that the reality of the eigenvalues is independent of the gauge choice --- Hermitian circulant matrices $c_{i}^*=c_{n-i}$ have real eigenvalues. Defining $\omega=e^{i\pi/3}$ the eigensystem of the 6-patch model is given by
\begin{align}
\lambda_s &= - (v_{0^\circ} + v_{180^\circ}) - 2 (v_{60^\circ} + v_{120^\circ}) ,
& \mathbf{\Delta}_s &= \frac{1}{\sqrt{6}}(1, \, 1, \, 1, \, 1, \, 1, \, 1)^T, \\[4pt]\hline
\lambda_p &= \vphantom{\Bigg|} - (v_{0^\circ} - v_{180^\circ}) - (v_{60^\circ} - v_{120^\circ}),
& \mathbf{\Delta}_{p,1} &= \frac{1}{\sqrt{6}}(1, \, \omega, \, \omega^2, \, \omega^3, \, \omega^4, \, \omega^5)^T, \\[-2pt]
&& \mathbf{\Delta}_{p,2} &= \frac{1}{\sqrt{6}}(1, \, \omega^{-1}, \, \omega^{-2}, \, \omega^{-3}, \, \omega^{-4}, \, \omega^{-5})^T, \\[4pt]\hline
\lambda_d &= \vphantom{\Bigg|} - (v_{0^\circ} + v_{180^\circ}) + (v_{60^\circ} + v_{120^\circ}),
& \mathbf{\Delta}_{d,1} &= \frac{1}{\sqrt{6}}(1, \, \omega^2, \, \omega^4, \, 1, \, \omega^2, \, \omega^4)^T, \\[-2pt]
&& \mathbf{\Delta}_{d,2} &= \frac{1}{\sqrt{6}}(1, \, \omega^{-2}, \, \omega^{-4}, \, 1, \, \omega^{-2}, \, \omega^{-4})^T, \\[4pt]\hline
\lambda_f &= \vphantom{\Bigg|} - (v_{0^\circ} - v_{180^\circ}) + 2 (v_{60^\circ} - v_{120^\circ}),
& \mathbf{\Delta}_f &= \frac{1}{\sqrt{6}}(1, \, -1, \, 1, \, -1, \, 1, \, -1)^T.
\end{align}
Here, the labels $s$, $p$, $d$, and $f$ are alluding to the similarity between the symmetries of the eigenstates and those of the angular harmonics. For both the $p$- and $d$-wave solutions we have a two-fold degeneracy. This implies that we can construct real valued linear combinations within the respective eigenspaces: 
    \begin{align}
        \begin{array}{c|c}
            \begin{aligned}
                \mathbf{\Delta}_{p,\cos} &= \frac{1}{\sqrt{3}}\left(1,\,\tfrac12,\,-\tfrac12,\,-1,\,-\tfrac12,\,\tfrac12\right)^T \\[6pt]
                \mathbf{\Delta}_{p,\sin} &=\left(0,\,\tfrac12,\,\tfrac12,\,0,\,-\tfrac12,\,-\tfrac12\right)^T
            \end{aligned}
        &
            \begin{aligned}
                \mathbf{\Delta}_{d,\cos} &= \frac{1}{\sqrt{3}}\left(1,\,-\tfrac12,\,-\tfrac12,\,1,\,-\tfrac12,\,-\tfrac12\right)^T \\[6pt]
                \mathbf{\Delta}_{d,\sin} &= \left(0,\,\tfrac12,\,-\tfrac12,\,0,\,\tfrac12,\,-\tfrac12\right)^T.
            \end{aligned}
        \end{array}
    \end{align}

Pictorially these solutions are represented as Fig.~\ref{supp:fig:6_patch_soln}(c). As mentioned earlier, the Fermi-surface projected interaction is even in momenta -- a consequence of working in the time-reversed gauge and the indistinguishability of fermions (see Sec.~\ref{supp:TRS_gauge_section}). This implies that $v_{0^{\circ}}=v_{180^{\circ}}$ and $v_{60^{\circ}}=v_{120^{\circ}}$ leading to the vanishing of eigenvalues for the odd-parity solutions $\lambda_p=\lambda_f=0$ as mentioned in the main text. This is in fact true for all generalizations of the patch models as shown in later subsections and is consistent with Refs.~\cite{Samokhin2015Symmetry,Samokhin2017Noncentrosymmetric1D,Samokhin2008Gap3D} that demonstrate that the intra-band gap function is necessarily of even parity in this gauge due to fermionic antisymmetry.

\subsection{Exact solution to the 12-patch model} \label{supp:12-patch}

As introduced in the main text [see Fig.~1 and Eq.~(1)], in this section we examine the solution to the 12-patch model taking into account zero-momentum scattering processes in the particle-particle channel. The 12 patches correspond to two points on each arc and we can label the points in the patch model by the chirality of the associated Weyl points. The linearized gap equation is
\begin{align}
    -\mathbf{V} \vec{\mathbf{\Delta}}^{(l)} = \lambda_{l} \vec{\mathbf{\Delta}}^{(l)}\quad\longrightarrow \quad
    -\begin{pmatrix}
        \mathbf{g} & \mathbf{w} \\
        \mathbf{w}^T & \mathbf{g}
    \end{pmatrix}\begin{pmatrix}
        \boldsymbol{\Delta}_+ \\
        \boldsymbol{\Delta}_-
    \end{pmatrix} = \lambda\begin{pmatrix}
        \boldsymbol{\Delta}_+ \\
        \boldsymbol{\Delta}_-
    \end{pmatrix}\,,\label{supp:eq:patch_model_gap}
\end{align}
with $\boldsymbol{\Delta}_{\pm}$ the gap function on points of positive/negative chirality. Here,
\begin{equation}
    \mathbf{g} = \textbf{circ}\big(g_{0^\circ},\,g_{60^\circ},\,g_{120^\circ},\,g_{180^\circ},\,g_{120^\circ},\,g_{60^\circ}\big)
\end{equation} 
characterizes intra-chirality scattering processes and is circulant and symmetric, whereas
\begin{equation}
    \mathbf{w} = \textbf{circ}\big(w_{0^\circ},\,w_{60^\circ}^1,\,w_{120^\circ}^2,\,w_{180^\circ},\,w_{120^\circ}^1,\,w_{60^\circ}^2\big)
\end{equation}
 describes inter-chirality scattering processes and is circulant but not symmetric in general. The interactions are depicted in Fig.~1(b) of the main text.

\subsubsection{Narrow arc limit, $w^1_{60^\circ}=w^2_{60^\circ}$}\label{supp:narrow_arc_limit}

To begin, we study the case of the narrow arc limit setting $w_{60^\circ}^{1}=w_{60^\circ}^{2}$ and $w_{120^\circ}^{1}=w_{120^\circ}^{2}$. Within this approximation the circulant inter-chirality interaction matrix $\mathbf{w}$ is also symmetric, i.e., $\mathbf{w}^T=\mathbf{w}$. This results in Eq.~\eqref{supp:eq:patch_model_gap} taking the form Eq.~\eqref{supp:eq:narrow_arc_patch_model_gap} which can be block diagonalized by defining the unitary operator $\mathbf{Q}$
\begin{align}
    -\begin{pmatrix}
        \mathbf{g} & \mathbf{w} \\
        \mathbf{w} & \mathbf{g}
    \end{pmatrix}\begin{pmatrix}
        \boldsymbol{\Delta}_+ \\
        \boldsymbol{\Delta}_-
    \end{pmatrix} &= \lambda\begin{pmatrix}
        \boldsymbol{\Delta}_+ \\
        \boldsymbol{\Delta}_-
    \end{pmatrix},\, \label{supp:eq:narrow_arc_patch_model_gap}
    \qquad
    \mathbf{Q}=\frac{1}{\sqrt{2}}
        \begin{pmatrix}
            \mathds{1} & \mathds{1}\\
            \mathds{1} & -\mathds{1}
        \end{pmatrix} \implies
        \mathbf{Q}^\dagger \mathbf{V} \mathbf{Q} = \begin{pmatrix}
            \mathbf{g}+\mathbf{w} & \mathbf{0}\\
            \mathbf{0} & \mathbf{g}-\mathbf{w}
        \end{pmatrix}.
\end{align}
Since $\mathbf{g}$ and $\mathbf{w}$ are symmetric circulant matrices, any linear
combination of them is also symmetric and circulant.  Consequently, the eigensystem of the two resulting $6\times 6$ blocks have the same structure as in Sec.~\ref{supp:circulant_primer}. The eigenvectors of the full $12$-patch model are obtained by taking tensor products of the 6-patch eigenvectors with $\frac{1}{\sqrt{2}}(1, \pm1)^T$ which correspond to aligned or anti-aligned combinations.

\begin{table}[h]
\centering
\begin{tabular}{|c|c|c|}
\hline
\multirow{2}{*}{\textbf{6-patch model}}
& \multicolumn{2}{c|}{\textbf{12-patch model}} \\
\cline{2-3}
& \textbf{aligned} & \textbf{anti-aligned} \\
\hline\hline
$s$-wave & $s\;(A_1)$   & $i\; (A_2)$ \\
$p$-wave & $p\;(E_1)$   & $h\;(E_1)$ \\
$d$-wave & $d\;(E_2)$   & $g\;(E_2)$ \\
$f$-wave & $f\;(B_2)$ & $f\;(B_1)$ \\
\hline
\end{tabular}
\caption{Correspondence between 6-patch and 12-patch pairing channels. Here, the letters in the parentheses denote irreps of $C_{6v}$ group (see Sec.~\ref{sec:effective_c6v}).}
\label{tab:table_12_patch}
\end{table}

\paragraph{Aligned sector, $\mathbf{g}+\mathbf{w}$:}
 We begin with the six aligned modes of the the $\mathbf{g+w}$ sector. These inherit the same form factors as their 6-patch counterparts. For the $f$-wave state we use the subscript $B_2$ and $B_1$ to distinguish the aligned and anti-aligned sectors respectively using their corresponding irreps in the $C_{6v}$ point group Tab.~\ref{tab:table_12_patch} (see Sec.~\ref{sec:effective_c6v}). The eigenvalues and eigenstates of the aligned sector are
\begin{align}
  \lambda_s &= -(g_{0^\circ}+w_{0^\circ}) - 2(g_{60^\circ}+w_{60^\circ}) - 2(g_{120^\circ}+w_{120^\circ}) - (g_{180^\circ}+w_{180^\circ}), \nonumber \\
  \mathbf{\Delta}_s &= \frac{1}{\sqrt{2}}(1,\,1)\otimes\frac{1}{\sqrt{6}}(1,\,1,\,1,\,1,\,1,\,1)^T,
\\[6pt]\hline\nonumber\\
  \lambda_p &= -(g_{0^\circ}+w_{0^\circ}) - (g_{60^\circ}+w_{60^\circ}) + (g_{120^\circ}+w_{120^\circ}) + (g_{180^\circ}+w_{180^\circ}), \nonumber \\
  \mathbf{\Delta}_{p,\,\cos} &=  \frac{1}{\sqrt{2}}(1,\,1)\otimes\frac{1}{\sqrt{3}}\left(1,\,\tfrac12,\,-\tfrac12,\,-1,\,-\tfrac12,\,\tfrac12\right)^T,\\[-2pt]
  \mathbf{\Delta}_{p,\,\sin} &= \frac{1}{\sqrt{2}}(1,\,1)\otimes\left(0,\,\tfrac12,\,\tfrac12,\,0,\,-\tfrac12,\,-\tfrac12\right)^T,
\\[6pt]\hline\nonumber\\
  \lambda_d &= -(g_{0^\circ}+w_{0^\circ}) + (g_{60^\circ}+w_{60^\circ}) + (g_{120^\circ}+w_{120^\circ}) - (g_{180^\circ}+w_{180^\circ}), \nonumber\\
  \mathbf{\Delta}_{d,\,\cos} &=  \frac{1}{\sqrt{2}}(1,\,1)\otimes\frac{1}{\sqrt{3}}\left(1,\,-\tfrac12,\,-\tfrac12,\,1,\,-\tfrac12,\,-\tfrac12\right)^T, \\[-2pt]
  \mathbf{\Delta}_{d,\,\sin} &= \frac{1}{\sqrt{2}}(1,\,1)\otimes\left(0,\,\tfrac12,\,-\tfrac12,\,0,\,\tfrac12,\,-\tfrac12\right)^T,
\\[6pt]\hline\nonumber\\
  \lambda_{f,\,B_2} &= -(g_{0^\circ}+w_{0^\circ}) + 2(g_{60^\circ}+w_{60^\circ}) - 2(g_{120^\circ}+w_{120^\circ}) + (g_{180^\circ}+w_{180^\circ}), \nonumber\\
  \mathbf{\Delta}_{f,\,B_2} &=  \frac{1}{\sqrt{2}}(1,\,1)\otimes\frac{1}{\sqrt{6}}(1,\,-1,\,1,\,-1,\,1,\,-1)^T.
\end{align}

\paragraph{Anti-aligned sector, $\mathbf{g}-\mathbf{w}$:}
For the anti-aligned $\mathbf{g-w}$ sector the symmetry factors of the gap eigenvectors do not correspond to their 6-patch counterparts in contrast to the case for the aligned solutions Tab.~\ref{tab:table_12_patch}. The eigenvectors and eigenvalues are given by

\begin{align}
  \lambda_{f,\,B_1} &= -(g_{0^\circ}-w_{0^\circ}) + 2(g_{60^\circ}-w_{60^\circ}) - 2(g_{120^\circ}-w_{120^\circ}) + (g_{180^\circ}-w_{180^\circ}), \nonumber\\
  \mathbf{\Delta}_{f,\,B_1} &=  \frac{1}{\sqrt{2}}(1,\,-1)\otimes\frac{1}{\sqrt{6}}(1,\,-1,\,1,\,-1,\,1,\,-1)^T,
\\[6pt]\hline\nonumber\\
  \lambda_g &= -(g_{0^\circ}-w_{0^\circ}) + (g_{60^\circ}-w_{60^\circ}) + (g_{120^\circ}-w_{120^\circ}) - (g_{180^\circ}-w_{180^\circ}), \nonumber\\
  \mathbf{\Delta}_{g,\,\cos} &=  \frac{1}{\sqrt{2}}(1,\,-1)\otimes\frac{1}{\sqrt{3}}\left(1,\,-\tfrac12,\,-\tfrac12,\,1,\,-\tfrac12,\,-\tfrac12\right)^T, \\[-2pt]
  \mathbf{\Delta}_{g,\,\sin} &= \frac{1}{\sqrt{2}}(1,\,-1)\otimes\left(0,\,\tfrac12,\,-\tfrac12,\,0,\,\tfrac12,\,-\tfrac12\right)^T,
\\[6pt]\hline\nonumber\\
  \lambda_h &= -(g_{0^\circ}-w_{0^\circ}) - (g_{60^\circ}-w_{60^\circ}) + (g_{120^\circ}-w_{120^\circ}) + (g_{180^\circ}-w_{180^\circ}), \nonumber \\
  \mathbf{\Delta}_{h,\,\cos} &=  \frac{1}{\sqrt{2}}(1,\,-1)\otimes\frac{1}{\sqrt{3}}\left(1,\,\tfrac12,\,-\tfrac12,\,-1,\,-\tfrac12,\,\tfrac12\right)^T,\\[-2pt]
  \mathbf{\Delta}_{h,\,\sin} &= \frac{1}{\sqrt{2}}(1,\,-1)\otimes\left(0,\,\tfrac12,\,\tfrac12,\,0,\,-\tfrac12,\,-\tfrac12\right)^T,
\\[6pt]\hline\nonumber\\
  \lambda_i &= -(g_{0^\circ}-w_{0^\circ}) - 2(g_{60^\circ}-w_{60^\circ}) - 2(g_{120^\circ}-w_{120^\circ}) - (g_{180^\circ}-w_{180^\circ}), \nonumber \\
  \mathbf{\Delta}_i &= \frac{1}{\sqrt{2}}(1,\,-1)\otimes\frac{1}{\sqrt{6}}(1,\,1,\,1,\,1,\,1,\,1)^T.
\end{align}
Fermionic indistinguishability in the time-reversed gauge dictates that the Fermi surface-projected interactions must be even in momenta (Sec.~\ref{supp:TRS_gauge_section}). This leads to the constraint mentioned in the main text: $(g,w)_{0^\circ}=(g,w)_{180^\circ}$ and $(g,w^j)_{60^\circ}=(g,w^j)_{120^\circ}$. Substituting these relations in the expressions above lead to all odd-parity solutions having vanishing eigenvalues.
\begin{figure}[h]
    \centering
    \includegraphics[width=\linewidth]{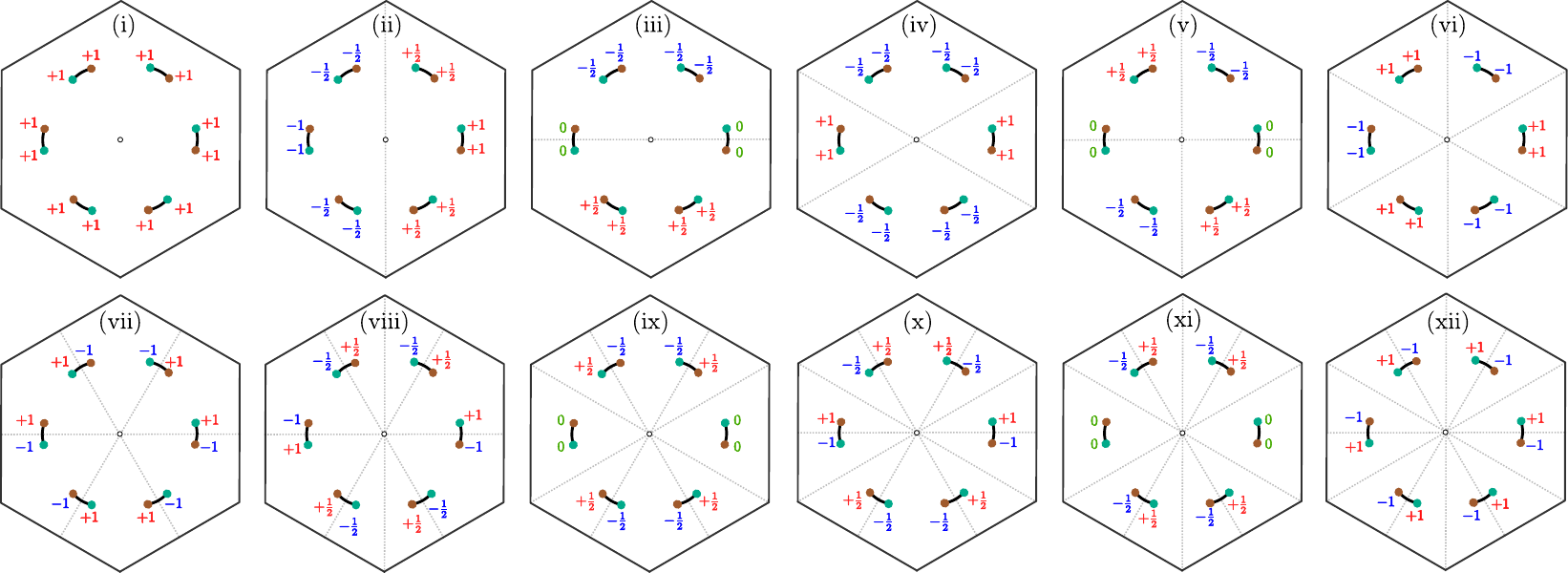}
    \caption{Pictorial representation of the gap eigenvectors within the \textbf{12-patch model}: (i) $s$-wave; (ii)-(iii) $p$-wave; (iv)-(v) $d$-wave; (vi-vii) $f$-wave; (viii-ix) $g$-wave; (x-xi) $h$-wave; (xii) $i$-wave. In the time-reversed gauge all odd-parity solutions have a vanishing eigenvalue.}
    \label{supp:fig:12_patch_soln}
\end{figure}

An intuitive way to understand this can be seen from considering Fig.~\ref{supp:fig:12_patch_soln} (vii) and (xii) corresponding to the $f_{B_1}$-wave and $i$-wave solutions respectively. A repulsive inter-chirality interaction $w_{0^\circ} > 0$ favors a sign changing gap on any given arc which both these states display. However, since $w_{0^\circ}=w_{180^\circ}$ this also favors a sign change between different chiralities on the opposite arc as well. This is only captured by the $i$-wave solution and not the $f_{B_1}$-wave solution.

\subsubsection{General case $w^1_{60^\circ}\neq w^2_{60^\circ}$}
\label{supp:general_12_patch}
In general $\mathbf{w} = \textbf{circ}\big(w_{0^\circ},\,w_{60^\circ}^1,\,w_{120^\circ}^2,\,w_{180^\circ},\,w_{120^\circ}^1,\,w_{60^\circ}^2\big)$ is not a symmetric matrix, meaning $\mathbf{w}\neq\mathbf{w}^T$. Consequently, the unitary transformation that block-diagonalized the narrow arc limit no longer suffices. Nevertheless, Eq.~\eqref{supp:eq:patch_model_gap} is still amenable to an exact solution. This is because $\mathbf{g}$, $\mathbf{w}$, and $\mathbf{w}^T$ are all $6 \times 6$ circulant matrices, they mutually commute and share the same eigenbasis spanned by the discrete Fourier transform vectors $\mathbf{u}_l = \frac{1}{\sqrt{6}}(1, \omega^l, \omega^{2l}, \dots, \omega^{5l})^T$ with $\omega=e^{i\pi/3}$ and $l \in \{0, 1, \dots, 5\}$. In this basis, the $12 \times 12$ pairing matrix reduces to six decoupled $2 \times 2$ blocks coupling the two chiralities:
\begin{equation}
    -\mathbf{V}_l = -\begin{pmatrix} g_l & w_l \\ w_l^* & g_l \end{pmatrix}\,,
\end{equation}
where $g_l$ and $w_l$ are the eigenvalues of the individual circulant blocks in the $l^{\rm th}$ Fourier sector Eq.~\eqref{supp:eq:circulant_eigsystem}. As such, the general eigenvalues for each block are given by $\lambda_{l\pm} = -(g_l \pm|w_l|)$. \\

\paragraph{Vanishing of odd-parity solutions:} Fermionic indistinguishability in the time-reversed gauge dictates that the Fermi surface-projected interactions must be even in momenta (Sec.~\ref{supp:TRS_gauge_section}). This imposes the constraints $(g,w)_{0^\circ}=(g,w)_{180^\circ}$ and $(g,w^j)_{60^\circ}=(g,w^j)_{120^\circ}$. Substituting these relations into the Fourier sum for the intra-chirality term:
\begin{align}
    g_l &= g_{0^\circ} + g_{60^\circ}\omega^l + g_{120^\circ}\omega^{2l} + g_{180^\circ}\omega^{3l} + g_{120^\circ}\omega^{4l} + g_{60^\circ}\omega^{5l} \nonumber \\
        &= g_{0^\circ}(1 + \omega^{3l}) + g_{60^\circ}(\omega^l + \omega^{2l} + \omega^{4l} + \omega^{5l}).
\end{align}
For odd $l=1,3,5$ we have $\omega^{3l} = e^{i\pi l} = -1$, ensuring that the first term vanishes. Additionally, since the sum of all roots of unity is zero $\sum_{n=0}^5 \omega^{nl} = 0$, the second term identically evaluates to zero as well (recall $1+\omega^{3l}=0$) and we have $g_l = 0$. Applying the same logic to the inter-chirality matrix $\mathbf{w}$, we similarly find $w_l=0$. Therefore, all odd-parity solutions strictly have a vanishing eigenvalue even in the general case: $\lambda_{p} = \lambda_{f} = \lambda_{h} = 0$.\\

\paragraph{$s$- and $i$-wave sectors:} For $l=0$ the Fourier coefficients evaluate to the purely real scalars:
\begin{align}
    g_{l=0} &= 2g_{0^\circ} + 4g_{60^\circ}, \\
    w_{l=0} &= 2w_{0^\circ} + 2w_{60^\circ}^1 + 2w_{60^\circ}^2.
\end{align}
Because $w_{l=0}$ lacks an imaginary component, no relative phase winding is induced between the chiralities. The eigenvectors are thus entirely identical to those derived in the narrow arc limit:
\begin{align}
    \lambda_{s} &= -2(g_{0^\circ} + 2g_{60^\circ}) - 2(w_{0^\circ} + w_{60^\circ}^1 + w_{60^\circ}^2), \quad &\mathbf{\Delta}_s &= \frac{1}{\sqrt{2}}(1,\,1)^T\otimes\frac{1}{\sqrt{6}}(1,\,1,\,1,\,1,\,1,\,1)^T, \\
    \lambda_{i} &= -2(g_{0^\circ} + 2g_{60^\circ}) + 2(w_{0^\circ} + w_{60^\circ}^1 + w_{60^\circ}^2), \quad &\mathbf{\Delta}_i &= \frac{1}{\sqrt{2}}(1,\,-1)^T\otimes\frac{1}{\sqrt{6}}(1,\,1,\,1,\,1,\,1,\,1)^T.
\end{align}

\paragraph{$d$- and $g$-wave sectors:} For the remaining even parity sectors $l=2,\,4$ the asymmetry $w_{60^\circ}^1 \neq w_{60^\circ}^2$ introduces an imaginary component to the inter-chirality coupling:
\begin{align}
    g_{l=2} &= 2g_{0^\circ} - 2g_{60^\circ}, \\
    w_{l=2} &= 2w_{0^\circ} - w_{60^\circ}^1 - w_{60^\circ}^2 + i\sqrt{3}(w_{60^\circ}^1 - w_{60^\circ}^2) \equiv |w_2|e^{i\varphi},
\end{align}
which manifests as a phase twist $\varphi$ defined as: 
\begin{equation}
    \tan\varphi = \frac{\sqrt{3}(w_{60^\circ}^1 - w_{60^\circ}^2)}{2w_{0^\circ} - w_{60^\circ}^1 - w_{60^\circ}^2}.
\end{equation}
The eigenvalues for these channels are determined by the magnitude of this complex coupling
\begin{align}
    \lambda_{d} &= -(2g_{0^\circ} - 2g_{60^\circ}) - \sqrt{(2w_{0^\circ} - w_{60^\circ}^1 - w_{60^\circ}^2)^2 + 3(w_{60^\circ}^1 - w_{60^\circ}^2)^2}\,,\nonumber\\
     \lambda_{g} &= -(2g_{0^\circ} - 2g_{60^\circ}) + \sqrt{(2w_{0^\circ} - w_{60^\circ}^1 - w_{60^\circ}^2)^2 + 3(w_{60^\circ}^1 - w_{60^\circ}^2)^2}.
\end{align}
To construct purely real eigenvectors, we can distribute the phase symmetrically $\pm \varphi/2$ between the positive and negative chirality patches, effectively applying a rotation within the $\mathbf{\Delta}_{d,\cos}\oplus \mathbf{\Delta}_{d,\sin}$ basis composed by taking the direct sum of the $d$-wave solutions in the 6-patch model. For the $d$-wave channel (the analogue of the aligned sector), the two degenerate eigenvectors are:
\begin{align}
    \mathbf{\Delta}_{d,1} &= \frac{1}{\sqrt{2}} \begin{pmatrix} \cos(\varphi/2) \mathbf{\Delta}_{d,\cos} - \sin(\varphi/2) \mathbf{\Delta}_{d,\sin} \\ \cos(\varphi/2) \mathbf{\Delta}_{d,\cos} + \sin(\varphi/2) \mathbf{\Delta}_{d,\sin} \end{pmatrix}, \\[6pt]
    \mathbf{\Delta}_{d,2} &= \frac{1}{\sqrt{2}} \begin{pmatrix} \sin(\varphi/2) \mathbf{\Delta}_{d,\cos} + \cos(\varphi/2) \mathbf{\Delta}_{d,\sin} \\ -\sin(\varphi/2) \mathbf{\Delta}_{d,\cos} + \cos(\varphi/2) \mathbf{\Delta}_{d,\sin} \end{pmatrix}.
\end{align}
Similarly, the two degenerate eigenvectors for the $g$-wave channel (analogue of the anti-aligned sector) take the form:
\begin{align}
    \mathbf{\Delta}_{g,1} &= \frac{1}{\sqrt{2}} \begin{pmatrix} -\cos(\varphi/2) \mathbf{\Delta}_{d,\cos} + \sin(\varphi/2) \mathbf{\Delta}_{d,\sin} \\ \cos(\varphi/2) \mathbf{\Delta}_{d,\cos} + \sin(\varphi/2) \mathbf{\Delta}_{d,\sin} \end{pmatrix}, \\[6pt]
    \mathbf{\Delta}_{g,2} &= \frac{1}{\sqrt{2}} \begin{pmatrix} -\sin(\varphi/2) \mathbf{\Delta}_{d,\cos} - \cos(\varphi/2) \mathbf{\Delta}_{d,\sin} \\ -\sin(\varphi/2) \mathbf{\Delta}_{d,\cos} + \cos(\varphi/2) \mathbf{\Delta}_{d,\sin} \end{pmatrix}.
\end{align}
 Note that in the narrow arc limit $\varphi\to0$ we recover the solutions derived in Sec.~\ref{supp:narrow_arc_limit}.

\subsection{Solution to the 18-patch model}\label{supp:18-patch}

For the sake of brevity, here we focus on the \emph{nodal $s$-wave} state mentioned in the main text. For the $18$-patch model, the linearized gap equation reads
\begin{align}
    -{\renewcommand{\arraystretch}{1.15}
    \setlength{\arraycolsep}{6pt}
    \begin{pmatrix}
        \mathbf{g} & \mathbf{b} & \mathbf{w} \\
        \mathbf{b}^T & \mathbf{c} & \mathbf{b} \\
        \mathbf{w}^T & \mathbf{b}^T & \mathbf{g}
    \end{pmatrix}
    \begin{pmatrix}
        \boldsymbol{\Delta}_+ \\
        \boldsymbol{\Delta}_0 \\
        \boldsymbol{\Delta}_-
    \end{pmatrix}}
    = \lambda
    {\renewcommand{\arraystretch}{1.15}
    \begin{pmatrix}
        \boldsymbol{\Delta}_+ \\
        \boldsymbol{\Delta}_0 \\
        \boldsymbol{\Delta}_-
    \end{pmatrix}}\,.
    \label{supp:Eq:18_patch_model}
\end{align}
Here, two additional circulant matrices appear: `$\mathbf{c}$' a symmetric matrix coupling all arc centers to each other and a circulant matrix `$\mathbf{b}$' coupling arc centers to the two chiralities. Following the procedure outlined in Sec.~\ref{supp:general_12_patch} we can block diagonalize the $18\times18$ pairing matrix into six decoupled $3\times3$ blocks that couple the two chiralities and a central patch within each Fourier sector of the 6-patch basis.

For the $l=0$ sector the Fourier coefficients evaluate to the purely real scalars:
\begin{align}
    g_{l=0} &= 2g_{0^\circ} + 4g_{60^\circ}, \qquad w_{l=0} = 2w_{0^\circ} + 2w_{60^\circ}^1 + 2w_{60^\circ}^2, \\
    c_{l=0} &= 2c_{0^\circ} + 4c_{60^\circ}, \qquad
    b_{l=0} = 2b_{0^\circ} + 2b_{60^\circ}^1 + 2b_{60^\circ}^2.
\end{align}
To find the eigenvalues and eigenvectors corresponding to this sector we simply need to diagonalize the following $3\times3$ matrix, 
\begin{align}
    -\begin{pmatrix}
        g & b & w \\
        b & c & b \\
        w & b & g
    \end{pmatrix}_{l=0}
\end{align}
and take a tensor product of the eigenvectors derived with the $l=0$ eigenvector of the 6-patch model.
\begin{align}
    \lambda_i &= -(g-w)_{l=0}, \quad\mathbf{\Delta}_i = \frac{1}{\sqrt{2}} \begin{pmatrix} 1 \\ 0 \\ -1 \end{pmatrix}
    \otimes\frac{1}{\sqrt{6}}(1,\,1,\,1,\,1,\,1,\,1)^T,\\
    \lambda_{s,1} &= -\frac{1}{2}(g+w+c -D)_{l=0}, \quad\mathbf{\Delta}_{s,1} \propto \begin{pmatrix} \lambda_{s,1}+c \\-2b \\ \lambda_{s,1}+c \end{pmatrix}_{l=0}\otimes\frac{1}{\sqrt{6}}(1,\,1,\,1,\,1,\,1,\,1)^T,\\
    \lambda_{s,2} &= -\frac{1}{2}(g+w+c +D)_{l=0}, \quad\mathbf{\Delta}_{s,2} \propto  \begin{pmatrix} \lambda_{s,2}+c \\-2b \\ \lambda_{s,2}+c \end{pmatrix}_{l=0}\otimes\frac{1}{\sqrt{6}}(1,\,1,\,1,\,1,\,1,\,1)^T.
\end{align}
Here, $D_{l=0} \equiv (\sqrt{(g + w - c)^2 + 8b^2})_{l=0}$ and is $\geq0$ which implies $\lambda_{s,1}\geq \lambda_{s,2}$, i.e., $\mathbf{\Delta}_{s,1}$ is the leading $s$-wave solution. Henceforth, the subscript $l=0$ is implied. On inspecting roots, we find that for $b<0$, $\mathbf{\Delta}_{s,1}$ corresponds to the fully gapped $s$-wave state whereas for $b\geq0$, $\mathbf{\Delta}_{s,1}$ corresponds to the nodal $s$-wave state (and vice-versa for $\mathbf{\Delta}_{s,2}$). To see this we examine the sign of the eigenvector components corresponding to the two end points (chiralities). Defining $x = -(g + w - c)_{l=0}$ we have
\begin{align}
    \lambda_{s,1}+c  = \frac{x + \sqrt{x^2 + 8b^2}}{2},\qquad\lambda_{s,2}+c  = \frac{x - \sqrt{x^2 + 8b^2}}{2}.
\end{align}
Since $\sqrt{x^2 + 8b^2}\geq|x|$ with equality only being the case if $b=0$, we have, $\lambda_{s,1}+c \geq0$ and $\lambda_{s,2}+c\leq0$ regardless of the sign of $x$. In the narrow arc limit all other sectors can be constructed in a similar manner.
\begin{figure}[t]
    \centering
    \includegraphics[width=\linewidth]{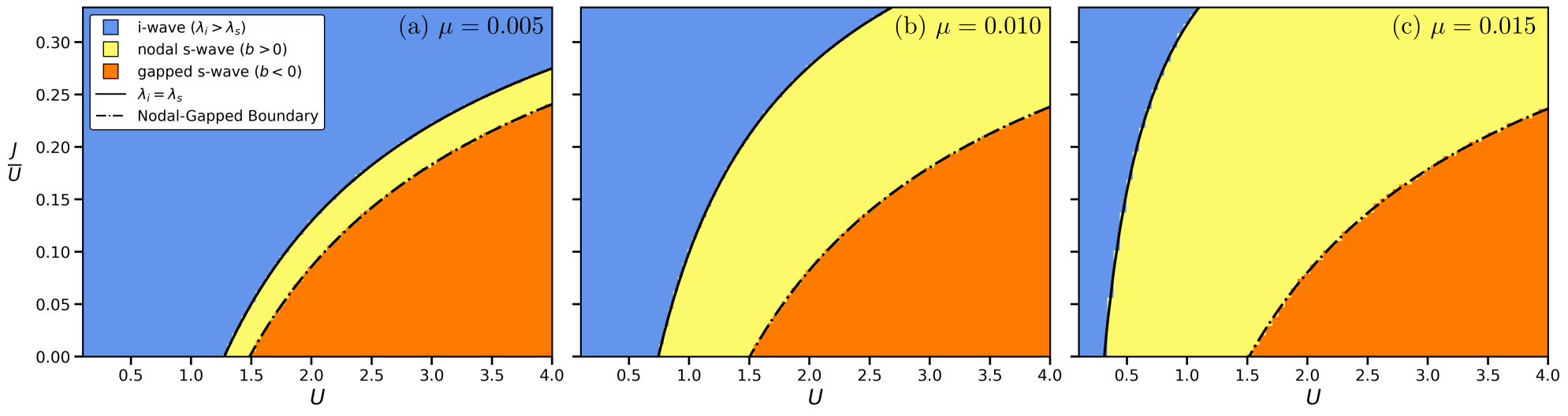}
    \caption{\label{supp:fig:18-patch_phase_diagram_KL} Semi-analytical phase diagram obtained from the analytical eigenvalues of the 18-patch model, with the scattering interactions extracted from the numerically calculated Kohn--Luttinger vertex at different chemical potentials (a) $\mu = 0.005$, (b) $\mu = 0.010$, (c) $\mu = 0.015$ in units of the in-plane intra-orbital hopping. }
\end{figure}

In Fig.~\ref{supp:fig:18-patch_phase_diagram_KL} we plot the phase diagram obtained from projecting the microscopic interactions of the Kohn-Luttinger calculation onto the phenomenological parameters of the 18-patch model. For $\mu=0.005$, where the arc exhibits almost no curvature, the phase diagram is nearly identical to the one obtained from the full microscopic calculation. As $\mu$ is increased, the arc begins to curve more, and the patch model deviates more from the microscopic calculation, as expected.

\section{Microscopic calculation details}

\subsection{Slab model construction}\label{supp:sec:slab_model}

The Hamiltonian of the tight-binding model is derived in Ref.~\cite{Vocaturo2024Electronic} and presented in the End Matter of the main text. Here we reproduce parts of it for convenience. The Bloch Hamiltonian reads
\begin{equation}
H(\mathbf{k}) =h_0(\mathbf{k})+\alpha\,h_1(\mathbf{k}) + \gamma\,\tau_x\otimes\sigma_0\,,
\end{equation}
where $\gamma \otimes \tau_x\sigma_0$ is the inversion symmetry-breaking term, $h_0(\mathbf{k})$ captures the dispersion and hopping between orbitals, and $h_1(\mathbf{k})$ introduces SOC and is explicitly given by
\begin{align}
        h_0(\mathbf{k}) &= \left\{m- t\left[\cos k_1-\cos k_2 -\cos (k_1 + k_2)\right] +\beta\cos k_3 \right\}\Gamma_1  +\left\{\beta\sin k_3  + \lambda[\sin k_1+\sin k_2 -\sin(k_1+k_2)]\right\}\Gamma_3 \\
     h_1(\mathbf{k}) &= (1-\cos k_3)[\sin k_1 \Gamma_2 + \sin k_2\Gamma_{2,1}
     -\sin(k_1+k_2)\Gamma_{2,2}]\,,
\end{align}
with $k_i = \mbfk \cdot \mathbf{a}_i$. All parameters are measured in units of the in-plane intra-orbital hopping, $t$, which is set to unity throughout the manuscript. The matrices $\Gamma$ represent orbital $\boldsymbol{\tau}$ and spin $\boldsymbol{\sigma}$ degrees of freedom, defined as
\begin{align}
    \Gamma_1 &= \tau_z\sigma_0 , \qquad
    \Gamma_2 = \tau_x\sigma_x,\qquad
    \Gamma_3 = \tau_y\sigma_0 \quad\nonumber\\
    \Gamma_{2,j} &=\mathcal{C}_3^j\Gamma_2\mathcal{C}_3^{-j}\;\;\left(\text{where  }\mathcal{C}_3\equiv\tau_0\exp\left\{-i\frac{\pi}{3}\sigma_z\right\} \right)
\end{align}
with $\Gamma_1$ corresponding to intra-orbital hoppings, $\Gamma_3$ to inter-orbital hoppings, and $\Gamma_2$ along with $\Gamma_{2,j}$ represent the spin-orbit coupling.

To construct the slab model we first need to decompose the Bloch Hamiltonian into the following form
\begin{equation}
    H(\mathbf{k}) = H(k_1, k_2, k_3) = H_{\text{layer}}(k_1, k_2) \,+\, e^{ik_3}T(k_1, k_2) \, +\, e^{-ik_3}T^\dagger(k_1, k_2). 
\end{equation}
To do this we introduce the following three functions,
\begin{align}
    \text{ In-plane mass term}:\quad & M(k_1, k_2) = m - \cos k_1 - \cos k_2 - \cos (k_1 +k_2) \nonumber\\
    \text{Inter-orbital hopping term}:\quad  & L(k_1, k_2) = \sin k_1 +\sin k_2 - \sin (k_1 + k_2) \nonumber\\
    \text{Spin-orbit coupling term}:\quad & S(k_1, k_2) = \sin k_1 \,\Gamma_2 + \sin k_2 \,\Gamma_{2,1} - \sin (k_1+k_2) \,\Gamma_{2,2}
\end{align}
We now have the constituent terms of the slab model,
\begin{align}
    H_{\text{layer}}(k_1, k_2) &= M(k_1, k_2)\,\Gamma_1 + \lambda\,L(k_1, k_2)\,\Gamma_3 \,+\,\alpha\,S(k_1, k_2)\,+\,\gamma \,\tau_x\sigma_0, \\
    T(k_1, k_2) &=  \frac{1}{2}\Big[\beta\,\Gamma_1\,-\,i\beta\,\Gamma_3\,-\,\alpha S(k_1, k_2)\Big]\\
    H_{\text{Slab}}(k_1, k_2) &= \begin{pmatrix}
        H_{\text{layer}} &  T & 0 & \cdots & 0 \\
         T^\dagger&   H_{\text{layer}}& T & \cdots & 0 \\
        0 &  T^\dagger & H_{\text{layer}} & \ddots & \vdots \\
        \vdots & \vdots & \ddots & \ddots & T \\
        0 & 0 & \cdots &  T^\dagger & H_{\text{layer}}
    \end{pmatrix}_{4L\times4L} 
\end{align}
We explicitly write out the diagonalization of the slab Hamiltonian at each $\mathbf{k}$-point in the following manner
\begin{align}
    E_{\text{slab}}^{\lambda\lambda}(\mathbf{k}) = \mathcal{U}^\dagger_{\lambda\alpha}(\mathbf{k}) H_{\text{slab}}^{\alpha\beta}(\mathbf{k}) \,\mathcal{U}_{\beta\lambda}(\mathbf{k})\,,
\end{align}
where $\lambda$ is a band index and $\alpha,\beta\in \{0,\,1\,...,\,L-1\}\times\{A,\, B\}\times\{\uparrow,\,\downarrow\}$. The unitary matrices are simply the eigenvectors of the slab Hamiltonian at each $\mbfk$-point. We make use of this unitary transformation to go from the orbital basis to the band basis in which the Green function is diagonal, allowing us to perform  Matsubara sums analytically.

\subsection{Surface susceptibility calculation}

The bare particle-hole susceptibility tensor for the slab model is given by
\begin{align}
 \chi^{\beta\gamma}_{\delta\alpha}(\mbfq,\iOn)&=-\frac{1}{\mathcal{V}}\int_0^{1/T}d\tau\,e^{i\Omega_n\tau}\sum_{\mbfk\mbfk'}\left\langle \mathcal{T}_{\tau}\,c^{\dagger}_{\mbfk,\beta}(\tau)\,c_{\mbfk+\mbfq,\gamma}(\tau)\,c^\dagger_{\mathbf{k}^\prime,\delta}(0)\,c_{\mathbf{k}^\prime-\mbfq,\alpha}(0) \right\rangle_0\\
    &= \frac{T}{\mathcal{V}}\sum_{\;\mbfk,\iwm}G^{\alpha\beta}(\mbfk,\iwm)\,G^{\gamma\delta}(\mbfk+\mbfq,\iwm+\iOn)\,.
\end{align}
Note that the static susceptibility is negative within this convention. Working in the band basis with band indices denoted by $m$ and $n$ allows for the evaluation of Matsubara summations. Analytically continuing the result yields
\begin{equation}
      \chi^{\beta\gamma}_{\delta\alpha}(\mbfq,\Omega)=\frac{1}{\mathcal{V}}\sum_{\substack{\mbfk \\ mn}}\mathcal{M}_{\alpha\beta}^m(\mathbf{k})\,\mathcal{M}_{\gamma\delta}^n(\mathbf{k}+\mbfq)\;\frac{n_F\left(\epsilon_{\mathbf{k},m}\right)-n_F\left(\epsilon_{\mathbf{k}+\mathbf{q},n}\right)}{\Omega+ \epsilon_{\mathbf{k},m}-\epsilon_{\mathbf{k}+\mathbf{q},n} + i0^+}.
      \label{Generalized susc expression}
\end{equation}
For notational convenience we define $\mathcal{M}_{\alpha\beta}^m(\mathbf{k})\equiv \mathcal{U}_{\alpha m}(\mathbf{k})\,\mathcal{U}^\dagger_{m\beta}(\mbfk)= \mathcal{U}_{\alpha m}(\mathbf{k})\,\mathcal{U}^*_{\beta m}(\mbfk)$. To calculate the surface susceptibility we take $\alpha,\beta,\gamma,\delta\in  \{A,\,B\}\times\{\uparrow,\,\downarrow\}$ for the top or bottom layer. For all subsequent calculations of the effective pairing vertex we utilize the static limit of the surface susceptibility $\chi^{\beta\gamma}_{\delta\alpha}(\mbfq) = \chi^{\beta\gamma}_{\delta\alpha}(\mbfq, \Omega \rightarrow 0) $. Physical susceptibilities are calculated using $\chi_{ij}(\mathbf{q})=\sum_{\alpha\beta\gamma\delta}(\mathcal{O}_i)_{\beta\gamma}\,\chi^{\beta\gamma}_{\delta\alpha}(\mathbf{q})\,(\mathcal{O}_j)_{\delta\alpha}$, with $\mathcal{O}_i$ some operator, e.g., spin or orbital density. Note that in the construction of the pairing vertex we make use of all components of the surface susceptibility tensor.

\begin{figure}[h]
    \centering
    \includegraphics[width=0.8\linewidth]{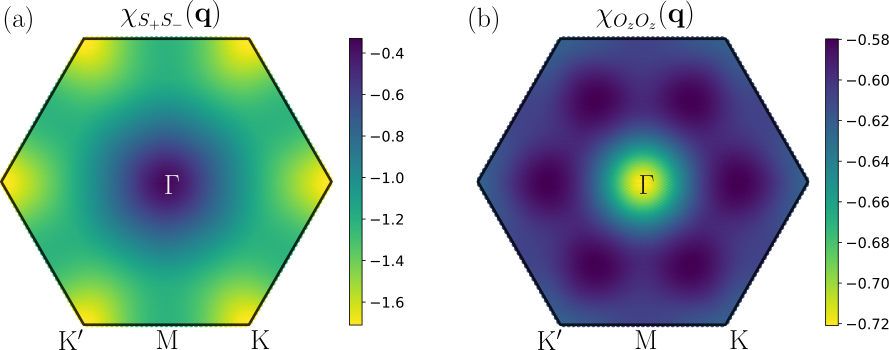}
    \caption{(a) Transverse spin-susceptibility with $S_{\pm}=\tau_0 \left(\sigma_x\pm i\sigma_y\right)$ (b) Orbital-susceptibility with $O_z= \tau_z\sigma_0$.}
\end{figure}

\subsection{Constructing the pairing vertex}
\begin{figure}[h]
    \centering
    \includegraphics[width=\linewidth]{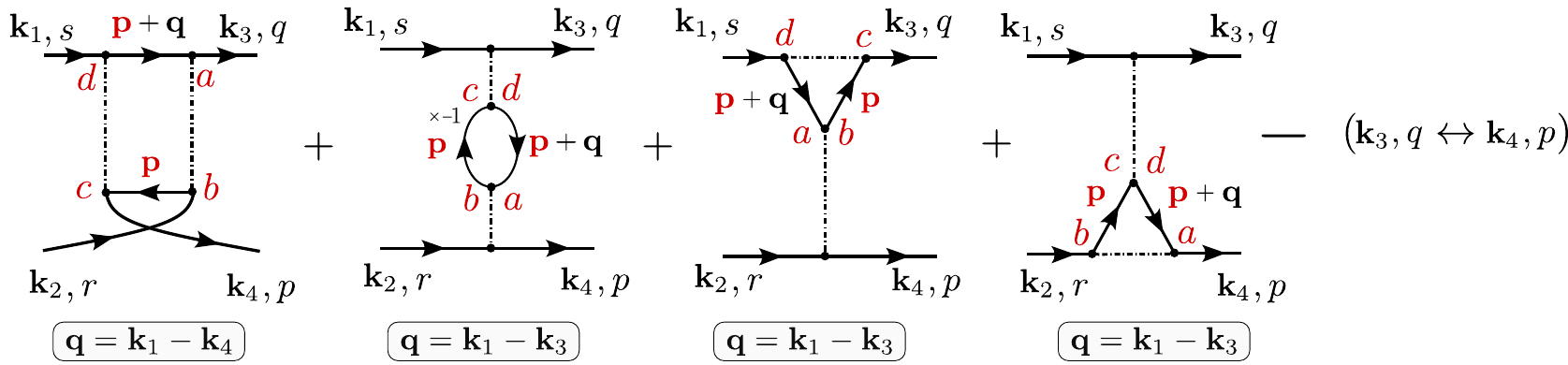}
    \caption{\textbf{The Kohn-Luttinger diagrams.} Each diagram depends on a certain combination of external momenta $\mathbf{q}$. The red font indicates the internal indices and momenta that are summed over and the dashed line represents the bare vertex $\Gamma$.} 
\label{SM:fig:KL_diag_with_indices}
\end{figure}
In this section we provide the explicit expression for the Kohn-Luttinger pairing vertex. For clarity, we provide the diagrams with explicit indices for the momenta and quantum numbers in Fig. \ref{SM:fig:KL_diag_with_indices}. Their evaluation leads to the two-particle irreducible Kohn-Luttinger vertex \cite{Maiti2013Superconductivity}

\begin{align}
    V^{qs}_{pr}(k_4, k_3 ; k_2,k_1)
    &= \frac{1}{2}\bigg[\Gamma^{qs}_{pr}
    - \overbrace{\Gamma^{qa}_{br}\,\chi^{ba}_{dc}(\mbfk_1-\mbfk_4)\,\Gamma^{ds}_{pc}}^{\rm cross}
    + \overbrace{\Gamma^{qs}_{dc}\,\chi^{ba}_{dc}(\mbfk_1-\mbfk_3)\,\Gamma^{ba}_{pr}}^{\rm bubble}
    \nonumber\\
    &\qquad
    - \underbrace{\Gamma^{qc}_{ds}\,\chi^{ba}_{dc}(\mbfk_1-\mbfk_3)\,\Gamma^{ba}_{pr}}_{\rm vertex_{1}}
    - \underbrace{\Gamma^{qs}_{dc}\,\chi^{ba}_{dc}(\mbfk_1-\mbfk_3)\,\Gamma^{br}_{pa}}_{\rm vertex_{2}}\bigg] - \,(\mbfk_3, q\longleftrightarrow\mbfk_4, p).
    \label{eq:KL_expansion}
\end{align}
As in the main text, $\Gamma^{qs}_{pr}$ labels the components of the on-site repulsive bare Hubbard-Hund-Kanamori interaction tensor and $\chi^{ba}_{dc}(\mathbf{q})$ is the non-interacting (static) susceptibility as defined in \eqref{Generalized susc expression}. To construct the zero-momentum transfer, irreducible, pairing vertex in the Cooper channel, we set $\mbfk_1=-\mbfk_{2}=\mbfk$ and $\mbfk_3=-\mbfk_{4}=\mbfk'$ which yields
\begin{align}
    {V}^{qs}_{pr}(\mathbf{k}',\,\mathbf{k})
    = \frac{1}{2}\Bigg(
    \Gamma^{qs}_{pr}
    - \Gamma^{ps}_{qr}
    + \Big[
        \Gamma^{qs}_{dc}\,\Gamma^{ba}_{pr}
        + \Gamma^{pa}_{br}\,\Gamma^{ds}_{qc}
        - \Gamma^{qc}_{ds}\,\Gamma^{ba}_{pr}
        - \Gamma^{qs}_{dc}\,\Gamma^{br}_{pa}
    \Big]\chi^{ba}_{dc}(\mathbf{k}-\mathbf{k}')
    \nonumber\\
    \qquad
    - \Big[
        \Gamma^{ps}_{dc}\,\Gamma^{ba}_{qr}
        + \Gamma^{qa}_{br}\,\Gamma^{ds}_{pc}
        - \Gamma^{pc}_{ds}\,\Gamma^{ba}_{qr}
        - \Gamma^{ps}_{dc}\,\Gamma^{br}_{qa}
    \Big]\chi^{ba}_{dc}(\mathbf{k}+\mathbf{k}')
    \Bigg).
    \label{SM:eqn:KL_antisym_vertex}
\end{align}
This irreducible vertex can be further projected onto the Fermi surface to construct $V_{\rm FS}(\mathbf{k}',\mathbf{k})$, which constitutes the pairing kernel for the linearized gap equation. 
    \begin{align}
    V_{\rm FS}(\mathbf{k}',\mathbf{k})
     &= \sum_{pqrs}
     \langle-\mathbf{k}'|_p
     \langle\mathbf{k}'|_q 
    \,{V}^{qs}_{pr}(\mathbf{k}',\mathbf{k})\,
    |-\mathbf{k}\rangle_r
     |\mathbf{k}\rangle_s \label{supp:projection_onto_FS}\\
    \lambda\,\Delta(\mathbf{k}) &= -\frac{1}{{L}_{FS}}\int_{\rm FS}\frac{{\rm d} {\mathbf{k}'}}{|v_{\rm F}(\mathbf{k'})|}\,{V}_{\rm FS}(\mathbf{k}',\mathbf{k})\,\Delta(\mathbf{k}'),\quad \text{with $\mbfk,\mbfk' \in$ FS and $|-\mathbf{p}\rangle\equiv\mathcal{T}|\mathbf{p}\rangle$}.\label{supp:Eq:gap_eqn}
    \end{align}

\section{Gauge choices for $\Delta(\mbfk)$ and singlet-triplet mixing}\label{supp:TRS_gauge_section}

In Eq.~\eqref{supp:projection_onto_FS} we construct the two-particle incoming and outgoing states using TRS to define $|-\mathbf{k}\rangle \equiv \mathcal{T}|\mathbf{k}\rangle$ and $|-\mathbf{k}'\rangle\equiv\mathcal{T}|\mathbf{k'}\rangle$ with $\mathcal{T}=i\tau_0\sigma_y\mathcal{K}$. This is purely a gauge choice that exploits a symmetry of the non-centrosymmetric WSMs under consideration, which preserve TRS. Adopting this gauge has three convenient advantages:
\begin{enumerate}
    \item The Fermi surface-projected interaction is entirely real.
    \item It allows for numerical evaluation and avoids running into any potential discontinuous phase issues.
    \item The projected interaction is even in momenta: \\
    $V_{FS}(\mathbf{k}',\mathbf{k})=V_{FS}(\pm\mathbf{k}',\pm\mathbf{k})\implies$ the gap function is even, $\Delta(\mbfk)=\Delta(-\mbfk)$~\cite{Sigrist1991Phenomenological, Samokhin2015Symmetry}.
\end{enumerate}
Another gauge choice commonly encountered in the literature for superconductivity of non-degenerate bands involves keeping track of a $\mathbf{k}$-dependent global phase accompanying the single-particle electronic states~\cite{Fu2008Superconducting,Maeland2025mechanism}. For the Fermi surface-projected interaction and the linearized gap equation this amounts to
\begin{align}
    \widetilde{V}_{\rm FS}(\mathbf{k}',\mathbf{k}) &= e^{i(\phi_{\mathbf{k}}-\phi_{\mathbf{k}'})}V_{\rm FS}(\mathbf{k}',\mathbf{k}),\label{supp:Eq:gauge_freedom_interaction}\\
    \lambda\,\widetilde{\Delta}(\mathbf{k}) &= -\frac{1}{{L}_{FS}}\int_{\rm FS}\frac{{\rm d} {\mathbf{k}'}}{|v_{\rm F}(\mathbf{k'})|}\,\widetilde{V}_{\rm FS}(\mathbf{k}',\mathbf{k})\,\widetilde{\Delta}(\mathbf{k}'),\\
    \implies \lambda\,\left[e^{-i\phi_{\mathbf{k}}}\widetilde{\Delta}(\mathbf{k})\right]&= -\frac{1}{{L}_{FS}}\int_{\rm FS}\frac{{\rm d} {\mathbf{k}'}}{|v_{\rm F}(\mathbf{k'})|}\,{V}_{\rm FS}(\mathbf{k}',\mathbf{k})\,\left[e^{-i\phi_{\mathbf{k}'}}\widetilde{\Delta}(\mathbf{k}')\right]\label{supp:Eq:winding_gap_eqn}.
\end{align}
On comparing Eq.~\eqref{supp:Eq:gap_eqn} and Eq.~\eqref{supp:Eq:winding_gap_eqn}, we have ${\Delta}(\mathbf{k})=e^{-i\phi_{\mathbf{k}}}\widetilde{\Delta}(\mathbf{k})$. In the following subsection \ref{supp:explicit_gauge} we examine how Eq.~\eqref{supp:Eq:gauge_freedom_interaction} arises by explicitly evaluating the Fermi-surface projected interaction in both gauge choices.

\subsection{Explicit evaluation of Fermi-surface projected Kohn-Luttinger interactions}\label{supp:explicit_gauge}

In this subsection, we illustrate the idea of \emph{singlet-triplet mixing} at the level the of the Fermi surface-projected pairing interaction $V_{\rm FS}(\mathbf{k},\mathbf{k}')$. We begin by deriving the effective interaction tensor $V^{qs}_{pr}(\mathbf{k},\mathbf{k}')$, obtained from the Kohn-Luttinger diagrams for a generic single-orbital, spinful tight-binding model without spin-orbit coupling, assuming a purely repulsive on-site Hubbard interaction. 

On establishing this tensor, we project it onto the Fermi-surface. We first treat the case of spin-degenerate Fermi surfaces, where the interaction naturally decomposes into singlet and triplet channels. We then turn to a helical-metal, i.e., a TRS preserving, non-degenerate spin-momentum locked Fermi surface more closely resembling the electronic surface states of a non-centrosymmetric WSM. 

In this setting, we contrast the two gauges discussed: one in which pairing occurs between time-reversed states $|\mathbf{k}\rangle$ and $ \mathcal{T}|\mathbf{k}\rangle$, and another that keeps track of a $\mathbf{k}$-dependent global phase of the single-particle electronic state $|\mathbf{k}\rangle$ and $|-\mathbf{k}\rangle$. This comparison makes explicit the origin of Eq.~\eqref{supp:Eq:gauge_freedom_interaction}. Finally, we state how these considerations extend to two-orbital spinful models with helical-metal like Fermi surfaces in which the repulsive interaction is taken to be the Hubbard-Hund-Kanamori interaction.

\subsubsection{Spin-degenerate Fermi surface}

Consider an arbitrary spin-degenerate Hamiltonian of the form $h(\mathbf{k})=\varepsilon_\mathbf{k}\sigma_0$ with the electronic dispersion satisfying $\varepsilon_\mathbf{k}=\varepsilon_{-\mathbf{k}}$. The Matsubara Green functions of such a system are given by
\begin{align}
    G^{\uparrow\uparrow}(\mathbf{k},\iwm) = G^{\downarrow\downarrow}(\mathbf{k},\iwm)=\frac{1}{\iwm-\xi_\mathbf{k}}, 
    \quad
    G^{\uparrow\downarrow}(\mathbf{k},\iwm) = G^{\downarrow\uparrow}(\mathbf{k},\iwm) = 0,\quad\text{with  $\xi_\mathbf{k} = \varepsilon_\mathbf{k} - \mu$}.
\end{align}
The particle-hole susceptibility tensor can be constructed using
\begin{equation}
    \chi^{\alpha\beta}_{\gamma\delta}(\mathbf{q},\iOn) = \frac{T}{\mathcal{V}}\sum_{\mathbf{k},\,\iwm} G^{\delta\alpha}(\mathbf{k},\,\iwm)\,G^{\beta\gamma}(\mathbf{k}+\mathbf{q},\iwm + \iOn). 
\end{equation}
The only non-zero components are  $\chi^{\uparrow\uparrow}_{\uparrow\uparrow}(\mathbf{q},\iOn),\, \chi^{\uparrow\downarrow}_{\downarrow\uparrow}(\mathbf{q},\iOn),\,\chi^{\downarrow\uparrow}_{\uparrow\downarrow}(\mathbf{q},\iOn)$ and $\chi^{\downarrow\downarrow}_{\downarrow\downarrow}(\mathbf{q},\iOn)$. In fact, all of these components are equal since the spin-up and spin-down propagators are identical, and so, we denote these components as $\chi(\mathbf{q},\iwm)$. Performing the Matsubara summation and taking the static limit gives 
\begin{equation}
    \chi(\mathbf{q})= \frac{1}{\mathcal{V}}\sum_{\mathbf{k}}\frac{n_F(\varepsilon_\mathbf{k})-n_F(\varepsilon_{\mathbf{k}+\mathbf{q}})}{\varepsilon_\mathbf{k} - \varepsilon_{\mathbf{k}+\mathbf{q}}+i0^+}\,\leq0.
    \label{supp:Eq:simpler_sus}
\end{equation}

The Kohn--Luttinger interaction tensor $V^{qs}_{pr}(\mathbf{k}',\mathbf{k})$ is constructed from Eq.~\eqref{SM:eqn:KL_antisym_vertex} using the repulsive Hubbard vertex $\Gamma^{qs}_{pr}$. Because Eq.~\eqref{SM:eqn:KL_antisym_vertex} produces an antisymmetrized effective interaction, the bare vertex $\Gamma^{qs}_{pr}$ may be taken in either antisymmetrized or non-antisymmetrized form; both conventions yield the same $V^{qs}_{pr}(\mathbf{k}',\mathbf{k})$.
In the following we choose the former taking the bare interaction tensor to be
\begin{align}
\Gamma^{qs}_{pr} =
\left(
\begin{array}{c|cccc}
 & \uparrow\uparrow & \uparrow\downarrow & \downarrow\uparrow & \downarrow\downarrow \\
\hline
\uparrow\uparrow & 0 & 0 & 0 & \frac{U}{4} \\
\uparrow\downarrow & 0 & 0 & -\frac{U}{4} & 0 \\
\downarrow\uparrow & 0 & -\frac{U}{4} & 0 & 0 \\
\downarrow\downarrow & \frac{U}{4} & 0 & 0 & 0
\end{array}
\right)
\quad
&\text{in the basis; rows $\to c^\dagger_q c_s$ and columns $\to c^\dagger_p c_r$.}
\end{align}
which originates from
$U c^\dagger_{\uparrow}c_{\uparrow} c^\dagger_{\downarrow}c_{\downarrow} \longleftrightarrow \frac{U}{4} c^\dagger_{\uparrow}c_{\uparrow} c^\dagger_{\downarrow}c_{\downarrow} -\frac{U}{4}\, c^\dagger_{\downarrow}c_{\uparrow} c^\dagger_{\uparrow}c_{\downarrow} - \frac{U}{4}c^\dagger_{\uparrow}c_{\downarrow}c^\dagger_{\downarrow}c_{\uparrow}+
\frac{U}{4}c^\dagger_{\downarrow}c_{\downarrow}c^\dagger_{\uparrow}c_{\uparrow}$.
Using Eq.~\eqref{SM:eqn:KL_antisym_vertex} yields the following anti-symmetrized effective Kohn-Luttinger pairing tensor
\begin{align}
    {
    \small
    {V}^{qs}_{pr}(\mathbf{k}',\mathbf{k}) = 
    \begin{pmatrix}
        \frac{U^2}{8}[\chi(\mathbf{k}-\mathbf{k}')-\chi(\mathbf{k}+{\mathbf{k}'})] & 0 & 0 & \frac{U}{4} - \frac{U^2}{8}\chi(\mathbf{k}+\mathbf{k}') \\[1.5ex]
    0 & 0 & -\frac{U}{4} + \frac{U^2}{8}[\chi(\mathbf{k}-{\mathbf{k}'})] & 0 \\[1.5ex]
     0 & -\frac{U}{4} + \frac{U^2}{8}[\chi(\mathbf{k}-{\mathbf{k}'})] & 0 & 0 \\[1.5ex]
     \frac{U}{4} - \frac{U^2}{8}[\chi(\mathbf{k}+\mathbf{k}')] & 0 & 0 &  \frac{U^2}{8}[\chi(\mathbf{k}-\mathbf{k}')-\chi(\mathbf{k}+{\mathbf{k}'})]
    \end{pmatrix}.
    \label{supp:eq:anti_sym_vertex}
    }
\end{align}

The Fermi surface---being spin-degenerate at all $\mathbf{k}$-points---allows for the possibility of constructing entangled incoming/outgoing states. Projecting the effective interaction tensor Eq.~\eqref{supp:eq:anti_sym_vertex} onto the singlet ($S=0$) and triplet ($S=1$) channels gives
\begin{alignat}{2}
    S=0,\,m_s=0 &:\quad & V_{s}(\mathbf{k}',\mathbf{k}) &= \left[ \frac{\langle\uparrow|_p\langle\downarrow|_q -\langle\downarrow|_p\langle\uparrow|_q}{\sqrt{2}}\right] {V}^{qs}_{pr}({\mathbf{k}'},\mathbf{k}) \left[ \frac{|\uparrow\rangle_r|\downarrow\rangle_s - |\downarrow\rangle_r|\uparrow\rangle_s}{\sqrt{2}}\right]\nonumber\\
    & & &= \frac{U}{2} - \frac{U^2}{8}\Big(\chi(\mathbf{k} - \mathbf{k}')\,+\,\chi(\mathbf{k} + \mathbf{k}')\Big)\,,
    \label{supp:eq:singlet_interaction} \\
    S=1,\,m_s=0,\pm1 &:\quad & V_{t}(\mathbf{k}',\mathbf{k}) &=  \frac{U^2}{8}\Big(\chi(\mathbf{k}-\mathbf{k}')-\chi(\mathbf{k}+{\mathbf{k}'})\Big)\,,
    \label{supp:eq:triplet_interaction}
\end{alignat}
recall that, in our convention, $\chi(\mathbf{q})\leq 0$. As expected, the singlet interaction is even in momenta, $V_{s}(\mathbf{k}',\mathbf{k})=V_{s}(\pm\mathbf{k}',\pm\mathbf{k})$, while the triplet interaction is odd, $V_{t}(\mathbf{k}',\mathbf{k})=-V_{t}(-\mathbf{k}',\mathbf{k})=-V_{t}(\mathbf{k}',-\mathbf{k})$. The parity can be verified using the relation $\chi(\mathbf{q})=\chi(-\mathbf{q})$. Notably, this property of the susceptibility does not require inversion symmetry; it relies strictly on the translational invariance of the lattice and can be derived via a change of momentum variables in Eq.~\eqref{supp:Eq:simpler_sus}. 

\subsubsection{Helical-metal Fermi surface in two distinct gauge choices}\label{supp:sec:Helical-metal}

\begin{figure}[b]
    \centering
    \includegraphics[width=0.9\linewidth]{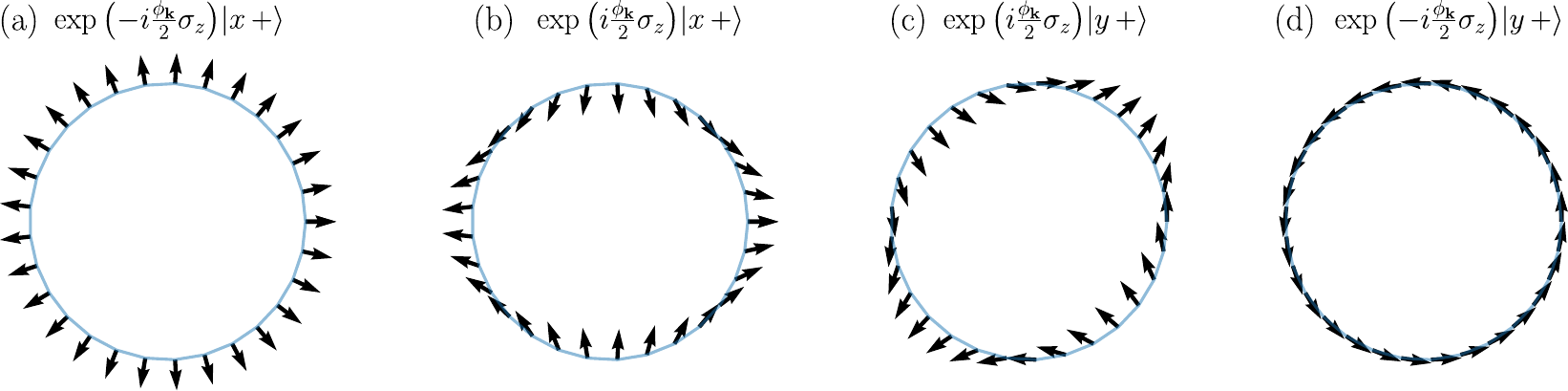}
    \caption{Examples of helical Fermi surfaces in 2D with their respective spin-textures.}
    \label{fig:helical_metal_FS}
\end{figure}

For the case of spin non-degenerate Fermi-surfaces we can no longer consider intra-band singlet/triplet Cooper pairs to be the zero-momentum incoming and outgoing two-particle electronic states. As an example we consider the idealized Fermi surface of a helical metal which preserves time-reversal symmetry as pictured in Fig.~\ref{fig:helical_metal_FS}.
The electronic state at each $\mathbf{k}$-point on the helical Fermi surface pictured in Fig.~\ref{fig:helical_metal_FS}(b) is given by
\begin{align}
    |\mathbf{k}\rangle = \exp{\left(i\frac{\phi_{\mathbf{k}}}{2}\sigma_z\right)}|x\,+\rangle=\frac{1}{\sqrt2}\begin{pmatrix}
        e^{i{\phi_{\mathbf{k}}}/{2}}\\ e^{-i{\phi_{\mathbf{k}}}/{2}}
    \end{pmatrix}\longrightarrow
    \frac{1}{\sqrt2}
    \begin{pmatrix}
        e^{i\phi_{\mathbf{k}}}\\ 1
    \end{pmatrix},\label{supp:eq:state_at_k}
\end{align}
with
\begin{align}
    e^{i\phi_{\mathbf{k}}} = \frac{k_x+ik_y}{\sqrt{k_x^2+k_y^2}}.\label{supp:eq:exp_i_phi_k}
\end{align}
The incoming zero-momentum Cooper pair is constructed by considering a two-particle state composed of the single-particle states at $\mathbf{k}$ and $-\mathbf{k}$. The electronic state at $-\mathbf{k}$ can be constructed  by either applying the time-reversal operator on the state at $\mathbf{k}$, i.e., $\mathcal{T}|\mathbf{k}\rangle$ with $\mathcal{T}=i\sigma_y\mathcal{K}$ or by simply substituting $-\mathbf{k}$ in Eq.~\eqref{supp:eq:state_at_k} and using $e^{i\phi_{-\mathbf{k}}} = e^{i(\phi_{\mathbf{k}} +\pi)}=-e^{i\phi_{\mathbf{k}}}$. As such, the Cooper pair can be written in two gauges $|\Psi_\mathbf{k}\rangle = \mathcal{T}|\mathbf{k}\rangle\otimes|\mathbf{k}\rangle$ or $|\tilde \Psi_\mathbf{k}\rangle = |-\mathbf{k}\rangle\otimes|\mathbf{k}\rangle$. We call the former choice the \emph{time-reversed gauge}. 
Projecting these states onto the interaction tensor Eq.~\eqref{supp:eq:anti_sym_vertex} using the time-reversed gauge yields the real-valued interaction
\begin{align}
    V_{FS}(\mathbf{k}',\mathbf{k}) &= \langle\Psi_\mathbf{k'}| {V}^{qs}_{pr}(\mathbf{k}',\mathbf{k}) |\Psi_\mathbf{k}\rangle = \frac{1}{2}\bigg(V_t(\mathbf{k}',\mathbf{k})\cos(\phi_{\mathbf{k}'}-\phi_{\mathbf{k}}) + V_{s}(\mathbf{k}',\mathbf{k})\bigg)\,, 
    \label{supp:eq:trs_gauge_V_FS}
\end{align}
with
\begin{align}
    |\Psi_\mathbf{k}\rangle &= \frac{1}{2}\left(e^{i\phi_k}|\uparrow\uparrow\rangle +|\uparrow\downarrow\rangle - |\downarrow\uparrow \rangle- e^{-i\phi_k}|\downarrow\downarrow\rangle\right).
\end{align}
In Eq.~\eqref{supp:eq:trs_gauge_V_FS}, $V_s(\mathbf{k}',\mathbf{k})$ and $V_t(\mathbf{k}',\mathbf{k})$ given in Eqs.~\eqref{supp:eq:singlet_interaction} and \eqref{supp:eq:triplet_interaction}.
The expression in Eq.~\eqref{supp:eq:trs_gauge_V_FS} manifestly contains both singlet and triplet projected interactions thereby exhibiting singlet-triplet mixing---forming the basis of our claim that this is a kinematic effect as seen by the interference factor $\cos(\phi_{\mathbf{k}'}-\phi_{\mathbf{k}})$. Eq.~\eqref{supp:eq:trs_gauge_V_FS} is even in momenta since $V_s(\mathbf{k}',\mathbf{k})=V_s(\pm\mathbf{k}',\pm\mathbf{k})$ and $V_t(\mathbf{k}',\mathbf{k})\cos(\phi_{\mathbf{k}'}-\phi_{\mathbf{k}})$ is a product of two functions that are odd in momenta 
\begin{equation}
    \because \cos(\phi_{\mathbf{k}'}-\phi_{\mathbf{-k}}) = \cos(\phi_{\mathbf{k}'}-\phi_{\mathbf{k}}-\pi) = - \cos(\phi_{\mathbf{k}'}-\phi_{\mathbf{k}}) \text{ and similarly for }\mathbf{k}'.
\end{equation}
The other gauge choice results in a Hermitian but complex-valued projected interaction given by 
\begin{equation}
    \widetilde{V}_{FS}(\mathbf{k}',\mathbf{k}) = e^{i(\phi_{\mathbf{k}}-\phi_{\mathbf{k}'})}V_{\rm FS}(\mathbf{k}',\mathbf{k}).
     \label{supp:eq:other_gauge_V_FS}
\end{equation}
Eq.~\eqref{supp:eq:other_gauge_V_FS} is rendered odd in momenta due to the appearance of the momentum-dependent global phase factors. The origin of these can be traced back to the gauge choice in the single-particle electronic state via $\mathcal{T}|\mathbf{k}\rangle = -e^{-i\phi_\mathbf{k}}|-\mathbf{k}\rangle$.

Using the interaction tensor in Eq.~\eqref{supp:eq:anti_sym_vertex} makes things simpler. The spin-orbit coupling required to produce such a Fermi surface would inherently generate off-diagonal Green functions $G^{\uparrow\downarrow}, G^{\downarrow\uparrow}$ and subsequently additional spin-flip terms in the susceptibility $\chi^{\alpha\beta}_{\gamma\delta}(\mathbf{q})$. However, treating the interaction and the non-degenerate Fermi surface separately serves as a proof of principle in illustrating the kinematic origin of singlet-triplet mixing. A consistent SOC treatment would generate more terms without altering the geometric interference effect highlighted here. These terms are included in the calculation presented in the main text.

\subsubsection{Hubbard-Hund-Kanamori interaction}

A similar construction can be carried out for the Hubbard-Hund-Kanamori interaction. We begin by explicitly writing out the anti-symmetrized Hubbard-Kanamori interaction in the basis (rows) $c_q^\dagger c_s$ and (columns) $c_p^\dagger c_r$. With $U'=U-2J$ and $U''=U-3J$,

{
\small
    \begin{equation}
        \Gamma^{qs}_{pr} = \frac{1}{4}
        \left(
        \begin{array}{c|cccc|cccc|cccc|cccc}
         & \multicolumn{4}{c|}{AA} & \multicolumn{4}{c|}{AB} & \multicolumn{4}{c|}{BA} & \multicolumn{4}{c}{BB} \\
         & \uparrow\uparrow & \uparrow\downarrow & \downarrow\uparrow & \downarrow\downarrow & \uparrow\uparrow & \uparrow\downarrow & \downarrow\uparrow & \downarrow\downarrow & \uparrow\uparrow & \uparrow\downarrow & \downarrow\uparrow & \downarrow\downarrow & \uparrow\uparrow & \uparrow\downarrow & \downarrow\uparrow & \downarrow\downarrow \\
        \hline
        A\uparrow, A\uparrow   & 0 & 0 & 0 & U  &   &   &   &   &   &   &   &   & U''& 0 & 0 & U' \\
        A\uparrow, A\downarrow & 0 & 0 & -U & 0 &   &   &   &   &   &   &   &   & 0 & 0 & -J & 0 \\
        A\downarrow, A\uparrow & 0 & -U & 0 & 0 &   &   &   &   &   &   &   &   & 0 & -J & 0 & 0 \\
        A\downarrow, A\downarrow& U & 0 & 0 & 0 &   &   &   &   &   &   &   &   & U' & 0 & 0 & U''\\
        \hline
        A\uparrow, B\uparrow   &   &   &   &   & 0 & 0 & 0 & J & -U''& 0 & 0 & J &   &   &   &   \\
        A\uparrow, B\downarrow &   &   &   &   & 0 & 0 & -J & 0 & 0 & 0 & -U'& 0 &   &   &   &   \\
        A\downarrow, B\uparrow &   &   &   &   & 0 & -J & 0 & 0 & 0 & -U'& 0 & 0 &   &   &   &   \\
        A\downarrow, B\downarrow&   &   &   &   & J & 0 & 0 & 0 & J & 0 & 0 & -U''&   &   &   &   \\
        \hline
        B\uparrow, A\uparrow   &   &   &   &   & -U''& 0 & 0 & J & 0 & 0 & 0 & J &   &   &   &   \\
        B\uparrow, A\downarrow &   &   &   &   & 0 & 0 & -U'& 0 & 0 & 0 & -J & 0 &   &   &   &   \\
        B\downarrow, A\uparrow &   &   &   &   & 0 & -U'& 0 & 0 & 0 & -J & 0 & 0 &   &   &   &   \\
        B\downarrow, A\downarrow&   &   &   &   & J & 0 & 0 & -U''& J & 0 & 0 & 0 &   &   &   &   \\
        \hline
        B\uparrow, B\uparrow   & U''& 0 & 0 & U'&   &   &   &   &   &   &   &   & 0 & 0 & 0 & U \\
        B\uparrow, B\downarrow & 0 & 0 & -J & 0 &   &   &   &   &   &   &   &   & 0 & 0 & -U & 0 \\
        B\downarrow, B\uparrow & 0 & -J & 0 & 0 &   &   &   &   &   &   &   &   & 0 & -U & 0 & 0 \\
        B\downarrow, B\downarrow& U'& 0 & 0 & U''&   &   &   &   &   &   &   &   & U & 0 & 0 & 0 \\
        \end{array}
        \right).
    \end{equation}
}
The Kohn-Luttinger vertex is constructed using Eq.~\eqref{SM:eqn:KL_antisym_vertex}. For simplicity, and to obtain analytical insight, we restrict our attention to 16 of the 256 components of the susceptibility tensor $\chi^{\beta\gamma}_{\delta\alpha}(\mathbf{q})$, where $\alpha,\beta,\gamma,\delta \in \{A,B\}\times\{\uparrow,\downarrow\}$. The selected components are those that do not involve Green functions with spin or orbital flips. This corresponds to considering only susceptibilities of the form $\chi^{\alpha\beta}_{\beta\alpha}(\mathbf{q})$, yielding 16 terms in total. We group these into two classes: (i) those for which the orbital indices within the combined labels $\alpha,\beta$ coincide, denoted $\chi_{\rm intra}(\mathbf{q})$, and (ii) those for which the orbital indices differ, denoted $\chi_{\rm inter}(\mathbf{q})$. Additionally, for analytical tractability we make a further simplification by ignoring fluctuation contributions to the orbital-exchange and pair-hopping terms and considering only their bare contributions $-J$ and $J$ respectively. Within these simplifications the anti-symmetrized Kohn-Luttinger vertex is thereby

{
\small
\begin{equation}
{V}^{qs}_{pr}(\mathbf{k}',\mathbf{k}) = \left(
\begin{array}{c|cccc|cccc|cccc|cccc}
 & \multicolumn{4}{c|}{AA} & \multicolumn{4}{c|}{AB} & \multicolumn{4}{c|}{BA} & \multicolumn{4}{c}{BB} \\
 & \uparrow\uparrow & \uparrow\downarrow & \downarrow\uparrow & \downarrow\downarrow & \uparrow\uparrow & \uparrow\downarrow & \downarrow\uparrow & \downarrow\downarrow & \uparrow\uparrow & \uparrow\downarrow & \downarrow\uparrow & \downarrow\downarrow & \uparrow\uparrow & \uparrow\downarrow & \downarrow\uparrow & \downarrow\downarrow \\
\hline
A\uparrow, A\uparrow   & T & 0 & 0 & D_U  &   &   &   &   &   &   &   &   & D_{U''}& 0 & 0 & D_{U'} \\
A\uparrow, A\downarrow & 0 & 0 & X_U & 0 &   &   &   &   &   &   &   &   & 0 & 0 & -\frac{J}{4} & 0 \\
A\downarrow, A\uparrow & 0 & X_U & 0 & 0 &   &   &   &   &   &   &   &   & 0 & -\frac{J}{4} & 0 & 0 \\
A\downarrow, A\downarrow& D_U & 0 & 0 & T &   &   &   &   &   &   &   &   & D_{U'} & 0 & 0 & D_{U''}\\
\hline
A\uparrow, B\uparrow   &   &   &   &   & 0 & 0 & 0 & \frac{J}{4} & X_{U''}& 0 & 0 & \frac{J}{4} &   &   &   &   \\
A\uparrow, B\downarrow &   &   &   &   & 0 & 0 & -\frac{J}{4} & 0 & 0 & 0 & X_{U'}& 0 &   &   &   &   \\
A\downarrow, B\uparrow &   &   &   &   & 0 & -\frac{J}{4} & 0 & 0 & 0 & X_{U'}& 0 & 0 &   &   &   &   \\
A\downarrow, B\downarrow&   &   &   &   & \frac{J}{4} & 0 & 0 & 0 & \frac{J}{4} & 0 & 0 & X_{U''}&   &   &   &   \\
\hline
B\uparrow, A\uparrow   &   &   &   &   & X_{U''}& 0 & 0 & \frac{J}{4} & 0 & 0 & 0 & \frac{J}{4} &   &   &   &   \\
B\uparrow, A\downarrow &   &   &   &   & 0 & 0 & X_{U'}& 0 & 0 & 0 & -\frac{J}{4} & 0 &   &   &   &   \\
B\downarrow, A\uparrow &   &   &   &   & 0 & X_{U'}& 0 & 0 & 0 & -\frac{J}{4} & 0 & 0 &   &   &   &   \\
B\downarrow, A\downarrow&   &   &   &   & \frac{J}{4} & 0 & 0 & X_{U''}& \frac{J}{4} & 0 & 0 & 0 &   &   &   &   \\
\hline
B\uparrow, B\uparrow   & D_{U''}& 0 & 0 & D_{U'}&   &   &   &   &   &   &   &   & T & 0 & 0 & D_U \\
B\uparrow, B\downarrow & 0 & 0 & -\frac{J}{4} & 0 &   &   &   &   &   &   &   &   & 0 & 0 & X_U & 0 \\
B\downarrow, B\uparrow & 0 & -\frac{J}{4} & 0 & 0 &   &   &   &   &   &   &   &   & 0 & X_U & 0 & 0 \\
B\downarrow, B\downarrow& D_{U'}& 0 & 0 & D_{U''}&   &   &   &   &   &   &   &   & D_U & 0 & 0 & T \\
\end{array}
\right).
\label{supp:Eq:Hubbard-Hund-Kanamori_Kohn-Luttinger-vertex}
\end{equation}
}
Where we make the following definitions
{
\small
\begin{align}
    T &= \frac{U^2 + (U')^2 + (U'')^2}{8} \left[\chi_{\rm intra}(\mathbf{k}-\mathbf{k}') - \chi_{\rm intra}(\mathbf{k}+\mathbf{k}')\right], \nonumber\\
    \nonumber\\
    D_U &= \frac{U}{4} - \frac{U^2}{8} \chi_{\rm intra}(\mathbf{k}+\mathbf{k}') + \frac{U' U''}{4} \chi_{\rm intra}(\mathbf{k}-\mathbf{k}'), \qquad
    X_U = -\frac{U}{4} + \frac{U^2}{8} \chi_{\rm intra}(\mathbf{k}-\mathbf{k}') - \frac{U' U''}{4} \chi_{\rm intra}(\mathbf{k}+\mathbf{k}'), \nonumber\\
    \nonumber\\
    D_{U'} &= \frac{U'}{4} - \frac{(U')^2}{8} \chi_{\rm inter}(\mathbf{k}+\mathbf{k}') + \frac{U U''}{4} \chi_{\rm intra}(\mathbf{k}-\mathbf{k}'),\qquad
    X_{U'} = -\frac{U'}{4} + \frac{(U')^2}{8} \chi_{\rm inter}(\mathbf{k}-\mathbf{k}') - \frac{U U''}{4} \chi_{\rm intra}(\mathbf{k}+\mathbf{k}'), \nonumber\\
    \nonumber\\
    D_{U''} &= \frac{U''}{4} - \frac{(U'')^2}{8} \chi_{\rm inter}(\mathbf{k}+\mathbf{k}') + \frac{U U'}{4} \chi_{\rm intra}(\mathbf{k}-\mathbf{k}'), \qquad
    X_{U''} = -\frac{U''}{4} + \frac{(U'')^2}{8} \chi_{\rm inter}(\mathbf{k}-\mathbf{k}') - \frac{U U'}{4} \chi_{\rm intra}(\mathbf{k}+\mathbf{k}').
\end{align}
}
For the Femi-surface projected interaction we simply extend Eq.~\eqref{supp:eq:state_at_k} by considering a constant orbital texture
\begin{equation}
     |\mathbf{k}\rangle = \begin{pmatrix}
         a\\b
     \end{pmatrix}\otimes\frac{1}{\sqrt2}
    \begin{pmatrix}
        e^{i\phi_{\mathbf{k}}}\\ 1
    \end{pmatrix},\quad\text{with $|a|^2+|b|^2=1$}.
\end{equation}
Constructing the two-particle Cooper pair state in the time-reversed gauge $|\Psi_\mathbf{k}\rangle = \mathcal{T}|\mathbf{k}\rangle\otimes|\mathbf{k}\rangle$ where $\mathcal{T}=i\tau_0\sigma_y\mathcal{K}$ and projecting onto Eq.~\eqref{supp:Eq:Hubbard-Hund-Kanamori_Kohn-Luttinger-vertex} yields the even in momenta, real-valued interaction
\begin{align}
            V_{\rm FS}(\mathbf{k}',\mathbf{k}) &=   \frac{|a|^4 + |b|^4}{2} \Big[ T \cos(\phi_{\mathbf{k}} - \phi_{\mathbf{k}'}) + (D_U - X_U) \Big]  \nonumber \\
            &\qquad + |ab|^2 \Big[ (D_{U''} + X_{U''}) \cos(\phi_{\mathbf{k}} - \phi_{\mathbf{k}'}) + (D_{U'} - X_{U'}) + J \Big].
            \label{supp:eq:final_VFS_real}
\end{align}
Note that setting $b\to0$ and the bare interactions $U',U'',J \to 0$ in Eq.~\eqref{supp:eq:final_VFS_real} recovers the single-orbital result
Eq.~\eqref{supp:eq:trs_gauge_V_FS}. However, in the presence of inter-orbital and Hund's repulsion Eq.~\eqref{supp:eq:final_VFS_real} differs from Eq.~\eqref{supp:eq:trs_gauge_V_FS} in an important manner. The term $D_U - X_U$ turns out to become attractive instead of repulsive like in the single-orbital case at higher bare repulsion $U$ (recall $U'=U-2J$ and $U''=U-3J$ )
\begin{align}
    D_U-X_U&=\frac{U}{2}+\left(\frac{U'U''}{4}-\frac{U^2}{8}\right)\big(\chi_{\rm intra}(\mathbf{k}+\mathbf{k}') + \chi_{\rm intra}(\mathbf{k}-\mathbf{k}') \big)\nonumber\\
    &\approx \frac{U}{2}+\frac{U^2}{8}\,\underbrace{\big(\chi_{\rm intra}(\mathbf{k}+\mathbf{k}') + \chi_{\rm intra}(\mathbf{k}-\mathbf{k}') \big)}_{\text{negative}}\quad\text{for $J\to0$.}
\end{align}
As such, terms of this kind contribute to the $s$-wave solution being the leading solution at high values of the bare repulsion $U$ as seen in the Fig. 4 of the main-text.

\subsection{Spin-representation $\Delta_{ss'}(\mathbf{k})$}
In this subsection, we project the scalar gap function calculated in the band basis $\Delta(\mathbf{k})$ for the single band crossing the Fermi-level onto the physical basis of the top layer to see how pairing is distributed among the the orbitals $m, n \in \{A, B\}$ and spins $s, s' \in \{\uparrow, \downarrow\}$. The projection onto the physical basis forms the $16$ component tensor:
\begin{align}
    \Delta_{ms,ns'}(\mathbf{k}) = \Delta(\mathbf{k}) \, u_{ms}(\mathbf{k}) \, u_{ns'}(-\mathbf{k}),
\end{align} 
where $u_{ms}(\mathbf{k})$ is an eigenvector component corresponding to state crossing the Fermi level at $\mathbf{k}$ and the state at $-\mathbf{k}$ is constructed using $\mathcal{T}=i\tau_0\sigma_y\mathcal{K}$.
Focusing on the spin-content of the Cooper pair wavefunction, we sum over the orbitals. This leads to the spin-representation $ \Delta_{ss'}(\mathbf{k})$ written as
\begin{align}
    \text{Spin representation: }& \Delta_{ss'}(\mathbf{k}) = \Delta(\mathbf{k})\sum_{mn}u_{ms}(\mathbf{k})\,u_{ns'}(-\mathbf{k}).
\end{align}

\begin{figure}[h]
    \centering
    \includegraphics[width=0.66\linewidth]{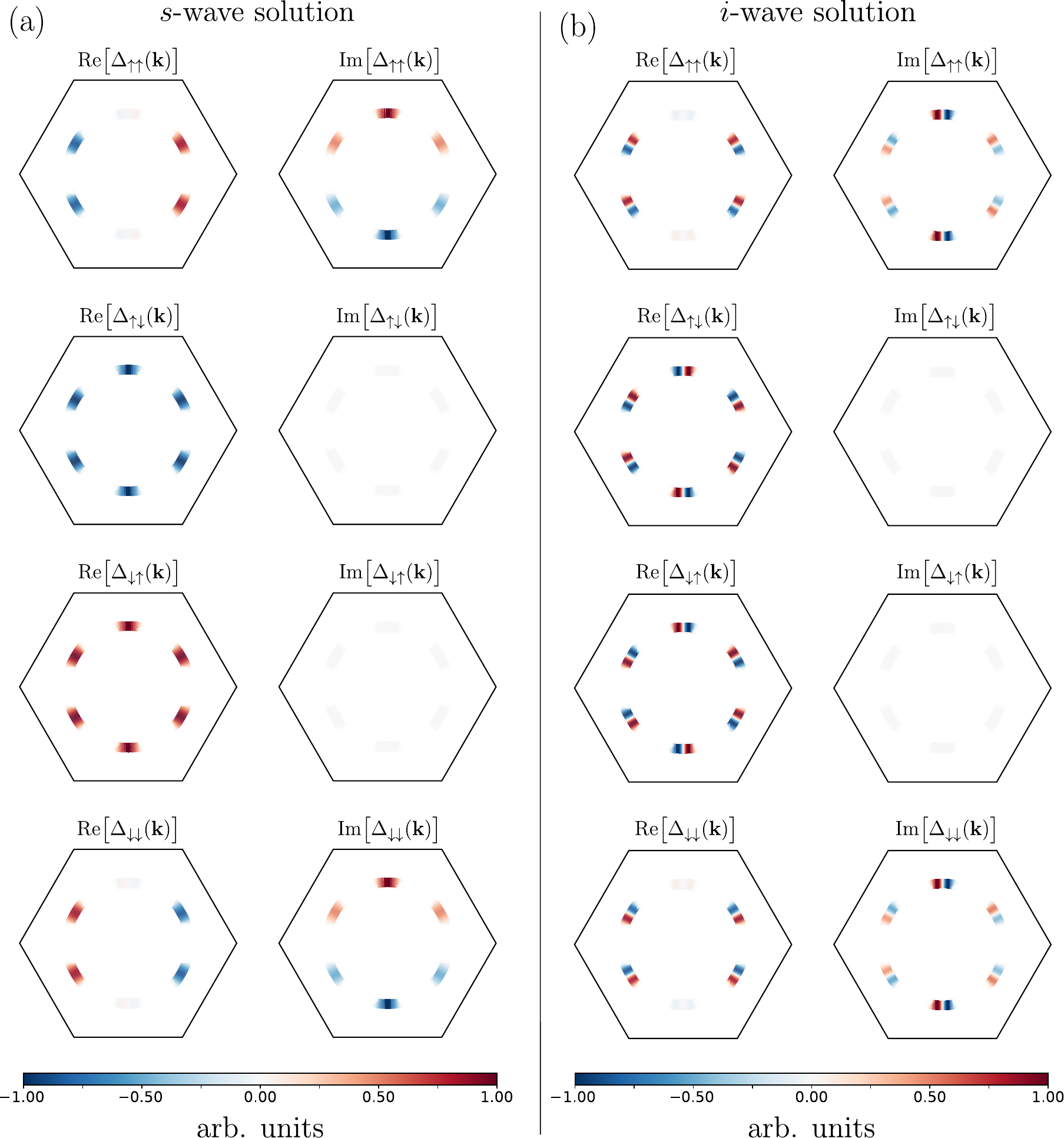}
    \caption{Spin-representation $\Delta_{ss'}(\mathbf{k})$ for the (a) $s$-wave and (b) $i$-wave solutions.}
    \label{supp:fig:spin_rep_s_and_i}
\end{figure}

In Fig.~\ref{supp:fig:spin_rep_s_and_i} we show the spin-representation of the gap functions in the $s$-wave and $i$-wave state. The gap components pictured can be understood using the example of a single-orbital helical metal. Consider the state in Eq.~\eqref{supp:eq:state_at_k}, using $\Delta_{ss'}(\mathbf{k})=\Delta(\mathbf{k})\,u_s(\mathbf{k})\,u_{s'}(-\mathbf{k})$ we have
\begin{equation}
    \Delta_{\uparrow\uparrow}(\mathbf{k}) = \frac{\Delta(\mathbf{k})}{2}e^{i\phi_\mathbf{k}},\qquad
    \Delta_{\uparrow\downarrow}(\mathbf{k}) = -\frac{\Delta(\mathbf{k})}{2},\qquad
    \Delta_{\downarrow\uparrow}(\mathbf{k}) = \frac{\Delta(\mathbf{k})}{2},\qquad
    \Delta_{\downarrow\downarrow}(\mathbf{k}) = -\frac{\Delta(\mathbf{k})}{2}e^{-i\phi_\mathbf{k}}.
    \label{supp:eq:spin_rep_helical_metal}
\end{equation}
From Fig.~\ref{supp:fig:spin_rep_s_and_i} and Eq.~\eqref{supp:eq:spin_rep_helical_metal} we find that the pairing function projected onto the spin-$\hat z$ basis is comprised of a singlet state and two equal spin pairing triplet states with opposite chirality. The opposite spin-pairing triplet state is absent. 
The singlet-triplet mixing evident in Fig.~\ref{supp:fig:spin_rep_s_and_i}(b), when the superconducting order parameter is written in the spin-$\hat{z}$ basis, is consistent with the state discussed in Refs.~\cite{Changdar2025Topological,Waje2025Ginzburg-Landau}.

\newpage
\section{Parameter dependencies of the microscopic phase diagram}

\subsection{Leading eigenvalues of the Kohn-Luttinger approach}

\begin{figure}[h]
    \centering
    \includegraphics[width=0.85\linewidth]{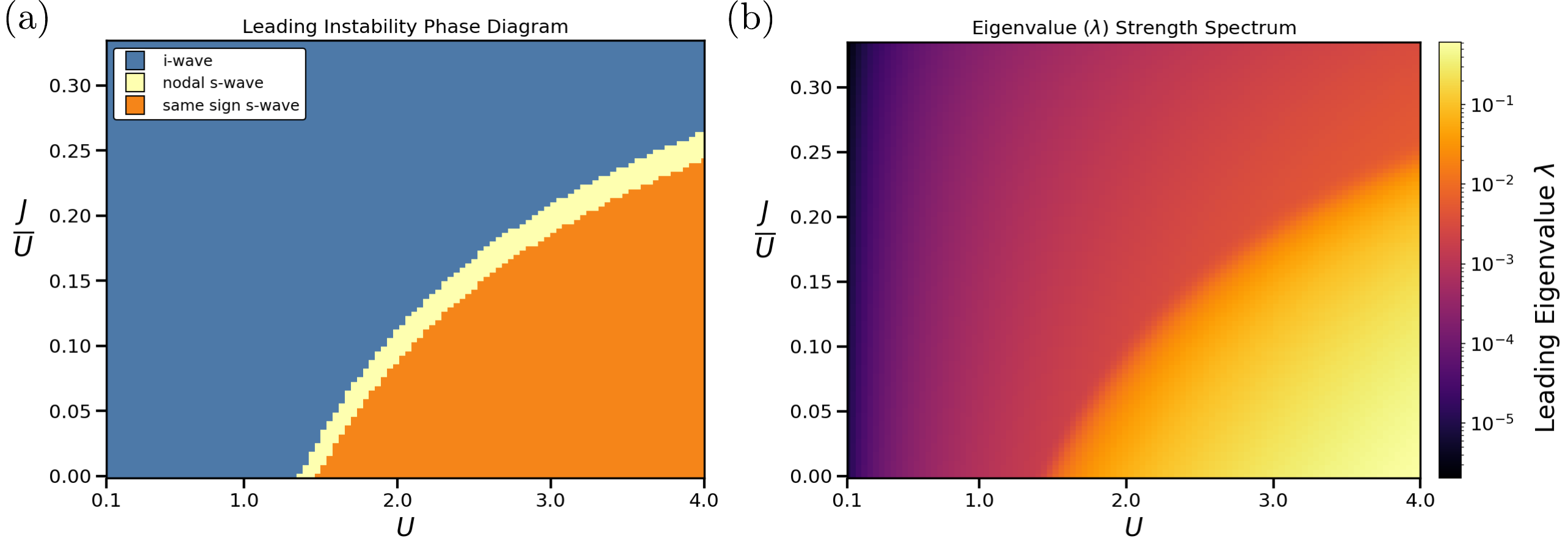}
    \caption{(a) Phase diagram for the base parameters at $T=0.001$ and $\mu = 0.005$. (b) Heatmap of the leading eigenvalues for the linearized gap equation calculation with the base parameters.}
    \label{supp:fig:base_params}
\end{figure}

In Fig.~\ref{supp:fig:base_params} we show how the eigenvalues develop as a function of the interactions $U$ and $J/U$. As expected, for small values of the interactions, the eigenvalues are small and increase as the interactions are increased. Fig.~\ref{supp:fig:base_params}(a) shows the phase diagram for $\mu=0.005$, also included in the main text, while Fig.~\ref{supp:fig:base_params}(b) shows a logarithmic heatmap of the leading eigenvalue.

\subsection{Phase diagrams for small changes in spin-orbit coupling and inversion breaking terms}

\begin{figure}[h]
    \centering
    \includegraphics[width=0.75\linewidth]{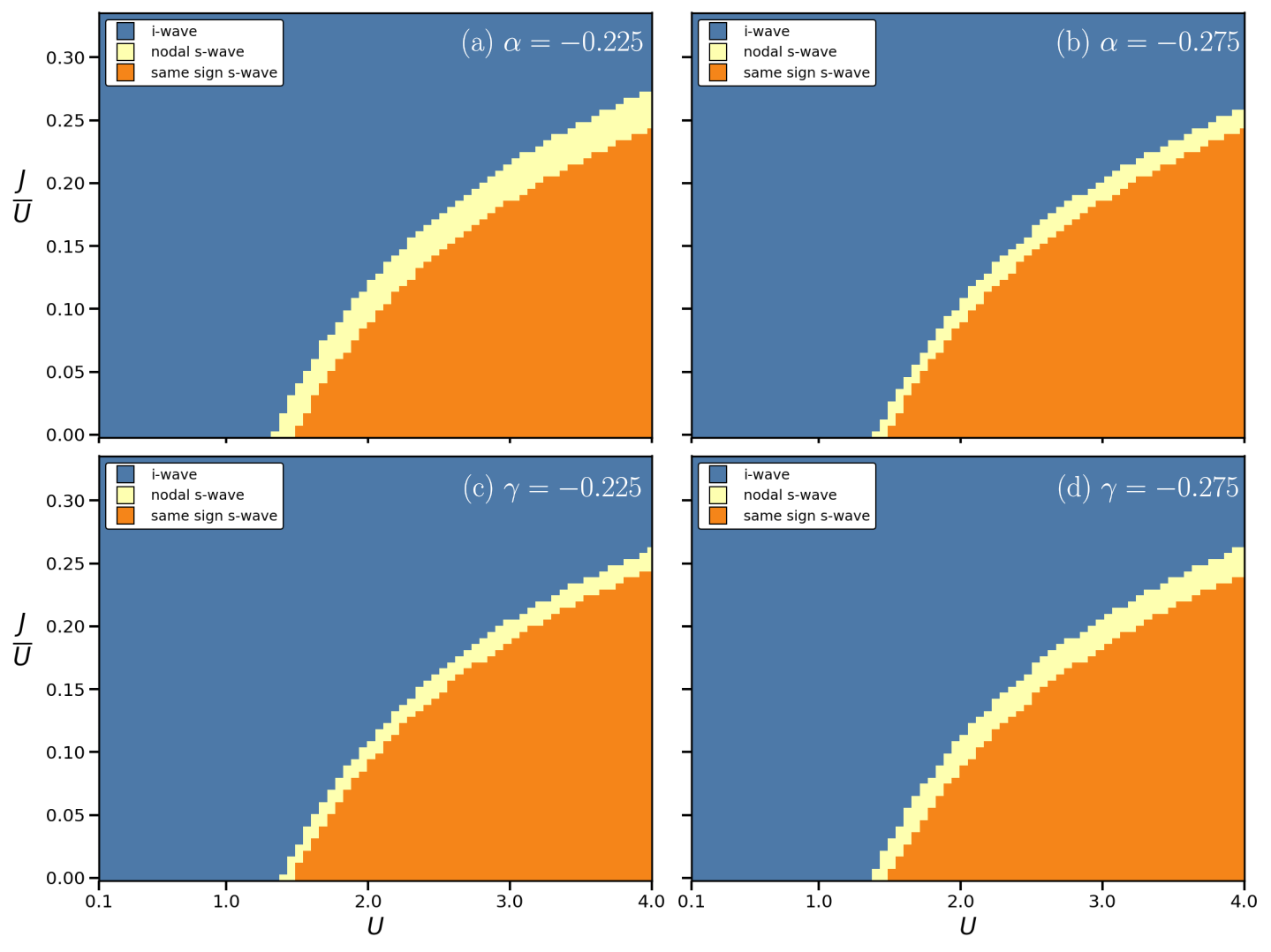}
    \caption{Phase diagrams of the linearized gap equation calculation performed at temperature $T=0.001$ and chemical potential $\mu=0.005$. (a) $\alpha=-0.225$  (b) $\alpha=-0.275$ (c) $\gamma=-0.225$ (d) $\gamma=-0.275$. In each of the figures all unmentioned parameters are set to their base values as listed in the End Matter.}
    \label{supp:fig:small_changes}
\end{figure}

As presented in the End Matter of the main text we define the base parameters of the tight binding model to be $m=0.4$, $\beta=-0.65$, $\lambda=1.2$, $\alpha=-0.25$, $\gamma=-0.25$ following Ref.~\cite{Vocaturo2024Electronic}. We find that varying the spin-orbit coupling $\alpha$ and inversion breaking $\gamma$ terms to within $\pm10\%$ of their respective base values does not result in significant changes to the phase diagrams, see Fig.~\ref{supp:fig:small_changes}. Decreasing the value of spin-orbit-coupling $\alpha$ seemingly increases the width of the region representing the nodal $s$-wave state whereas for the inversion breaking term $\gamma$ the opposite is observed.

\clearpage
\newpage
\subsection{Phase diagram for $\mu<0$}
\begin{figure}[h]
    \centering
    \includegraphics[width = \linewidth]{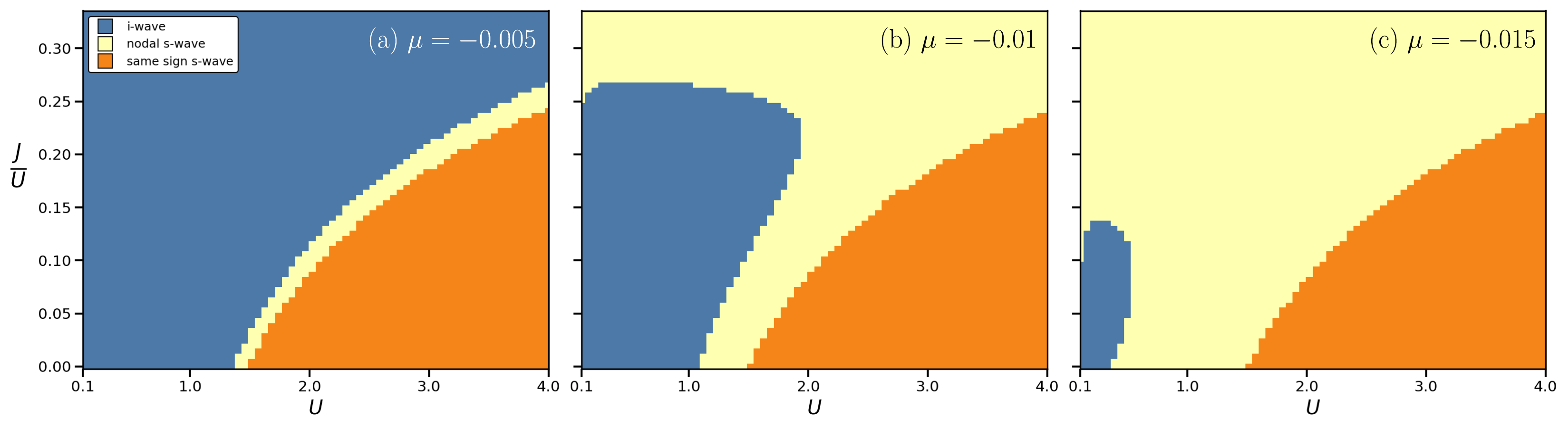}
    \caption{Phase diagrams for the base parameters mentioned in the end-matter of the main text for (a) $\mu=-0.005$ (b) $\mu=-0.01$ (c) $\mu=-0.015$}
    \label{supp:fig:negative_chem_pot}
\end{figure}

As mentioned in the main-text, in Fig.~\ref{supp:fig:negative_chem_pot} we show that the phase diagrams for chemical potentials $\mu<0$ display the same behavior.

\subsection{$g$-wave order parameter on increasing arc curvature via third nearest neighbour hopping }
The shapes of the Fermi arcs found in ARPES studies and first-principles based calculations of PtBi$_2$ exhibit significant dependence on the surface termination considered \cite{Mathisen2026Fermiology, Vocaturo2024Electronic}. Two kinds of surface terminations have been observed, the first being the Kagome-Type (KT) and the second being the Decorated Honeycomb (DH). The former surface termination is characterized by the presence of flatter Fermi arcs whereas the latter displays arcs that are curved in a horseshoe-like manner. While the minimal tight-binding model used in our calculations cannot distinguish between these two kinds of surface terminations, a recent study found that adding a relatively large in-plane third-nearest neighbor hopping, $t''$, results in arcs that mimic the curved Fermi arcs of the DH termination~\cite{Vocaturo2026Engineering}. This is achieved by adding
\begin{equation}
    h^{\rm third}_{\rm NN}(\mathbf{k}) = -t'' \big(\cos 2k_1 + \cos2k_2 + \cos2(k_1+k_2)\,\big)\Gamma_1
    \label{supp:eq:third_nn}
\end{equation}
to the tight-binding model discussed in Sec.~\ref{supp:sec:slab_model}

\begin{figure}[h]
    \centering
    \includegraphics[width=\linewidth]{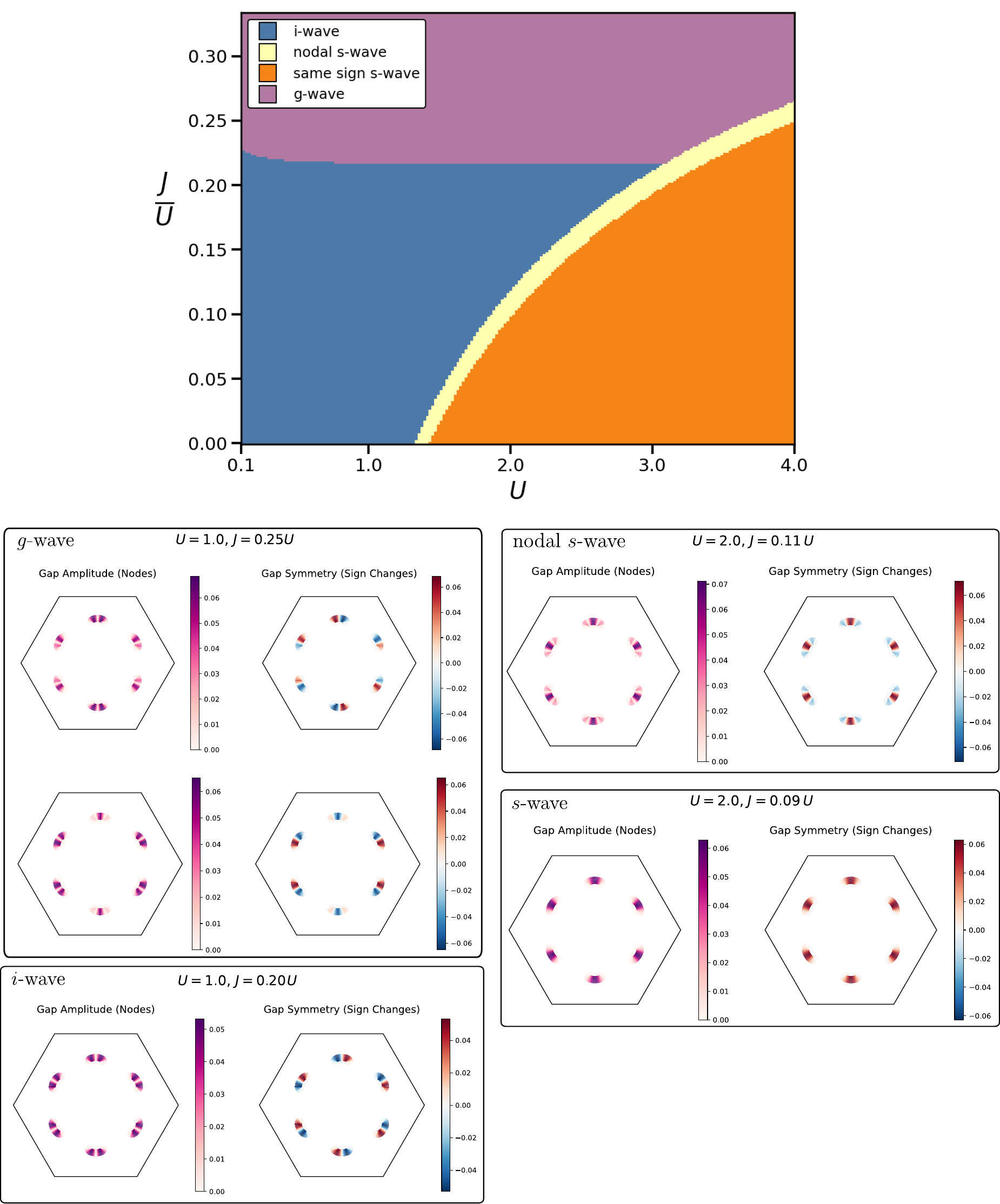}
    \caption{Phase diagram of the linearized gap equation calculation at temperature $T=0.001$ and chemical potential $\mu=0.02$ along with some representative solutions. The parameters used in the calculation $m=0.4$, $\beta=-0.65$, $\lambda=1.2$, $\alpha=-0.35$, $\gamma=-0.25$ and $t''=0.35$.}
    \label{supp:fig:g-wave_phase_diagram}
\end{figure}

\begin{figure}[h]
    \centering
    \includegraphics[width=\linewidth]{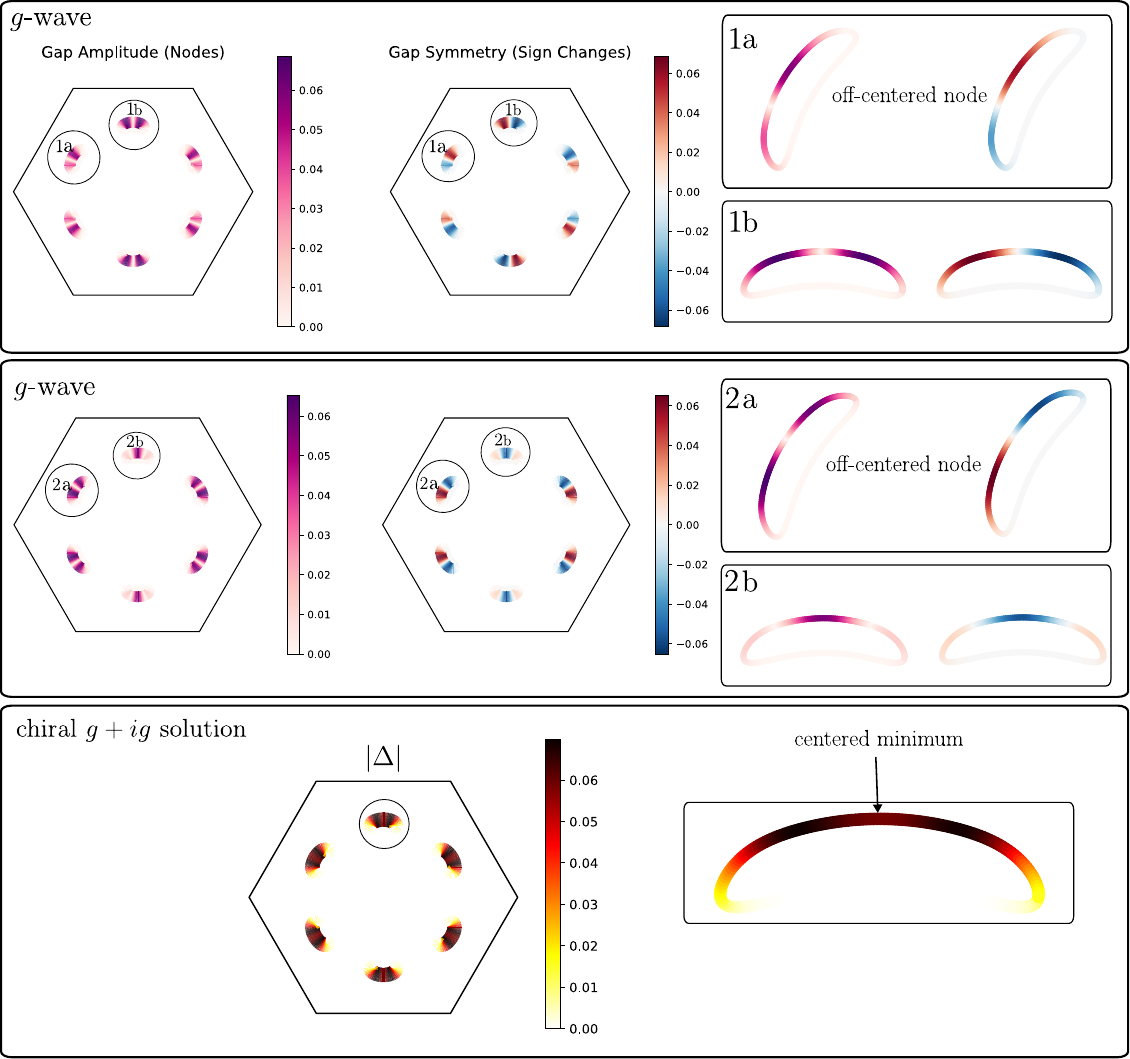}
    \caption{The top and middle panels depict the degenerate real-valued $g$-wave solutions; the insets depict zoomed-in versions of the gap on the Fermi arcs: 1a and 2a exhibit off-centred nodes. The bottom panel shows the absolute value of the time-reversal symmetry breaking chiral $g+ig$ state which exhibits a minima at the Fermi arc center.}
    \label{supp:fig:g-wave_arcs}
\end{figure}

We find that the increased arc curvature as a consequence of adding this term results in the appearance of a two-fold degenerate $g$-wave solution in the phase diagram pictured in Fig.~\ref{supp:fig:g-wave_phase_diagram}. In Fig.~\ref{supp:fig:g-wave_arcs} we show zoomed in pictures of the $g$-wave solution along with the chiral $g+ig$ solution. We find that some Fermi arcs of the non-chiral states exhibit off-centered nodes. The chiral $g+ig$ combination is rotationally symmetric and exhibits a minimum at the center of each Fermi arc. 


\end{document}